\documentclass{aa}  

\usepackage{graphicx}
\usepackage{txfonts}
\usepackage{upquote} % keeps correct quotes in verbatim so user can copy-paste queries in appendix
\usepackage[]{hyperref}
\usepackage{array}

\def\deg{\ensuremath{^\circ}}

\newcommand\gdr[1]{\gaia~DR#1}
\newcommand\egdr[1]{\gaia~EDR#1}

\newcommand{\gaia}{Gaia\xspace}
\newcommand{\hip}{Hipparcos\xspace}

\def\gmag{\ensuremath{G}\xspace}

\def\a0{\ensuremath{A_{\rm 0}}\xspace}

\providecommand{\M}{\ensuremath{\,{\rm M}}\xspace}
\providecommand{\Mcomp}{\ensuremath{\,{\rm \tilde{M}}_{\rm c}}\xspace}

\providecommand{\Mjup}{\ensuremath{\,{\rm M}_{\rm J}}\xspace}
\providecommand{\chisq}{\ensuremath{\,{\chi^2}}\xspace}

\newcommand{\typeSBone}{`SB1'\xspace}

\newcommand{\typeOrb}{`Orbital'\xspace}
\newcommand{\typeOrbAlt}{`OrbitalAlternative'\xspace}
\newcommand{\typeOrbAltVal}{`OrbitalAlternativeValidated'\xspace}
\newcommand{\typeOrbAltPVal}{`OrbitalAlternative[Validated]'\xspace}
\newcommand{\typeOrbAltStar}{`OrbitalAlternative*'\xspace}
\newcommand{\typeOrbTar}{`OrbitalTargetedSearch'\xspace}
\newcommand{\typeOrbTarVal}{`OrbitalTargetedSearchValidated'\xspace}
\newcommand{\typeOrbTarPVal}{`OrbitalTargetedSearch[Validated]'\xspace}
\newcommand{\typeOrbTarStar}{`OrbitalTargetedSearch*'\xspace}

\newcommand{\nssSolutionType}{\texttt{nss\_solution\_type}\xspace}
\newcommand{\nssTwoBodyOrbit}{\texttt{nss\_two\_body\_orbit}\xspace}
\newcommand{\nssAccelerationAstro}{\texttt{nss\_acceleration\_astro}\xspace}
\newcommand{\nssNonLinearSpectro}{\texttt{nss\_non\_linear\_spectro}\xspace}
\newcommand{\nssVimFl}{\texttt{nss\_vim\_fl}\xspace}

\usepackage{xcolor}

\newcommand\MOD[1]{{#1}}

\usepackage[normalem]{ulem} % for \sout support in intermediate version of paper

\begin{document} 

    \title{\gaia DR3 astrometric orbit determination\\ with Markov Chain Monte Carlo and Genetic Algorithms}
    \subtitle{Systems with stellar, substellar, and planetary mass companions}
  
   \titlerunning{\gaia DR3 astrometric orbit determination with MCMC and Genetic Algorithms}

   %Markov Chain Monte Carlo and Genetic Algorithms

   \author{B.~Holl\inst{\ref{inst1},\ref{inst2}}
            \fnmsep\thanks{Corresponding author: B. Holl (\href{mailto:berry.holl@unige.ch}{\tt berry.holl@unige.ch})},
          A.~Sozzetti\inst{\ref{inst3}},
          %\andhttps://www.overleaf.com/project/61766034d55467ae3026dfcb
          J.~Sahlmann\inst{\ref{inst4}},
          P.~Giacobbe\inst{\ref{inst3}},
          D.~S\'egransan\inst{\ref{inst1}},
          N.~Unger\inst{\ref{inst1}},
          J.-B.~Delisle\inst{\ref{inst1}},
          %\and 
          D.~Barbato\inst{\ref{inst1},\ref{inst3}},
           M.G.~Lattanzi\inst{\ref{inst3}},
           R.~Morbidelli\inst{\ref{inst3}},
%          \WORKSUGGESTION{[author to be added by torino]}
            \and 
          D.~Sosnowska\inst{\ref{inst1}}
          }

    \institute{
    Department of Astronomy, University of Geneva, Chemin Pegasi 51, CH-1290 Versoix, Switzerland\label{inst1}%\\
    \and 
    Department of Astronomy, University of Geneva, Ch. d'Ecogia 16, CH-1290 Versoix, Switzerland\label{inst2}
    \and 
    INAF - Osservatorio Astrofisico di Torino, Via Osservaorio 20, I- 10025 Pino Torinese, Italy\label{inst3}
    \and 
    RHEA Group for the European Space Agency (ESA), European Space Astronomy Centre (ESAC),\\ Camino Bajo del Castillo s/n, 28692 Villanueva de la Ca\~nada, Madrid, Spain\label{inst4}
    }
    \authorrunning{Holl et al.}

% \abstract{}{}{}{}{} 
% 5 {} token are mandatory
 
  \abstract
   {%\WORKSUGGESTION{(1) Let's use British English throughout (I set the latex spell checker also to this). (2) I removed all `DU437' occurrences, though should we mention the name somewhere? }
   Astrometric discovery of sub-stellar mass companions orbiting stars is exceedingly hard due to the required sub-milliarcsecond precision, limiting the application of this technique to only a few instruments on a target-per-target basis as well as the global astrometry space missions \hip and \gaia. %(which was able to tap into the brown-dwarf regime). but only with the onset of \gaia it will be possible to open up the astrometric low-mas companion regime down to planetary masses.
   The third \gaia data release (\gdr{3}) includes the first \gaia astrometric orbital solutions, whose sensitivity in terms of estimated companion mass extends down into the planetary-mass regime. 
   %The astrometric space mission \hip was able to tap into the brown-dwarf regime, but only with the onset of \gaia it will be possible to open up the astrometric low-mas companion regime down to planetary masses.
   %specialised ground-based telescope systems
   %Astrometric detections of sub-stellar mass components have been pretty rare up till now due to the extreme angular precision requirements (order $\sim$mas), only found on specialised ground-based telescope systems or dedicated space-based missions like \hip and \gaia.
   %Though there is currently a plenitude of planet and brown dwarf companion detections using the radial velocity and transit detection methods, astrometry has so far remained an underrepresented method as the needed precision is very hard to achieve from ground and dedicated astrometric space missions like \hip and \gaia are exceedingly rare.}
   }
   {%The third \gaia data release (\gdr{3}) includes the first \gaia astrometric orbital solutions, extending into the sub-stellar mass regime down to planetary mass. 
   We present the contribution of the `exoplanet pipeline' to the \gdr{3} sample of astrometric orbital solutions by describing the methods used for fitting the orbits, the identification of significant solutions, and their validation. We then present an overview of the statistical properties of the solution parameters. %, to identify and fit the companion orbits, the validation of the published sample, and discussion of the orbital parameters. %Precise companion mass estimates require additional information and are presented in \cite{DR3-DPACP-100}.
   %, showcasing what is, and will be, possible with \gaia astrometry.
   %The long history and of astrometric planet detection might finally come to fruition. ref to planet and BD papers with estimates.
   }
   {Using both a Markov Chain Monte Carlo and Genetic Algorithm we fit the 34~months of \gdr{3} astrometric time series with a single Keplerian astrometric-orbit model that has 12 free parameters and an additional jitter term, and retain the solutions with the lowest~\chisq. Verification and validation steps are taken using significance tests, internal consistency checks using the \gaia radial velocity measurements (when available), as well as literature radial velocity and astrometric data, leading to a subset of candidates that are labelled as `validated'.}
   {
   % total: 1162
   % 629 OrbitalAlternative(Validated) + 533 
   % OrbitalTargetedSearch(Validated).
   % OrbitalAlternativeValidated = 10, 
   % OrbitalTargetedSearchValidated = 188
   % mp_MJ_lt20_bool && kept = 17      of which 9 Validated
   % mp_MJ_20_120_bool && kept = 52    of which 29 Validated
   % mp_MJ_Gt120_bool && kept = 1093   of which 160 Validated
   We determined astrometric-orbit solutions for 1162 sources and 198 solutions have been assigned the `validated' label. 
   % COMMENTED OUT AND MOVED TO CONCLUSION, TOO DETAILED FOR HERE WITHOUT CONTEXT:
   %They are subdivided in the \gdr{3} \nssTwoBodyOrbit archive table into four \nssSolutionType: 629~\typeOrbAltPVal of which 10 validated and 533 \typeOrbTarPVal of which 188 validated.
   Precise companion mass estimates require external information and are presented elsewhere. %in \citet{DR3-DPACP-100}.
      %Though no calibrated companion mass \Mcomp estimates are derived in this paper, the number of candidates (and validated numbers in parenthesis), can be roughly divided into: 
    To broadly categorise the different mass regimes in this paper we use the pseudo-companion mass \Mcomp assuming a solar-mass host %(\Mhost=1\Msun)
    and define three solution groups:
   17 (9 validated) solutions with companions in the planetary-mass regime (\Mcomp~<~20~\Mjup), 52 (29 validated) in the brown dwarf regime (20~\Mjup~$\leq$~\Mcomp $\leq 120$~\Mjup), and 1093 (160 validated) in the low mass stellar companion regime (\Mcomp~>~120~\Mjup). 
   From internal and external verification and validation we estimate the level of spurious/incorrect solutions in our sample to be of the order of $\sim5\%$ and $\sim10\%$ in the \typeOrbAlt and \typeOrbTar candidate sample, respectively.
   %Based on verification and validation steps we estimate a degree of contamination of $\sim5\%$ and $\sim10\%$ in the \typeOrbAlt and \typeOrbTar sample, respectively. 
   }
   {We demonstrate that \gaia is able to confirm and sometimes refine known orbital companion orbits as well as identify new candidates, providing us with a positive outlook of the expected harvest from the full mission data in future data releases. 
}

%   ‘OrbitalAlternative’  (619),  ‘OrbitalAlternativeVali-dated’ (10), ‘OrbitalTargetedSearch’ (385), and ‘OrbitalTarget-edSearchValidated’ (177).

    \keywords{
    %Choose 6 max:
    -- astrometry 
    -- planets and satellites: detection
   %-- Techniques: photometric 
    %-- stars: general 
    %-- space vehicles: instruments
    % ADD BINARIES <--- lookup keyword
    %-- Stars: oscillation 
    %-- Stars: solar-type 
    %-- Stars:variables: general
    %-- Galaxy: stellar content
    %-- Stars: general, statistics, oscillations, activity, rotation, solar-type, starspots, supergiants % subdwarfs,
     %     Stars:variables: general, Cepheids, RR Lyrae, delta Scuti  %T Tauri, Herbig Ae/Be, S Doradus , Hertzsprung-Russell and C-M diagrams , horizontal-branch,  AGB and post-AGB, carbon, white dwarfs, Wolf-Rayet
          %-- binaries: eclipsing 
     %     -- Surveys -- Methods: analytical, statistical, data analysis
      %    -- Techniques: photometric
       %   -- Astronomical databases: Catalogs, Surveys
       %   Galaxy: stellar content, bulge, center, disk, halo %,general
    -- (stars:) brown dwarfs
    -- (stars:) binaries: general  
    -- Catalogs
    -- Techniques: radial velocities 
    }

   \maketitle
%

%------------------------------------------------------------------
% INTRODUCTION
%------------------------------------------------------------------

\section{Introduction \label{sec:intro}}

%{\bf Alessandro's suggested text:}

%\TODO{[add labels of the two different samples and point out sub-stellar companions in Orbital pipeline]}

The third \gaia data release \citep[\gdr{3},][]{DR3-DPACP-185} \MOD{is the first release that includes} non-single star (NSS) solutions \citep{DR3-DPACP-100, DR3-doc-NSS}.
The main astrometric NSS processing, which we will refer to as the `binary pipeline`, is described in \citet{Halbwachs:2022}. It analysed sources failing a single-star model %(based on a number of selection criteria) 
using a cascade of double-star models of increasing complexity, up to the determination of a full orbital solution for one companion. An alternative NSS processing module being the subject of this paper, which we dub as the `exoplanet pipeline', was designed with the two-fold goal of a) modeling higher-complexity NSS signals, such as those produced by multiple companions, and b) providing further insight in the regime of low-amplitude signals, such as those produced by sub-stellar companions, i.e.\ exoplanets and brown dwarfs, around nearby stars.  In  \gdr{3} we do not provide results for multiple companions due to the limited amount of available observations to constrain the solution. The per-source computational effort is higher for the exoplanet pipeline and the default channel for NSS processing is therefore the `binary pipeline`.

The design of the exoplanet pipeline takes advantage of some of the lessons learned from Doppler searches for planets. In particular, the modeling of complex, low-amplitude planetary  signals can be prone to ambiguities in the interpretation of the results, with well-known cases in the recent literature of disagreement on the actual values of the orbital elements of a given companion, or on the number of companions, also depending on the details of the treatment of noise sources in the RV measurements. A non-exhaustive list of `controversies' in radial velocity (RV) surveys includes the planetary systems around $\alpha$ Cen B (Dumusque et al. 2012; Hatzes 2013; Rajpaul et al. 2016), $\tau$ Ceti (Pepe et al. 2011; Tuomi et al. 2013; Feng et al. 2017a), GJ 667C (Anglada-Escud\'e et al. 2012, 2013; Delfosse et al. 2013; Feroz \& Hobson 2014; Robertson \& Mahadevan 2014), GJ 581 (Vogt et al. 2009; Baluev 2013; Robertson et al. 2014, 2015; Anglada Escud\'e \& Tuomi 2015; Hatzes 2016; Trifonov et al. 2018), GJ 176 (Endl et al. 2008; Butler et al. 2009; Forveille et al. 2009), HD 41248 (Jenkins et al. 2013; Jenkins \& Tuomi 2014; Santos et al. 2014; Feng e al. 2017b; Faria et al. 2019), Barnard's Star (Ribas et al. 2018; Lubin et al. 2021), Kapteyn's Star (Anglada Escud\'e et al. 2014; Robertson et al. 2015b; Anglada Escud\'e et al. 2016), GJ 3998 (Affer et al. 2016; Dodson-Robinson et al. 2022), Lalande 21185 (Butler et al. 2017; Diaz et al. 2019; Stock et al. 2020; Rosenthal et al. 2021; Hurt et al. 2022), BD $-06^\circ$ 1339 (Lo Curto et al. 2013; Simpson et al. 2022), and HD 219134 (Motalebi et al. 2015; Vogt et al. 2015; Gillon et al. 2017). 

The above considerations prompted us to a methodological approach that implements two different algorithms for alternative orbit fitting of Gaia DR3 astrometry. The exoplanet pipeline was applied for processing of two datasets: the first contained a large number of sources for which none of the models attempted in the `binary pipeline' could successfully improve upon the single-star fit based on the adopted thresholds on goodness-of-fit and significance statistics. These are labelled as \MOD{either \typeOrbAlt or \typeOrbAltVal, with the union of the two sets labelled as \typeOrbAltPVal (or shorthand \typeOrbAltStar)}. The second constituted a much smaller collection of high-visibility sources, either because of their intrinsic nature, or because of already known sub-stellar and low-mass stellar companions around them. These are labelled as \MOD{either \typeOrbTar or \typeOrbTarVal, with the union of the two sets labelled as \typeOrbTarPVal (or shorthand \typeOrbTarStar)}.
%a given source was assigned a stochastic solution, corresponding to a single-star solution with added uncorrelated noise. 

In this paper we provide an overview of the exoplanet pipeline, describe in details the functioning of the two orbit-fitting algorithms, and discuss the main characteristics of the orbital solution results obtained for the two processing experiments described above, which have been included in the \gdr{3} archive of NSS solutions. It is important to point out that `Orbital' solutions compatible with sub-stellar mass companions can also be found in the output of the `binary pipeline' \citep[cf.][]{Halbwachs:2022, DR3-DPACP-100}.

Our paper is organised as follows: 
in Sect~\ref{sec:dataProperties} and \ref{sec:orbitalModel}we shortly discuss the properties of the astrometric data and the model fitted to it. Section~\ref{sec:method} describes the algorithms used to derive the orbital solutions and the procedure to select the best solutions. The input source selection and solution filtering procedures are detailed in Sect.~\ref{sec:sourceSelection}, 
%followed by the source in Sect.~\ref{sec:solFilter}, 
with result verification and validation in Sect.~\ref{sec:results}. We conclude in Sect.~\ref{sec:conclusion}, with additional details regarding the 
%genetic algorithm operators and reference solution data in Appendices \ref{sec:genAlgoOperators} and \ref{sec:refSolParams}, respectively.
reference solution data, acronyms list, and \gaia archive queries in Appendices \ref{sec:refSolParams}, \ref{sec:acronyms}, and \ref{sec:archiveQueries}, respectively.

\section{Properties of the astrometric data \label{sec:dataProperties}}
 
 \begin{figure}[h]
  \includegraphics[width=0.98\columnwidth]{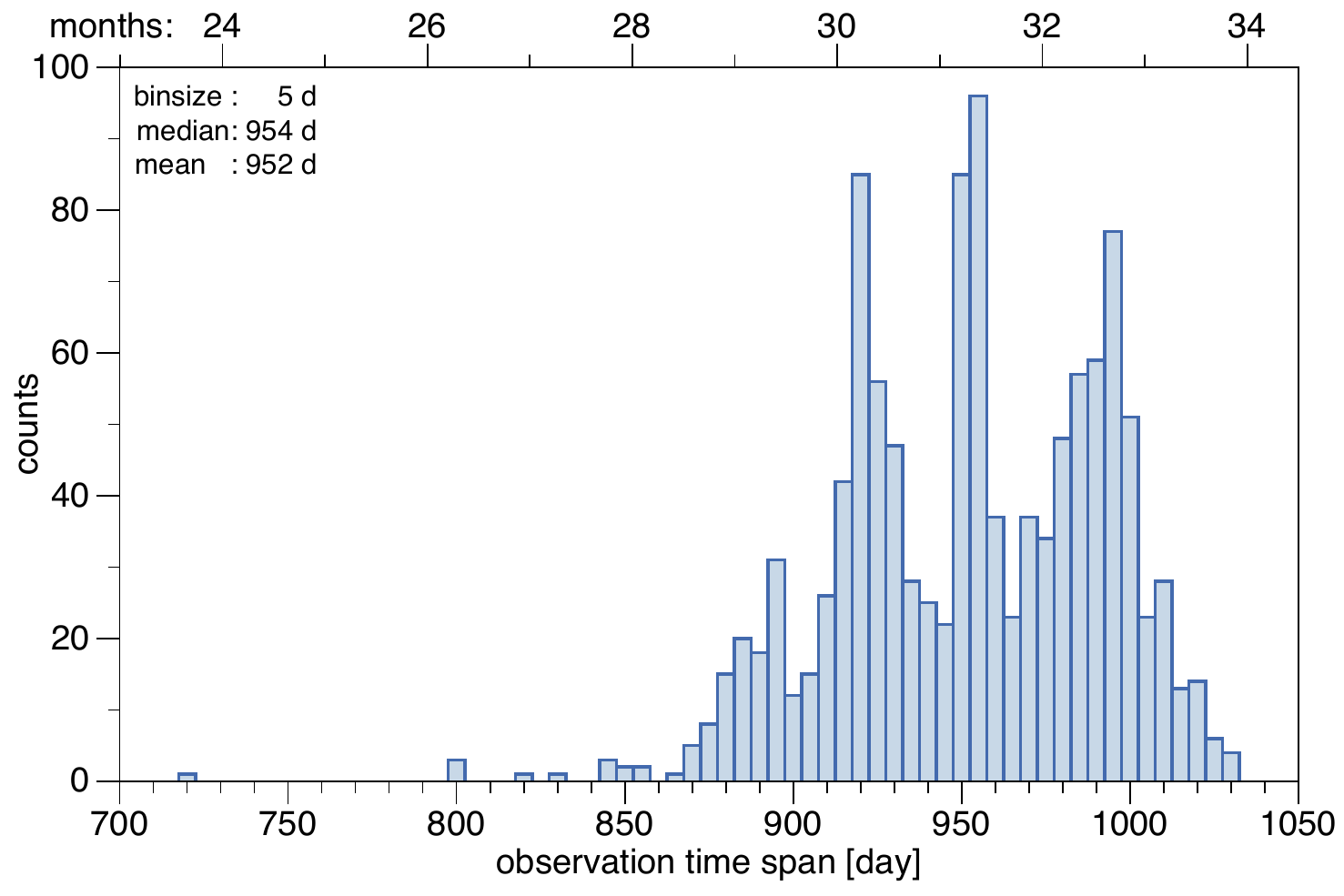}
  \vspace{-0.2cm}
\caption{Histogram of the observation time spans. 
%\WORKSUGGESTION{[Could add sub selection of validated only.]} 
}
\label{fig:obsDuration}
\end{figure} 

\begin{figure}[h]
  \includegraphics[width=0.98\columnwidth]{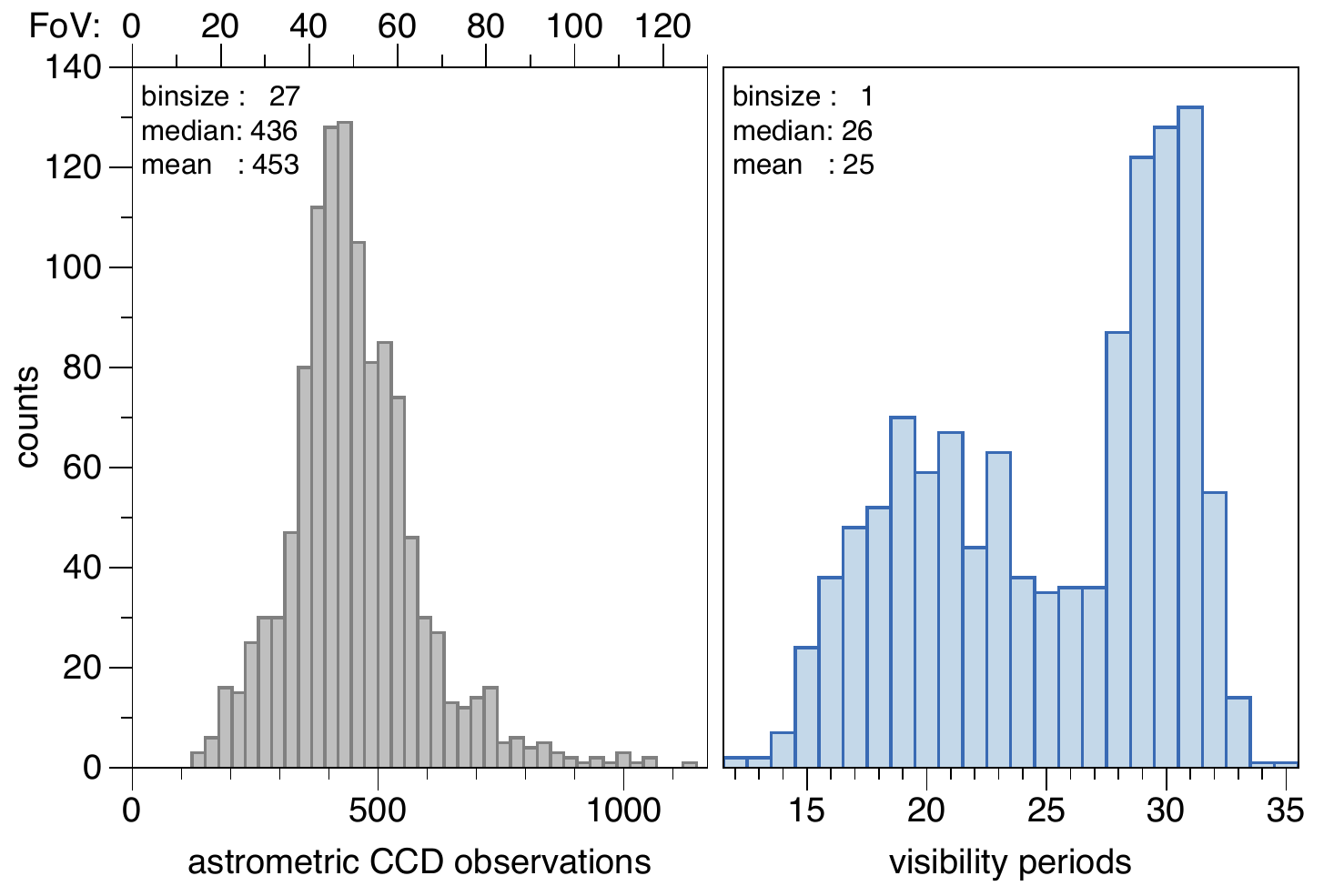}
  \vspace{-0.2cm}
\caption{Histogram of number of CCD observations and visibility periods. The number of CCD observations is divided by nine to provide the approximate number of FoV observations above the left panel.}
\label{fig:numObsHist}
\end{figure} 

%\TODO{[Short concise description, we could refer to documentation (e.g. subsection 1 of `Astrometric binaries') for details as this is not the main focus of this paper.]}

%\WORKSUGGESTION{[LOW PRIORITY: Berry (or any other volunteer): can write short (placeholder) description, but perhaps wait for above mentioned documentation update first]}

The input data spans about 34~months as shown in Fig.~\ref{fig:obsDuration}. 
It consists of time series of along-scan abscissa measurements $w$
with respect to the reference position $(\alpha_0, \delta_0)$ derived in the \egdr{3} Astrometric Global Iterative Solution \cite[AGIS, see][]{2021A&A...649A...2L} together with the associated scan angles and parallax factors.
 %(also called Local Plane Coordinates, or LPCs for short). The procedure is described in detail in
%The data consists of astrometric along-scan centroid-angle time series in Local Plane Coordinates (LPCs) computed with respect to the reference position $(\alpha_0, \delta_0)$ derived in the \egdr{3} Astrometric Global Iterative Solution \cite[AGIS, see][]{2021A&A...649A...2L}. The procedure is described in detail in 
Details as well as several pre-processing steps are  described in \cite{Halbwachs:2022} and section 7.2.2 of the DR3 NSS documentation \citep{DR3-doc-NSS} which include per-FoV\footnote{One field-of-view (FoV) passage of a source across the \gaia focal plane generally produces 8 or 9 individual CCD transits, each corresponding to the passage of the source image across an astrometric-field (AF) detector \citep{GaiaCollaboration:2016aa}.} CCD outlier rejection and modification of the measurements of stars with $\varpi>5$~mas (i.e., within 200~pc) to take into account the perspective acceleration (to identify the latter see \texttt{flags} in Sect.~\ref{ssec:archiveModelParams}). Figure~\ref{fig:numObsHist} shows the number of observations per source after outlier rejection. As expected the minimum number of visibility periods\footnote{
A `visibility period' groups observations separated from other groups by at least 4 days which (usually) assures that scan-angle and parallax factors have changed by a significant amount due to the evolution of the scanning law, and therefore is a better measure of independent `epochs' than simply counting CCD observations or FoV transits.} is not less than about 13, i.e., the number of parameters we are solving for with a single Keplerian orbit. Fitting for a second orbit would require an additional 7 parameters and thus (at least) 20 visibility periods. For this and various other reasons the general attempt to fit for more than one Keplerian was not made for \gdr{3}. %will be left to \gdr{4} and onwards, where at least double the amount of observations and time-span will be available.

\begin{figure}[h]
  \includegraphics[width=0.98\columnwidth]{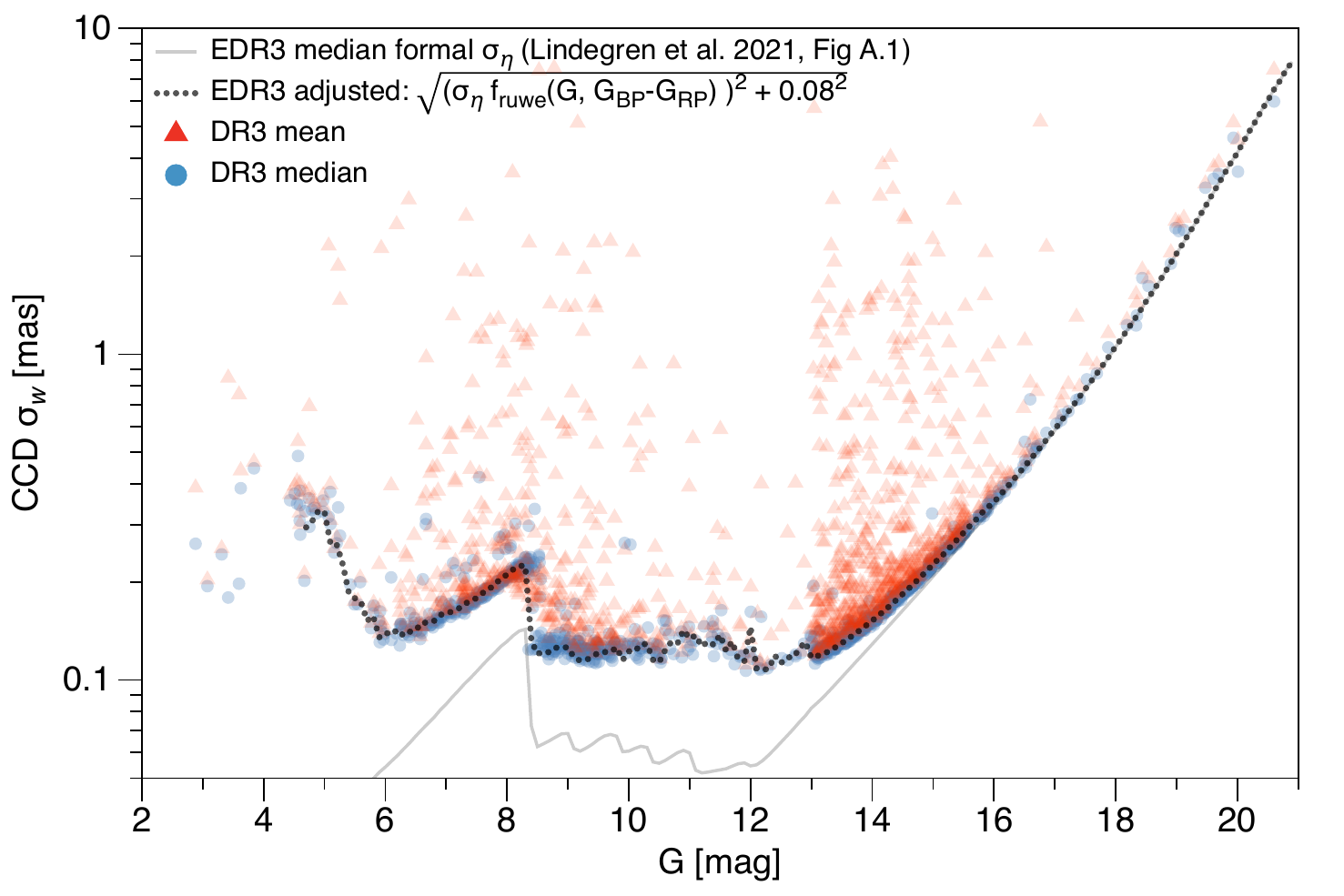}  
  \vspace{-0.2cm}
\caption{Mean and median CCD AL-scan abscissa uncertainty $\sigma_w$ per source.}
\label{fig:astroObsUncert}
\end{figure}

 The mean and median of the per-CCD uncertainties of each source are shown in Fig.~\ref{fig:astroObsUncert}.
 The outlier rejection applied in the `binary pipeline` pre-processing step did not remove all strong outliers in abscissa value and/or uncertainty, which generally is the reason for the difference between the mean and median. In the orbit figures that we show later (Figs. \ref{fig:orbit_1d_summary_DR3_2603090003484152064_companion0} - \ref{fig:orbit_1d_summary_DR3_4698424845771339520_companion0}), the persisting outliers are usually not displayed.
 We compare these data with the EDR3 median formal uncertainty $\sigma_\eta$ of \cite{2021A&A...649A...2L} (their fig.~A.1 red median line), here drawn as grey line. For the estimation of the CCD AL-scan abscissa uncertainty $\sigma_w$ in the astrometric NSS pipeline, the $\sigma_\eta$ was first inflated by\footnote{The renormalisation function value that transforms the AGIS unit weight error (uwe) into renormalised unit weight error (ruwe), see for details the definition of the \gaia archive  \href{https://gea.esac.esa.int/archive/documentation/GEDR3/Gaia_archive/chap\_datamodel/sec\_dm\_main\_tables/ssec\_dm\_gaia\_source.html\#gaia_source-ruwe}{\texttt{ruwe}} parameter.}
 $f_\text{ruwe}(G, G_\text{BP}-G_\text{RP})$, after which a (typical value of) 0.08~mas calibration noise was added in quadrature. The result is shown as the dotted line in Fig.~\ref{fig:astroObsUncert}, nicely matching the lower envelope of the median $\sigma_w$ of our sources.
 
 %In the next data releases, both the modelling of calibration effects and uncertainty estimation of the abscissa measurements will keep improving, as well as any potential outlier rejection procedure. 
 However, the abscissa data used for \gdr{3} is generally affected by some level of under/over estimation of uncertainties and potential biases, which can lead to unrealistic good or bad goodness-of-fit statistics. To mitigate the effects of such undefined noise contributions we fitted for an additional jitter term similar to the AGIS \texttt{astrometric\_excess\_noise}, as explained in Sect.~\ref{sec:orbitalModel}.
 %\WORKSUGGESTION{[BH: does the following paragraph fit here?]} 
 In fact, it proved difficult to define general filtering criteria to distinguish spurious and significant solution. This task required additional procedures where many sources were individually checked against literature data (in this release leading to those labelled `validated'), see Section~\ref{sec:results}.

 %\WORKSUGGESTION{[Should we say something about modelling errors and uncertainty estimation of the observations? E.g. that not all calibration effects have yet been properly modelling in AGIS, thus resulting in various systematic and possible under/over estimates of the uncertainties leading to unrealistic goodness of fit values? Perhaps there is a discussion on this already somewhere else? See also last paragraph of next section.]}

%------------------------------------------------------------------
% Orbital model
%------------------------------------------------------------------
\section{Astrometric model \label{sec:orbitalModel}}
% \TODO{[Here we should see if we simply put a very concise formulation and refer to another DPAC paper or the documentation for more details, or refer to it completely (I would probably prefer the former).]} 
% \TODO{[Give formula with symbols explained, that are used in rest of this paper.]}
%\WORKSUGGESTION{BH: need to mention Jitter term here (if it was not yet mentioned)}

\subsection{Mathematical description\label{ssec:mathModelDescr}}

As discussed in Sect.~\ref{sec:dataProperties}, the input data for the exoplanet pipeline are the Gaia along-scan abscissa measurements $w$. For a binary system and neglecting all noise considerations, these can be modelled by the combination of a single-source model $w_\mathrm{ss}$, describing the standard astrometric motion of the system's barycentre, and a Keplerian model $w_\mathrm{k1}$.\\
The single-source model can be written as 
\begin{equation}\label{eq:single_source_model}
w_\mathrm{ss} = ( \Delta\alpha^{\star} + \mu_{\alpha^\star} \, t ) \, \sin \psi + ( \Delta\delta + \mu_\delta \, t ) \, \cos \psi + \varpi \, f_\varpi, 
\end{equation}
where $\Delta\alpha^{\star}=\Delta\alpha \cos{\delta}$ and $\Delta\delta$ are small offsets in equatorial coordinates from some fixed reference point ($\alpha_0$, $\delta_0$), $\mu_{\alpha^\star}$ and $\mu_\delta$ are proper motions in these coordinates, $t$ is time since reference time J2016.0, $\varpi$ is the parallax, $f_\varpi$ is the parallax factor, and $\psi$ is the scan angle. 
%In practice, coordinate zero-points $\alpha^{\star}_0$ and $\delta_0$ have been subtracted from the abscissa and the local right ascension ($a=\alpha^{\star}-\alpha^{\star}_0$) and declination $d=\delta-\delta_0$ coordinates are used as parameters instead.
The Gaia scan angle is defined as having a value of $\psi=0$ when the field-of-view  is moving towards local North, and $\psi=90\degr$ towards local East\footnote{\url{https://www.cosmos.esa.int/web/gaia/scanning-law-pointings}}. This is not the same convention as used for Hipparcos \citep[e.g.][]{Leeuwen:2007kx}.

The astrometric motion corresponding to a Keplerian orbit of a binary system has generally seven independent parameters. These
are the period $P$, the epoch of periastron passage $T_0$, the eccentricity $e$, the inclination $i$, the ascending node $\Omega$, the argument of periastron $\omega$, and the semi-major axis of the photocentre $a_0$.
The Thiele-Innes coefficients $A, B, F, G$, which linearise part of the equations are defined as:

\begin{eqnarray}
%\begin{split}
% \begin{array}{ll}
A &=&  \ \ \, a_0 \; (\cos \omega \cos \Omega - \sin \omega \sin \Omega \cos i)   \\
B &=&  \ \ \, a_0 \; (\cos \omega \sin \Omega + \sin \omega \cos \Omega \cos i)  \\
F &=& -a_0 \; (\sin \omega \cos \Omega + \cos \omega \sin \Omega \cos i)  \\
G &=&  -a_0 \; (\sin \omega \sin \Omega - \cos \omega \cos \Omega \cos i) 
% \end{array}
%\end{split}
\label{eq:cu4nss_astrobin_orbital_ABFG}
\end{eqnarray}
The elliptical rectangular coordinates $X$ and $Y$ are functions of eccentric anomaly $E$ and eccentricity: 
\begin{eqnarray}
E - e \sin E &=& \frac{2\pi}{P} (t-T_0)\\
X &=& \cos E - e\\
Y &=& \sqrt{1-e^2} \sin E
\end{eqnarray}
The single Keplerian model can then be written as 
\begin{equation}\label{eq:k1_model}
w_\mathrm{k1} = (B \, X + G \, Y) \sin \psi + (A \, X + F \, Y) \cos \psi.
\end{equation}
The combined model $w^\mathrm{(model)}$ for the Gaia along-scan abscissa is

\begin{equation}\label{eq:abscissa1}
\begin{split}
w^\mathrm{(model)} =&\, w_\mathrm{ss} + w_\mathrm{k1} \\
 =&\, ( \Delta\alpha^{\star} + \mu_{\alpha^\star} \, t ) \, \sin \psi + ( \Delta\delta + \mu_\delta \, t ) \, \cos \psi + \varpi \, f_\varpi \\
 &+\, (B \, X + G \, Y) \sin \psi + (A \, X + F \, Y) \cos \psi.
\end{split}\end{equation}

This model has been extensively used for modelling the Hipparcos epoch data of non-single stars \citep[e.g.][]{Sahlmann:2011fk}.

More details on the modelling of non-single star data in the Gaia pipelines can be found in the DR3 NSS documentation \cite{DR3-doc-NSS} and in \citet{Halbwachs:2022}, where in the latter the instantaneous scan angle is described as the coordinate-derivative of the abscissa, e.g.\ $\sin \psi = \frac{\partial w}{\partial \Delta\alpha^{\star}}$.

%\WORKSUGGESTION{[BH: added this paragraph mentioning the jitter term]}
Finally, to account for potentially unmodelled signals or modelling errors, we additionally fit for a jitter term which is added in quadrature to the provided uncertainties of the along-scan abscissa, bringing the total number of fitted parameters to~13: five linear parameters for the single-source model $w_\mathrm{ss}$, seven for the single Keplerian model $w_\mathrm{k1}$ (of which the four $A$,$B$,$F$,$G$ are linear), and one non-linear jitter term $\sigma_\mathrm{jit}$.

Symmetric parameter uncertainty estimates are obtained by reconstructing the covariance matrix directly from the Jacobians of all parameters for all observations. No scaling of these formal covariances was performed, which might potentially suffer from over/under estimations e.g. due to unmodelled signals (that might be partially absorbed by the jitter term).

\subsection{Conversion of Thiele-Innes parameters to Campbell elements\label{ssec:TItoCampbell}}
The conversion of the Thiele-Innes parameters ($A,B,F,G$) to Campbell or geometric elements ($a_0,\omega,\Omega,i$) %\footnote{Campbell elements: semi-major axis of the photocentre orbit $a_0$, inclination angle $i$, longitude of the ascending node $\Omega$, and argument of the periastron $\omega$.} 
is in principle straight-forward \citep[e.g.][]{Halbwachs:2022} but several caveats exist that are related to the amplitudes of the $A,B,F,G$ co-variance terms which sometimes seem to be over-estimated, in particular for solutions with poorly-constrained eccentricities \citep[cf.][]{DR3-DPACP-100,DR3-DPACP-127}. In Sect.~\ref{sssec:campbElEst} we discuss some examples. 
%show the difference between the conversion using linear error propagation (taking into account the co-variances) and Monte-Carlo resampling for \typeOrbTarPVal solutions. %of the $A$, $B$, $F$, $G$ covariances. 
Unless specifically mentioned, all figures with Campbell element values in this paper use the linear error propagation.
Regarding the semi-major axis, in this work we always assume the companion is sufficiently non-luminous that the semi-major axis of the photocentre and that of the observed host star are the same, i.e., $a_1=a_0$, and will only refer to $a_1$.

%\WORKSUGGESTION{[copied from end of 4.1, to refactor with above:]} Standard conversion formulae and analytical error propagation calculations then allow to move back from the Thiele-Innes parameters representation to the Campbell orbital elements semi-major axis of the photocenter orbit $a_1$, inclination angle $i$, longitude of the ascending node $\Omega$, and argument of the periastron $\omega$
 
% \WORKSUGGESTION{(AS: SEE APPENDIX? BERRY, ARE THE CONVERSION FORMULAE IN THE DU437 CODE THE SAME AS IN HALBWACHS ET AL.? SHOULD WE JUST REFER TO THEM?).
%BH: I think we should indeed simply refer to the other paper for them}

\subsection{Archive model parameter fields\label{ssec:archiveModelParams}}
Table~\ref{tab:archiveTabFields} provides an overview of all solution parameters and additional fields populated for our sources in the \texttt{nss\_two\_body\_orbit} \gdr{3} archive table.
%\WORKSUGGESTION{[expand with details about e.g. the flags and efficiency.]}

The \texttt{goodness\_of\_fit} is the F2, or so-called ‘gaussianized chi-square’ \citep{1931PNAS...17..684W}, which should approximately follow a normal distribution with zero mean value and unit standard deviation for good fits.

The \texttt{flags} field only has integer values 0, 64, and 192. Value 0 translates to no bit-flags being set. Value 64 translates to bit 6 being set: which means that a mean RV value was available, and value 192 translates to bit 6 and 7 being set, where bit 7 indicates that mean RV was used for perspective acceleration correction of the local plane coordinates. In our published sample there are 164 sources with flags value 0, 571 sources with flags value 64, and 427 sources with flags value 192 (see result of query provided in Appendix~\ref{sec:archiveQueries} for finer details). 
%lags	nss_solution_type	counts
%0	OrbitalTargetedSearchValidated	2
%0	OrbitalTargetedSearch	14
%0	OrbitalAlternative	148
%64	OrbitalAlternativeValidated	10
%64	OrbitalTargetedSearchValidated	24
%64	OrbitalTargetedSearch	70
%64	OrbitalAlternative	467
%192	OrbitalAlternative	4
%192	OrbitalTargetedSearchValidated	162
%192	OrbitalTargetedSearch	261

The \texttt{astrometric\_n\_obs\_al} provides the number of CCD observations that were available before outlier rejection. In the archive the \texttt{astrometric\_n\_good\_obs\_al} should represent the number after outlier rejection, but erroneously ended up to report the same value as \texttt{astrometric\_n\_obs\_al}. Figure~\ref{fig:numObsHist} shows the correct number of filtered (`good') observations. For completeness we document here that the number of rejected CCD observations varies between zero and 269, with the majority having zero to three observations rejected, and the vast majority less than 12 observations rejected. 

%\WORKSUGGESTION{[BH: say this about efficiency since it is in the archive?]}
We did not make use of the \texttt{efficiency} parameter in our verification, which incidentally is 0 for a large fraction of our sample. 

\gaia photometric time series are available for 76 sources in our sample, i.e., sources in \texttt{gaia\_source} with \texttt{has\_epoch\_photometry=true}. Of these, 75 overlap with the variable source catalogue \citep{DR3-DPACP-162} and one source (367906215676634752) was released as part of the Gaia Andromeda Photometric Survey \citep{DR3-DPACP-142}, which can be identified in \texttt{gaia\_source} by \texttt{phot\_variable\_flag}=\texttt{VARIABLE} and \texttt{in\_andromeda\_survery}=\texttt{true}, respectively. In this \gdr{3} no \gaia astrometric time series were made public.

\begin{table}
\caption{\label{tab:archiveTabFields} \texttt{gaia\_dr3.nss\_two\_body\_orbit} table fields filled for the 1162 sources from the exoplanet pipeline described in this paper. Parameter names link directly to the 
\href{https://gea.esac.esa.int/archive/documentation/GDR3/Gaia_archive/chap\_datamodel/sec\_dm\_non--single\_stars\_tables/ssec\_dm\_nss\_two\_body\_orbit.html}{online data model documentation}.}
\centering                                      % used for centering table
\begin{tiny}
\begin{tabular}{llll}

\hline\hline
\gdr{3} table field name & unit & symbol & notes \\
\hline
\href{https://gea.esac.esa.int/archive/documentation/GDR3/Gaia_archive/chap\_datamodel/sec\_dm\_non--single\_stars\_tables/ssec\_dm\_nss\_two\_body\_orbit.html#nss\_two\_body\_orbit-solution\_id}{\texttt{solution\_id}}           & 	        &                               &   \\
\href{https://gea.esac.esa.int/archive/documentation/GDR3/Gaia_archive/chap\_datamodel/sec\_dm\_non--single\_stars\_tables/ssec\_dm\_nss\_two\_body\_orbit.html#nss\_two\_body\_orbit-source\_id}{\texttt{source\_id}}	            & 	        &                               &   \\
\href{https://gea.esac.esa.int/archive/documentation/GDR3/Gaia_archive/chap\_datamodel/sec\_dm\_non--single\_stars\_tables/ssec\_dm\_nss\_two\_body\_orbit.html#nss\_two\_body\_orbit-nss\_solution\_type}{\texttt{nss\_solution\_type}}    & 	        &                               & four types\tablefootmark{a} \\
\href{https://gea.esac.esa.int/archive/documentation/GDR3/Gaia_archive/chap\_datamodel/sec\_dm\_non--single\_stars\_tables/ssec\_dm\_nss\_two\_body\_orbit.html#nss\_two\_body\_orbit-ra}{\texttt{ra}}                     & deg       & $\alpha_\star$                &   \\
\href{https://gea.esac.esa.int/archive/documentation/GDR3/Gaia_archive/chap\_datamodel/sec\_dm\_non--single\_stars\_tables/ssec\_dm\_nss\_two\_body\_orbit.html#nss\_two\_body\_orbit-ra\_error}{\texttt{ra\_error}}              & mas       & $\sigma_{\alpha_\star}$          &   \\
\href{https://gea.esac.esa.int/archive/documentation/GDR3/Gaia_archive/chap\_datamodel/sec\_dm\_non--single\_stars\_tables/ssec\_dm\_nss\_two\_body\_orbit.html#nss\_two\_body\_orbit-dec}{\texttt{dec}}                    & deg       & $\delta$                      &   \\
\href{https://gea.esac.esa.int/archive/documentation/GDR3/Gaia_archive/chap\_datamodel/sec\_dm\_non--single\_stars\_tables/ssec\_dm\_nss\_two\_body\_orbit.html#nss\_two\_body\_orbit-dec\_error}{\texttt{dec\_error}}             & mas       & $\sigma_\delta$               &   \\
\href{https://gea.esac.esa.int/archive/documentation/GDR3/Gaia_archive/chap\_datamodel/sec\_dm\_non--single\_stars\_tables/ssec\_dm\_nss\_two\_body\_orbit.html#nss\_two\_body\_orbit-parallax}{\texttt{parallax}}               & mas       & $\varpi$                      &   \\
\href{https://gea.esac.esa.int/archive/documentation/GDR3/Gaia_archive/chap\_datamodel/sec\_dm\_non--single\_stars\_tables/ssec\_dm\_nss\_two\_body\_orbit.html#nss\_two\_body\_orbit-parallax\_error}{\texttt{parallax\_error}}        & mas       & $\sigma_\varpi$               &   \\
\href{https://gea.esac.esa.int/archive/documentation/GDR3/Gaia_archive/chap\_datamodel/sec\_dm\_non--single\_stars\_tables/ssec\_dm\_nss\_two\_body\_orbit.html#nss\_two\_body\_orbit-pmra}{\texttt{pmra}}                   & mas/yr    & $\mu_{\alpha^\star}$          &   \\
\href{https://gea.esac.esa.int/archive/documentation/GDR3/Gaia_archive/chap\_datamodel/sec\_dm\_non--single\_stars\_tables/ssec\_dm\_nss\_two\_body\_orbit.html#nss\_two\_body\_orbit-pmra\_error}{\texttt{pmra\_error}}            & mas/yr    & $\sigma_{\mu_{\alpha^\star}}$ &   \\ 
\href{https://gea.esac.esa.int/archive/documentation/GDR3/Gaia_archive/chap\_datamodel/sec\_dm\_non--single\_stars\_tables/ssec\_dm\_nss\_two\_body\_orbit.html#nss\_two\_body\_orbit-pmdec}{\texttt{pmdec}}                  & mas/yr    & $\mu_{\delta}$                &   \\
\href{https://gea.esac.esa.int/archive/documentation/GDR3/Gaia_archive/chap\_datamodel/sec\_dm\_non--single\_stars\_tables/ssec\_dm\_nss\_two\_body\_orbit.html#nss\_two\_body\_orbit-pmdec\_error}{\texttt{pmdec\_error}}           & mas/yr    & $\sigma_{\mu_{\delta}}$       &   \\ 
\href{https://gea.esac.esa.int/archive/documentation/GDR3/Gaia_archive/chap\_datamodel/sec\_dm\_non--single\_stars\_tables/ssec\_dm\_nss\_two\_body\_orbit.html#nss\_two\_body\_orbit-a\_thiele\_innes}{\texttt{a\_thiele\_innes}}       & mas       & $A$                           &   \\
\href{https://gea.esac.esa.int/archive/documentation/GDR3/Gaia_archive/chap\_datamodel/sec\_dm\_non--single\_stars\_tables/ssec\_dm\_nss\_two\_body\_orbit.html#nss\_two\_body\_orbit-a\_thiele\_innes\_error}{\texttt{a\_thiele\_innes\_error}} & mas       & $\sigma_A$                    &   \\
\href{https://gea.esac.esa.int/archive/documentation/GDR3/Gaia_archive/chap\_datamodel/sec\_dm\_non--single\_stars\_tables/ssec\_dm\_nss\_two\_body\_orbit.html#nss\_two\_body\_orbit-b\_thiele\_innes}{\texttt{b\_thiele\_innes}}       & mas       & $B$                           &   \\
\href{https://gea.esac.esa.int/archive/documentation/GDR3/Gaia_archive/chap\_datamodel/sec\_dm\_non--single\_stars\_tables/ssec\_dm\_nss\_two\_body\_orbit.html#nss\_two\_body\_orbit-b\_thiele\_innes\_error}{\texttt{b\_thiele\_innes\_error}} & mas       & $\sigma_B$                    &   \\
\href{https://gea.esac.esa.int/archive/documentation/GDR3/Gaia_archive/chap\_datamodel/sec\_dm\_non--single\_stars\_tables/ssec\_dm\_nss\_two\_body\_orbit.html#nss\_two\_body\_orbit-f\_thiele\_innes}{\texttt{f\_thiele\_innes}}       & mas       & $F$                           &   \\
\href{https://gea.esac.esa.int/archive/documentation/GDR3/Gaia_archive/chap\_datamodel/sec\_dm\_non--single\_stars\_tables/ssec\_dm\_nss\_two\_body\_orbit.html#nss\_two\_body\_orbit-f\_thiele\_innes\_error}{\texttt{f\_thiele\_innes\_error}}& mas       & $\sigma_F$                    &   \\
\href{https://gea.esac.esa.int/archive/documentation/GDR3/Gaia_archive/chap\_datamodel/sec\_dm\_non--single\_stars\_tables/ssec\_dm\_nss\_two\_body\_orbit.html#nss\_two\_body\_orbit-g\_thiele\_innes}{\texttt{g\_thiele\_innes}}       & mas       & $G$                           &   \\
\href{https://gea.esac.esa.int/archive/documentation/GDR3/Gaia_archive/chap\_datamodel/sec\_dm\_non--single\_stars\_tables/ssec\_dm\_nss\_two\_body\_orbit.html#nss\_two\_body\_orbit-g\_thiele\_innes\_error}{\texttt{g\_thiele\_innes\_error}}& mas       & $\sigma_G$                    &   \\
\href{https://gea.esac.esa.int/archive/documentation/GDR3/Gaia_archive/chap\_datamodel/sec\_dm\_non--single\_stars\_tables/ssec\_dm\_nss\_two\_body\_orbit.html#nss\_two\_body\_orbit-period}{\texttt{period}}                 & d         & $P$                           &   \\
\href{https://gea.esac.esa.int/archive/documentation/GDR3/Gaia_archive/chap\_datamodel/sec\_dm\_non--single\_stars\_tables/ssec\_dm\_nss\_two\_body\_orbit.html#nss\_two\_body\_orbit-period\_error}{\texttt{period\_error}}          & d         & $\sigma_P$                    &   \\
\href{https://gea.esac.esa.int/archive/documentation/GDR3/Gaia_archive/chap\_datamodel/sec\_dm\_non--single\_stars\_tables/ssec\_dm\_nss\_two\_body\_orbit.html#nss\_two\_body\_orbit-t\_periastron}{\texttt{t\_periastron}}          & d         & $T_0$                         & since J2016.0  \\
\href{https://gea.esac.esa.int/archive/documentation/GDR3/Gaia_archive/chap\_datamodel/sec\_dm\_non--single\_stars\_tables/ssec\_dm\_nss\_two\_body\_orbit.html#nss\_two\_body\_orbit-t\_periastron\_error}{\texttt{t\_periastron\_error}}   & d         & $\sigma_{T_0}$                &   \\
\href{https://gea.esac.esa.int/archive/documentation/GDR3/Gaia_archive/chap\_datamodel/sec\_dm\_non--single\_stars\_tables/ssec\_dm\_nss\_two\_body\_orbit.html#nss\_two\_body\_orbit-eccentricity}{\texttt{eccentricity}}           &           & $e$                           &   \\
\href{https://gea.esac.esa.int/archive/documentation/GDR3/Gaia_archive/chap\_datamodel/sec\_dm\_non--single\_stars\_tables/ssec\_dm\_nss\_two\_body\_orbit.html#nss\_two\_body\_orbit-eccentricity\_error}{\texttt{eccentricity\_error}}    &           & $\sigma_e$                    &   \\
\href{https://gea.esac.esa.int/archive/documentation/GDR3/Gaia_archive/chap\_datamodel/sec\_dm\_non--single\_stars\_tables/ssec\_dm\_nss\_two\_body\_orbit.html#nss\_two\_body\_orbit-astrometric\_n\_obs\_al}{\texttt{astrometric\_n\_obs\_al}} & 	        &                               &   \\
\href{https://gea.esac.esa.int/archive/documentation/GDR3/Gaia_archive/chap\_datamodel/sec\_dm\_non--single\_stars\_tables/ssec\_dm\_nss\_two\_body\_orbit.html#nss\_two\_body\_orbit-astrometric\_n\_good\_obs\_al}{\texttt{astrometric\_n\_good\_obs\_al}} &    &                               &   \\
\href{https://gea.esac.esa.int/archive/documentation/GDR3/Gaia_archive/chap\_datamodel/sec\_dm\_non--single\_stars\_tables/ssec\_dm\_nss\_two\_body\_orbit.html#nss\_two\_body\_orbit-bit\_index}{\texttt{bit\_index}}             &           &                               & 8191\tablefootmark{b} \\
\href{https://gea.esac.esa.int/archive/documentation/GDR3/Gaia_archive/chap\_datamodel/sec\_dm\_non--single\_stars\_tables/ssec\_dm\_nss\_two\_body\_orbit.html#nss\_two\_body\_orbit-corr\_vec}{\texttt{corr\_vec}}              &           &                               & \tablefootmark{c} \\
\href{https://gea.esac.esa.int/archive/documentation/GDR3/Gaia_archive/chap\_datamodel/sec\_dm\_non--single\_stars\_tables/ssec\_dm\_nss\_two\_body\_orbit.html#nss\_two\_body\_orbit-obj\_func}{\texttt{obj\_func}}             &           & $\chi^2$                      &   \\
\href{https://gea.esac.esa.int/archive/documentation/GDR3/Gaia_archive/chap\_datamodel/sec\_dm\_non--single\_stars\_tables/ssec\_dm\_nss\_two\_body\_orbit.html#nss\_two\_body\_orbit-goodness\_of\_fit}{\texttt{goodness\_of\_fit}}      &           & F2                            &   \\	
\href{https://gea.esac.esa.int/archive/documentation/GDR3/Gaia_archive/chap\_datamodel/sec\_dm\_non--single\_stars\_tables/ssec\_dm\_nss\_two\_body\_orbit.html#nss\_two\_body\_orbit-efficiency}{\texttt{efficiency}}	%\WORKSUGGESTION{[in dr3int7,del?]}
&           &                               &  \{0, [0.26--0.44]\} \\
\href{https://gea.esac.esa.int/archive/documentation/GDR3/Gaia_archive/chap\_datamodel/sec\_dm\_non--single\_stars\_tables/ssec\_dm\_nss\_two\_body\_orbit.html#nss\_two\_body\_orbit-significance}{\texttt{significance}}	        &           &  $a_1/\sigma_{a_1}$           &   \\
\href{https://gea.esac.esa.int/archive/documentation/GDR3/Gaia_archive/chap\_datamodel/sec\_dm\_non--single\_stars\_tables/ssec\_dm\_nss\_two\_body\_orbit.html#nss\_two\_body\_orbit-flags}{\texttt{flags}}                 &           &                               & 
\{0, 64, 192\} \\
%\tablefootmark{d}  \\
\href{https://gea.esac.esa.int/archive/documentation/GDR3/Gaia_archive/chap\_datamodel/sec\_dm\_non--single\_stars\_tables/ssec\_dm\_nss\_two\_body\_orbit.html#nss\_two\_body\_orbit-astrometric\_jitter}{\texttt{astrometric\_jitter}}    & mas       &  $\sigma_\text{jit}$          &   \\    
\hline

\end{tabular}
\tablefoot{
\tablefoottext{a}{The four types we described: \typeOrbAltPVal and \typeOrbTarPVal}
\tablefoottext{b}{Always value 8191, i.e. in bits flagging the 12~orbital parameters (excluding \texttt{astrometric\_jitter}) }
\tablefoottext{c}{vector form of the upper triangle of the correlation matrix (column-major ordered) of the 12 solved parameters (excluding \texttt{astrometric\_jitter}).}
%\tablefoottext{d}{Quality flag 0=no flags set, 64, and 192.}
}
\end{tiny}
\end{table}

%------------------------------------------------------------------
% Processing method
%------------------------------------------------------------------
\section{Processing procedure \label{sec:method}}
%\TODO{[From documentation `orbital model':] Two different methods are used to search for orbital parameter fits in the astrometric data: (1) an MCMC model and a (2) Genetic algorithm. Published results in \gdr{3} are only for single-companion models, though the codes are designed to efficiently search for higher multiplicity models too.}

%\WORKSUGGESTION{[(Berry or Alessandro) Short intro referring to the three sections that follow.]}

Astrometric orbit modelling requires solving a highly non-linear least squares problem with a minimum of 12 parameters (Sect.~\ref{sec:orbitalModel}). The orbital motion is usually seen as a small perturbation of the standard stellar motion, with a magnitude that can be orders of magnitude smaller than those of parallax and proper motion. 
% original:
%This consideration, along with the complications due to the intrinsic nature of the Gaia astrometric time-series (with intrinsically high sensitivity only in one dimension and with the two-dimensional information recovered only through the combination of different scanning directions \FEEDBACK{JSA: I don't think "intrinsic" is the correct term here. For DR3 the AC astrometry of gclass=0,1 sources is not well calibrated, hence we don't use it. In the future this may change. I suggest to rephrase.}) 
% alternative:
%This consideration, along with the intrinsic nature of \gaia astrometric observation being (much) more precise along the scan direction,
This
motivated the original design of the `exoplanet'-element of the non-single-star (NSS) processing pipeline (see
DR3 NSS documentation \cite{DR3-doc-NSS} for the full NSS pipeline structure).
In particular, it was recognised that orbit modelling in the limit of low signal amplitudes %\WORKSUGGESTION{(TBD) [BH: was this meant to be rephrased?]} 
would benefit from 
%an aggressive approach,
a more in-depth (computationally expensive) parameter search, which translated in the adoption of two independent orbit fitting algorithms exploiting different philosophies, which are described in turn below. Both algorithms are executed in parallel, and a standard recipe based on Bayesian model selection is utilised to select the best-fit solution, as further detailed below. 

%------------------------------------------------------------------
% Method SUB-SECTION: MCMC
%------------------------------------------------------------------
\subsection{Differential Evolution Markov Chain Monte Carlo \label{sec:method_mcmc}}
%\WORKSUGGESTION{[Alessandro: type away ;)]}

The first orbit fitting code is a hybrid implementation of a Bayesian analysis based on the differential evolution Markov chain Monte Carlo (DE-MCMC) method \citep{TerBraak2006,Eastman2013}. An earlier version of the code had been extensively tested in \citet{2008A&A...482..699C}, while its upgrade has been recently used in \citet{Drimmel2021}. In this scheme we take advantage of the four ($A$, $B$, $F$, $G$) Thiele-Innes constants representation (see Sect. \ref{ssec:mathModelDescr}. See also e.g., \citealt{Binnendijk1960,Wright2009}) to partially linearize the problem. Within this dimensionality reduction scheme, only three non-linear orbital parameters must be effectively explored using the DE-MCMC algorithm (e.g., \citealt{2008A&A...482..699C,Wright2009,Mendez2017,Drimmel2021}), namely $P$, $T_0$, and $e$. The fourth model parameter explored the DE-MCMC way is an uncorrelated astrometric jitter term $\sigma_\mathrm{jit}{}$. At each step of the DE-MCMC analysis, the resulting linear system of equations is solved in terms of the Thiele-Innes constants using simple matrix algebra, QR decomposition being the method of choice. The final likelihood function used in the DE-MCMC analysis is then: 
%\begin{equation}
%\ln \mathcal{L} = -\frac{1}{2} \left( 
%\sum_{j=1}^{{N_{astr}}}\frac{\left(w_j^{(obs)} - w_j^{(model)}\right)^2} {\sigma_{w,j}^2+\sigma_\mathrm{jit}^2}+
%\sum_{j=1}^{N_{astr}}\left(\ln\left[{\sigma_{w,j}^2 +\sigma_\mathrm{jit}^2}\right] \right )
%\right ) \label{eq:mcmcLikelihood}
%\end{equation}
\begin{align} 
-\ln \left( \mathcal{L} \right) \ = \ 
&\frac{1}{2} \ 
\sum_{j=1}^{{N_{astr}}}\frac{\left(w_j^{(obs)} - w_j^{(model)}\right)^2} {\sigma_{w,j}^2+\sigma_\mathrm{jit}^2}+ \nonumber \\
& \frac{1}{2} \ \sum_{j=1}^{N_{astr}}\ln\left({\sigma_{w,j}^2 +\sigma_\mathrm{jit}^2}\right)  \label{eq:mcmcLikelihood}
\end{align}

%\WORKSUGGESTION{ Damien : the above equation has an error in the denominator of the $\chi^2$. Here is the correction. \\
%Did you drop the the $\Big[ N^{obs} \ln{(2\pi)}\Big] $ term because it's a constant term? It should be present when defining the likelihood but it can be dropped in the code.
%\begin{equation}
%-\ln \mathcal{L} =\frac{1}{2} 
%\sum_{j=1}^{N_{astr}} \frac{\left(w_j^{(obs)} - %w_j^{(model)}\right)^2}{\sigma_{w,j}^2  +\sigma_\mathrm{jit}^2}} +
%\frac{1}{2} \sum_{j=1}^{N_{astr}}\ln\left({\sigma_{w,j}^2 %+\sigma_\mathrm{jit}^2}\right) 
%\end{equation}
%}

The DE-MCMC analysis is carried out with a number of chains equal to twice the number of free
parameters. A period search is first performed in order to identify statistically more probable periodicities. Given the nature of the astrometric dataset, the direct application of publicly available tools for the periodogram analysis of unevenly sampled time-series (e.g. the Generalized Lomb-Scargle periodogram, \citealt{2009A&A...496..577Z}) is not possible. For any given source, the DE-MCMC module draws a large sample of initial trial periods for sinusoidal signals projected along the scan directions of the time-series, based on a uniform grid up to twice the observations time span. A sparsely sampled selection of periods corresponding to local $\chi^2$ minima becomes the seed for the $P$ parameter initialization of the DE-MCMC chains. Uniform priors in the ranges [$-P/2$,$P/2$] and [0,10] mas are used for $T_0$ and $\sigma_\mathrm{jit}$, respectively. 
Finally, starting values for $e$ are drawn from a Beta distribution, following \citet{Kipping2013}. 

Convergence and good mixing of the chains are checked based on the Gelman–Rubin statistics (e.g., \citealt{Ford2006}). The medians of posterior distributions are taken as the final parameters. In order to comply with the choice of the main non-single-star processing chain \citep{DR3-DPACP-163}, we did not adopt the standard approach for computing the $1\sigma$ uncertainties on model parameters, i.e. evaluating the $\pm34.13$ per cent intervals from the posterior distributions, which typically results in asymmetric error-bars, but rather provided symmetric estimates of the uncertainties by reconstructing the covariance matrix directly from the Jacobians of all parameters for all observations. 

%------------------------------------------------------------------
% Method SUB-SECTION: GA
%------------------------------------------------------------------
\subsection{Genetic Algorithm\label{sec:method_ga}}

 The implementation of the Genetic Algorithm for \gaia (MIKS-GA) is a direct adaptation of \textsc{yorbit}, a tool used to search for exoplanets in radial velocity time series \citep{2011A&A...535A..54S,2016A&A...588A.145H,2017A&A...608A.129T, 2019A&A...631A.125K, 2022MNRAS.tmp..156T}. An implementation of the MIKS-GA algorithm has been successfully used to discover brown-dwarf binaries from the orbital solutions identified in high-precision astrometric timeseries obtained with ground-based telescopes \citep{2013A&A...556A.133S, Sahlmann:2015ac, 2020MNRAS.495.1136S}.

 Genetic Algorithms (GA) are a class of optimization algorithms that are loosely based of Darwin's theory of evolution by natural selection \citep{Holland:1975,Jong88learningwith}. In GA, a population of chromosomes is initialised and evolves by applying a set of genetic operators (such as crossover, recombination, mutation and selection) until the best genotype dominates the population. Such algorithms are particularly well suited for highly non-linear model with irregular sampling provided that the genetic operators are fine-tuned to the specificity of the problem.
 
Here, a chromosome consists of the non-linear parameters needed to model a merit function  $\mathcal{M}$ defined as the sum of the log-likelihood and the log-prior. 
We further assume that the residuals of the astrometric data to the model are drawn from independent realisations of a Normal distribution of zero mean with a variance composed of the astrometric uncertainty plus an additional jitter term which accounts for anything in the data that can't be modelled by the analytical astrometric model. It results in the following expression of the merit function

\begin{equation}
\mathcal{M}(\{\boldsymbol{t,\psi,w,\sigma_w}\};\{P,e,M_0,\sigma_{\rm Jit}\}) = -\ln\left( \mathcal{L}\right) -\ln\left( \mathcal{P}\right)
\label{eq:gaMerit}
\end{equation}
and the log-likelihood :
\begin{align} 
-\ln\left( \mathcal{L}\right) \ = \  
 &\frac{1}{2} \sum_{j=1}^{N_{astr}} \frac{\left(w_j^{(obs)} - w_j^{(model)}\right)^2}{\sigma_{w,j}^2 +\sigma_\mathrm{jit}^2} + \nonumber \\
&\frac{1}{2} \sum_{j=1}^{N_{astr}}\ln\left({\sigma_{w,j}^2 +\sigma_\mathrm{jit}^2}\right) + \frac{N_{astr}}{2} \ln\left( 2\pi \right)
\label{eq:gaLikelihood}
\end{align}
The priors expression is the product of uniform distributions for the mean anomaly  $M_0$ computed at J2016.0 (JD~2457388.5) in \gdr{3}, for the frequency (between 2.5~d and two times the observation time span), for the log of the jitter term (between [0.005, 2$\sigma_\text{5p,res}$]~mas, where  $\sigma_\text{5p,res}$ is the standard deviation of the residuals of a 5-parameter astrometric fit to the observations),
and of a truncated normal distribution for the eccentricity (with $\mu=0$, $\sigma=0.3$, truncated over $[0, 0.985]$)
to penalise highly eccentric orbit solutions that commonly arise in low Signal-to-Noise time series with irregular sampling.
\begin{align} 
 \mathcal{P}=  
 \mathcal{TN}_e(0, 0.3, 0, 0.985) . U_f(\nu_{\rm min},\nu_{\rm max}). U_{M_0}(0,360) . \nonumber \\ 
 U_{\sigma_{\rm jit}}(-2.30,\log{(2\sigma_\text{5p,res})}) 
\label{eq:gaPrior}
\end{align}

\subsubsection{Initialisation phase}
The initialisation phase of the chromosomes' population is based on a frequency analysis of the \gaia astrometric time series and on the analytical determination of orbital elements using Fourier analysis \citep{Delisle2022}. To do so, a least-square periodogram \citep{1976Ap&SS..39..447L, 1982ApJ...263..835S} of the \gaia astrometric time series is built (\cite{Delisle2022}, see \MOD{Eq.~}\ref{eq:gaPeriodogram}), comparing for each frequency, the {\it chi-square} of a circular orbit model plus the parallactic motion  ($\chi^2_{9p}$)  to  the {\it chi-square} of the parallactic motion only ($\chi^2_{5p}$).
\begin{equation}
z_{\rm GLS}(\nu) = \frac{\chi^2_\text{5p} - \chi^2_\text{9p}(\nu)}{\chi^2_\text{5p}}
\label{eq:gaPeriodogram}
\end{equation}

The first step consists in drawing a set of frequencies from a log-uniform law $\log{\Big(\mathcal{V}\Big)} \sim U\Big(\log{(\nu_{\rm min})},\log{(\nu_{\rm max,init})}\Big)$ (between 2.5~d and the observation time span),  where  frequencies with a lower significance according to the SDE statistic \citep[Signal Detection Efficiency, see][]{2000ApJ...542..257A, 2002A&A...391..369K} are redrawn until selected. This procedures discards from the initial population,   periodic signals with lower probability, improving the efficiency of the GA.

The second step of the initialisation concerns the eccentricity and the mean anomaly. For 50\% of the chromosomes,
the eccentricity is drawn according to $\sqrt{e} \sim U(0, \sqrt{0.985})$ while the mean anomaly is uniformly drawn according to  $\M_0 \sim U(0,2\pi)$. For the remaining 50\% of the chromosomes, the eccentricity and the mean anomaly are analytically derived using the signal frequency decomposition described in \cite{Delisle2022} and drawn accordingly to $N\Big( \hat{e}, \sigma_{\hat{e}} \Big)$ and $N\Big( \hat{M_0}, \sigma_{\hat{M_0}} \Big)$.
Finally, the astrometric jitter $\sigma_{\rm jit}$ of each chromosome is drawn from a log-uniform distribution
$\log{\Big(\Sigma_0\Big)} \sim U\Big(\log{(0.005\,{\rm mas})},\log{(2 \sigma_\text{5p,res})}\Big)$.

The last stage of the initialisation phase of the GA consists in evaluating the merit function $\mathcal{M}$ of each chromosome in the population. 

%\WORKSUGGESTION{[BH: here we use `chromosome', in appendix we say `genome': decide on terminology]}

%\WORKSUGGESTION{[DSEG: suggests periodogram figure: population after  init in a P/ecc diagram and/or Distribution of the merit function. BH: not sure I have time to add it; DPAC referees: would you like to see an example?]}

\subsubsection{Evolution}
The evolution of the  population is done by randomly drawing chromosomes from the population and by applying to them genetic operators. The efficiency of the GA is improved by applying the genetic operators  on the non-linear parameters of the model only while the linear parameters are obtained through a linear regression.\\\\
Drawing Process : The first step consists in drawing a  %random local sub-population of 5x5 chromosomes (over the 60x60 full population)
local random sub-population of 5x5 to maximum 7x7 chromosomes (from the 80x80 full population)
%[INFO: width and height uniform drawn from the range [5,7], thus containing minimum 25 and maximum 49 genomes.]
over which several operators are applied, among which these four are  especially effective :
\begin{itemize}
    \item Full Crossover : Two chromosomes (mother \& father) are drawn from the sub-population from which  a child genome is breed with the frequency, mean anomaly, eccentricity and astrometric jitter randomly drawn from the mother \& father. The child chromosomes replaces the worst chromosomes in a local sub-population according to the merit function. 
    \item Harmony Mutation : A chromosome is randomly drawn from the sub-population and a new frequency is drawn among possible harmonics. The mutated chromosomes replaces the worst chromosome in a local sub-population according to the merit function.  This operator is efficient to find the period of eccentric orbits where the fundamental frequency is not always dominant.
    \item Alias Mutation : A chromosome is randomly drawn from the sub-population and a new frequency is randomly drawn from the aliases spectral window frequencies. The mutated chromosomes replaces the worst chromosome in a local sub-population according to the merit function. This operator is efficient to find the true fundamental frequency of unevenly sampled data.
    \item Simplex Mutation : The best chromosome is selected from the sub-population and is improved using a Nelder-Mead Simplex algorithm and replaces the worst chromosome in a local sub-population according to the merit function. This operator is efficient at the end of the evolution and allows to reach convergence towards the best merit function.
    
\end{itemize}
The computation time allocated to each genetic operator depends on its ability to improve the merit function at different stages of evolution. In order to avoid the population converging to a local maximum, a minimum computation time is assigned to all operators.

\subsubsection{Termination}
% BH: gaia.cu4.du437.ExtrasolarPlanetsPipelineFacadeImplProc.gaPopulationConvergenceThreshold = 0.95
Termination  is reached once 95 \% of the population has converged towards the maximum merit function or that the total computing time reaches 60 seconds. 
\subsection{Pipeline execution and best solution choice \label{sec:method_execution}}

The sequential pipeline execution of the DE-MCMC and GA orbit fitting modules is as follows: 

\begin{enumerate}

%\item \WORKSUGGESTION{[COPY FROM DOCUMENTATION:] Each algorithm is allowed to determine its best-fit parameters for a maximum of 60~s of single-threaded wall-clock CPU computation time.}
\item both modules are  executed \MOD{independently}, until convergence is achieved on an optimized best-fit configuration or for a maximum execution time of 60\,sec. Each algorithm produces the best-fit parameter solution based on their internal likelihood merit functions: Eq.~\ref{eq:mcmcLikelihood} for DE-MCMC and Eq.~\ref{eq:gaLikelihood} for GA;
%\item \WORKSUGGESTION{[COPY FROM DOCUMENTATION:] Finally, the best-fitting model parameters are determined by evaluating the Bayesian Information Criterion (\citealt{Schwarz1978}) metric: $\mathrm{BIC} = k\ln(n) -2\ln \mathcal{L}$ (where $k$ is the number of parameters estimated by the model and $n$ is the number of data points). The solution with the lowest BIC is selected as the best-fit solution. The published solution therefore originates from either the DE-MCMC or GA approaches, but the information on which was chosen for a given source is not published; }
\item when both modules have obtained convergence, the selection of the statistically preferred solution is made based on the Bayesian Information Criterion \citep[BIC,][]{Schwarz1978}: $\mathrm{BIC} = k\ln(n) -2\ln \mathcal{L}$ (where $k$ is the number of parameters estimated by the model and $n$ is the number of data points). The adopted best-fit model is the one with the lowest BIC. It is published if it passes the subsequent export filters (see Sect.~\ref{sec:sourceSelection}). No information is published on which module provided a particular source solution.  
%\FEEDBACK{[Need agreement from everyone on following sentence before passing the paper on to DPAC for evaluation:]}
%Overall, the vast majority of published solutions were from GA, which is expected since it is a maximum likelihood estimator.
For the \gdr{3} processing the GA always provided the maximum likelihood %it could find 
while the DE-MCMC provided the more conservative median of its posterior distribution around the best solution. This imbalanced solution comparison led to the vast majority of published solutions to be provided by GA; something we intend to improve upon in \gdr{4}.

\
% From checking the solutionType for kept sources (+1 for ga and -1 for mcmc):
% I count 1 from mcmc and 1161 from ga
%\WORKSUGGESTION{[BH: add statment about rough division over the algos that provided best solution]}

%Finally, the solutions of the two algorithms are compared on equal footing using the Bayesian Information Criterion (\citealt{Schwarz1978}) metric: $\mathrm{BIC} = k\ln(n) -2\ln \mathcal{L}$ (where $k$ is the number of parameters estimated by the model and $n$ is the number of data points). The solution with the lowest BIC is published if it passes the export filters (see Sect.~\ref{sec:solFilter}). No information is published on which method provided a particular solution.
%\item \WORKSUGGESTION{[COPY FROM DOCUMENTATION:] Symmetric estimates of the parameter uncertainties are obtained by reconstructing the covariance matrix directly from the Jacobians of all parameters for all observations.
\end{enumerate}

%\TODO{[From documentation `processing steps':] The modeling fitting process follows the following steps:
%\begin{enumerate}
%\item in both algorithms a (sparely sampled) periodogram is computed from which the initial MCMC chains and Genetic Algorithm population are seeded;
%\item for a maximum of 60~s each algorithm is allowed to optimize its best parameter fit to the astrometric data (single-threaded without GPU optimization);
%\item after the fit it made, a $\Delta$BIC comparison is made to choose the best fitting model of each algorithm, and the best one is chosen. Note that the archive does not contain information regarding the algorithm origin of the published result.
%\end{enumerate}}

%\WORKSUGGESTION{[Alessandro and/or Berry)]}

%------------------------------------------------------------------
% Source selection
%------------------------------------------------------------------
%\section{Source selection \label{sec:sourceSelection}}
\section{Source selection and solution filtering\label{sec:sourceSelection}}

%------------------------------------------------------------------
% Source selection SUB-SECTION: OrbitalAlternative
%------------------------------------------------------------------
%\subsection{Input stochastic solutions: \typeOrbAlt \label{sec:sourceSelection_orbAlt}}
\subsection{Stochastic solutions: \typeOrbAlt \label{sec:sourceSelection_orbAlt}}

\subsubsection{Input source list}
%\paragraph{\textbf{Input source list}} 
The \gaia non-single-star (NSS) processing pipeline tests out a variety of astrometric solution models \citep{DR3-DPACP-163,DR3-doc-NSS} 
%\WORKSUGGESTION{(see which? THIS SHOULD BE HALBWACHS ET AL. 2022, AND/OR DOCUMENTATION, POURBAIX ET AL. AS CITED IN PVP100, I THINK)} 
and if none fits the data to a satisfactory degree 
\MOD{the single-star solution from \egdr{3} was retained, in which the excess noise will have absorbed any unmodelled (presumably `stochastic') signal. Though the exoplanet pipeline is too computationally intensive to be run on all sources that pass through the NSS chain (see Sect.~\ref{sec:method}), it is however run on this sample of 2,457,530 presumed stochastic sources to look for difficult to detect orbital signals with both the DE-MCMC and GA algorithms. This}
%That
sample is mostly composed of faint and distant sources.

\subsubsection{Solution filtering}
%\paragraph{\textbf{Solution filtering}} 
%[\WORKSUGGESTION{This section is 99\% a copy of the documentation, need `reworking'.]}
Force-fitting the sample of sources that had received a stochastic solution returned a vast majority of solutions of dubious quality, primarily due to known aliasing effects with scanning law periodicities \citep{DR3-DPACP-164}. An aggressive filtering strategy was therefore applied to the output of the exoplanet pipeline in order to provide a sub-sample of candidate solutions for which the likelihood of retaining spurious solutions would be minimized. We filtered solutions utilizing the following constraints:
%\TODO{[BH: removed: "... provide a sub-sample, small as it might be, of statistically solid solutions."]} 
%In particular, orbital solutions were retained, which simultaneously fulfilled the following constraints:

\begin{itemize}
    \item fractional difference in parallax between the one fitted by the exoplanet pipeline and that in the original EDR3 single-star solution $< 5\%$;
    \item statistical significance of the derived semi-major axis of the orbit $a_1/\sigma_\mathrm{a_1} > 20$ (uncertainty derived from the Thiele-Innes parameters using linear error propagation);
    \item ratio of the EDR3 astrometric excess noise to the uncorrelated jitter term fitted by the exoplanet pipeline $> 20$;
    \item number of individual FoV transits $>36$; 
    %\WORKSUGGESTION{[BH: Confirmed, previously I looked at the Targeted sample instead..]}
    \item EDR3 parallax of the source $>0.1$ mas.
%    \WORKSUGGESTION{[OUR BH: a detail, but which parallax, the one we derived or the AGIS one?]}
\end{itemize}

Overall, the sample that survived the filtering process is composed of 629 orbital solutions. The selected solutions have a period distribution mostly free of the doubtful spurious values as discussed in Sect.~\ref{sssec:results_spuriousOrbits}. The filtered sample is published in \gdr{3} with the \nssSolutionType \typeOrbAltPVal, and it underwent careful inspection for verification and validation purposes, as described in Sect.~\ref{sec:results}).

%------------------------------------------------------------------
% Source selection SUB-SECTION: OrbitalTargetedSearch
%------------------------------------------------------------------
\subsection{Input source list: `OrbitalTargetedSearch' \label{sec:sourceSelection_orbTargetedSearch}}

\subsubsection{Input source list}
%\paragraph{\textbf{Input source list}} 
The modules of the exoplanet pipeline have been in development from a time before the launch of \gaia and were verified mostly with the help of simulated data. Only from DR3 the number of epochs and calibration noise-level were sufficient to fit meaningful orbits with the exoplanet pipeline. In order to test its performance with real data we compiled source lists that would serve the following purposes: 

%The modules of the exoplanet pipeline have been developed before the launch of \gaia and verified mostly with the help of simulated data. To test its performance with real data, when those were becoming available, we compiled source lists that would serve the following purposes: 
% In broad terms, the following selection criteria were applied:
% of sources for DU437 testing and verification purposes.
% Several purposes:
\begin{itemize}
    \item Pipeline testing, verification, and validation, e.g.\ for demonstrating that the orbits of known exoplanets can be detected.
    \item Sample the properties of Gaia astrometric time-series in different regimes, e.g.\ bright and faint, and investigate how that influences pipeline performance.
\end{itemize}

As we progressed in understanding the performances of the non-single star pipelines and when the input source selection for the binary pipeline was finalised \citep{Halbwachs:2022}, it was decided to perform a dedicated (additional) run of the exoplanet pipeline on a pre-defined source list, hence the orbital name suffix \typeOrbTar. 
%targete became clear that the 
% \WORKSUGGESTION{[We have to discuss how to best phrase this]}

Starting from the previously defined list of test sources, we therefore compiled a more extensive sample for the targeted search. We identified sources for which information on the presence/absence of exoplanets and substellar companions was available in the literature, typically these were stars included in observational planet-search programs. The list included: 
\begin{itemize}
    % \item DEDREQ-695 list described \href{https://issues.cosmos.esa.int/gaia/browse/DEDREQ-695?focusedCommentId=170665&page=com.atlassian.jira.plugin.system.issuetabpanels\%3Acomment-tabpanel#comment-170665}{here}
    \item Sources in the Nasa Exoplanet Archive\footnote{\url{https://exoplanets.nasa.gov/discovery/exoplanet-catalog/}}. These are hosts of confirmed and candidate exoplanets discovered with various observation techniques.
    \item Sources in planet-search programmes, predominantly using spectrographs for precision radial-velocity measurements. This included the samples of e.g.\ HIRES \citep{2017AJ....153..208B}, CORALIE \citep{Udry:2000kx}, HARPS \citep{Mayor:2003cs}, SOPHIE \citep{Bouchy:2009rt}, and HARPS M-dwarfs \citep{Bonfils:2013aa}.  
    % - direct imaging (NACO-ISPY)\\
    \item Sources in known astrometric binaries from the Hipparcos binary solutions compiled in Table F1 of \citet{Leeuwen:2007kx}.
\end{itemize}
All the source samples above consist predominantly of bright stars ($G\lesssim 10$). We complemented these with sources that promised compelling scientific outcomes in the case of orbit detection, and that may otherwise have not been processed with an orbit-fitting pipeline. For example, the binary pipeline only processed sources with $\gmag<19$, regardless of distance. To probe fainter sources that yet remain within a distance horizon that in principle allows the detection of signals caused by sub-stellar companions, we included the ultra-cool dwarf sample of \citet{Smart:2019aa} and metal-polluted white dwarfs within 20 pc from the \citet{Sion:2014aa} and \citet{Giammichele:2012aa} compilations. The total number of unique sources selected for the targeted search was 19\,845.

We obtained \gdr{2} source identifiers of these targets either directly from the respective catalog, from Simbad \citep{Wenger:2000ve}, or from a positional crossmatch with the \gdr{2} catalog. The corresponding \gdr{3} identifiers where then submitted for the dedicated processing run. 

\subsubsection{Solution filtering}\label{ssec:sol_filter}
%\paragraph{\textbf{Solution filtering}} 
%\WORKSUGGESTION{[JBD+BH+NU: it should be clearly stated in this section that the filtering was done not very consistent and based on much visual inspection, and to convey that users should NOT perform any statistical analyses on our results.]}

In-depth scrutiny of the resulting solutions was performed on multiple levels, in an attempt to retain the most sensible orbits. The different steps taken in order to filter out implausible solutions, which we detail below, have an important degree of heterogeneity, and our final choices translate in a complex selection function. 

Our approach to solution filtering was three-fold. We first defined various indicators of the statistical significance of the solutions, then we fine-tuned their threshold values with an iterative process, and lastly performed a selection of different sub-samples of solutions based on different choices of subsets of the indicators. The statistical filters included an extensive model comparison with alternative, less-complex models to safeguard detection of bona-fide candidates, distance-dependent threshold for the orbit significance, relative agreement between the fitted parallax and the AGIS parallax, an upper limit to the derived value of the mass function, a constraint on closed orbits, and variable thresholds for the value of the ratio of the AGIS astrometric excess noise to the Keplerian jitter term in the solution. %This is 

We used two sets of statistical filters, in each set the filters were applied in conjunction and then the full list was constructed with the targets that passed the filters of either one OR the other set.

%The resulting solutions were scrutinised to retain only sensible orbits. As described in Sections \ref{sssec:results_bicModelComp} and \ref{sssec:validationRv} \WORKSUGGESTION{[revise this list when done with description below, can also reference the general sections, if filtering extended beyond these two subsections]} this included evaluating various statistical indicators for the significance of the solution, and visual inspection of the orbits. The number of sources that passed this stage is 

%Choosing good statistical indicators proved difficult and thus a second step was necessary where solutions were looked at individually against literature data or cross matched with independent Gaia solutions to either validate these solutions or remove spurious solutions (see Sect. \ref{sec:results_validation}). From the 742 targets that passed the statistical filtering, 533 sources remained and those are published in \gdr{3} with the \nssSolutionType \typeOrbTarPVal. Of those, 188 sources were validated with the \nssSolutionType \typeOrbTarVal.

% \FEEDBACK{NU: please check if it's ok like this.}

Set 1:

\begin{itemize}
    \item $\mathrm{BIC}_{\mathrm{Kep}} - \mathrm{BIC}_{\mathrm{5}} < -30$
    \item $\mathrm{BIC}_{\mathrm{Kep}} - \mathrm{BIC}_{\mathrm{7}} < -30$
    \item $\varpi > \sigma_{jit}$
    \item $|\Delta\varpi| < 0.5 \ a_1$
    \item $a_1 / \sigma_{a_1} > 2$
    \item $P < \Delta T$
    \item $f(\mathcal{M}) < 0.02$
    \item $\sigma_{jit} < \max(0.1, 2\sigma_{jit,agis})$
    \item $\sigma_{STD} < 1.5\sigma_{MAD}$
\end{itemize}

Set 2:  %\FEEDBACK{Torino Filters}

\begin{itemize}
%    \item $|\Delta\varpi| < 0.05$~mas \FEEDBACK{NU: looking at the gitlab I found a comment from Paolo that says that the difference in parallaxes was required to be less than 5 \% (\textbf{not} less than 0.05 mas). If that's right remove this item and keep the next one}
    \item $|\Delta\varpi| < 0.05 \, \varpi$
    \item $a_1/\sigma_{a_1} > 5 $
    \item $\sigma_{jit,agis}/\sigma_{jit} > 5 $ %\FEEDBACK{NU: same thing here. From the comment of Paolo he says the ratio was 1.5 (\textbf{not} 5). If that's right remove this item and keep the next one}
%    \item \FEEDBACK{$\sigma_{jit,agis}/\sigma_{jit} > 1.5 $}
\end{itemize}
%\FEEDBACK{NU: Torino should check if these were all the filters they used.}

where $\mathrm{BIC}_{\mathrm{Kep}}$ is the BIC for the Keplerian solution, and $\mathrm{BIC}_{\mathrm{5}}$ and $\mathrm{BIC}_{\mathrm{7}}$ are the BIC for the 5 and 7 parameter solution respectively\footnote{\citet{Ranalli:2018aa} presents an independent approach of using BIC for model selection.}.
$\varpi$ is the parallax from the Keplerian solution, %\FEEDBACK{JSA: I think we should use $\varpi$ instead of $\pi$, yet we need to distinguish between AGIS and DU437 parallax}
$|\Delta\varpi|$ is the absolute difference between the parallax of the Keplerian solution and the AGIS parallax, $a_1$ and $\sigma_{a_1}$ are respectively the semi-major axis and its uncertainty, $P$ is the period of the companion, $\Delta T$ is the time span of observation for each target, $f(\mathcal{M})$ is the mass function of the primary/companion system and is calculated as $f(\mathcal{M}) = \nu^2 a_1^3/G $ (where $\nu = 2\pi / P$ , is the orbital frequency), $\sigma_{jit}$ is the astrometric excess noise as calculated by the exoplanet pipeline, $\sigma_{jit,agis}$ is the astrometric excess noise from the 5-parameter AGIS solution, $\sigma_{STD}$ is the weighted standard deviation of the residuals of the Keplerian solution, and $\sigma_{MAD}$ is 1.4826 times the Median Absolute Deviation (MAD) of the residuals of the Keplerian solution. 

We complemented the statistical filtering approach with visual inspection of individual orbits (%\FEEDBACK{JSA: remove "subjective as the criterion might be"} subjective as the criterion might be, 
see Sect \ref{sssec:orbit_visualisation}). In this way we looked for symptoms of spurious results due to e.g. important numbers of outliers and/or correlated residuals even for cases of statistically robust solutions. Finally, we used a threshold (10\%) in the relative agreement between the fitted value of $P$ and that from existing literature data and independent Gaia solutions as additional discriminant, in an attempt to recover bona-fide solutions that otherwise might have been discarded based on too-stringent statistical filtering. The final number of sources with solutions accepted for publications in \gdr{3} is 533, of which 188 were validated with the \nssSolutionType \typeOrbTarVal (see Sect.  \ref{sec:results_validation}).

The difficulties we faced in converging on a coherent approach for the identification of well-defined classes of robust solutions and spurious orbits are illustrative of the challenges inherent in the \gdr{3} NSS processing, particularly in the limit of low astrometric SNR (and correspondingly low companion mass) for bright stars, which are still affected by limitations in the error model, as well as the generally low number of visibility periods (see Sect. \ref{sec:dataProperties}). We caution users against performing detailed statistical analyses with this sample of orbital solutions. 

\section{Results \label{sec:results}}

In total the exoplanet pipeline populates 1162 orbits in the \gdr{3} table \nssTwoBodyOrbit into four \nssSolutionType: \typeOrbAlt (619), \typeOrbAltVal (10), \typeOrbTar (345), and \typeOrbTarVal (188).

Calibrated companion mass estimates require additional non-astrometric information and thus are out of the scope of this paper; they are presented in \citet{DR3-DPACP-100}. 

%As this paper is limited to astrometric data only for the parameter estimates we cannot make accurate mass estimates given that our solution will only be able to provide \WORKSUGGESTION{[insert mass function formula]}. Proper mass estimates for our targets are found in \cite{DR3-DPACP-100}, though here   we cannot combine non-astrometric information into the analyses we cannot discuss the this paper 
  
To ease the interpretation of the \typeOrbAltPVal and \typeOrbTarPVal sub-sample \MOD{figures}, they are always shown side by side: the former on the left and the latter on the right. For brevity we will in the text below refer to these two categories as \typeOrbAltStar and \typeOrbTarStar to refer to the union sample of the non-validated and validated solutions in both categories.

%The fitted parameter distributions of these categories is summarised in the figures that follow: %Figs.~\ref{fig:sumFig1}, \ref{fig:a1SumFigHist}, \ref{fig:skyPlots} and \ref{fig:agisExcessNoise} where the 
%in general the left panel(s) show the combined 629 \typeOrbAltPVal solutions and the right panel(s) the combined 533 \typeOrbTarPVal solutions. 

\subsection{General overview}

%\paragraph{\textbf{Sky distribution}} 
\subsubsection{Sky distribution} 

\begin{figure*}[ht!]
  \includegraphics[width=0.98\textwidth]{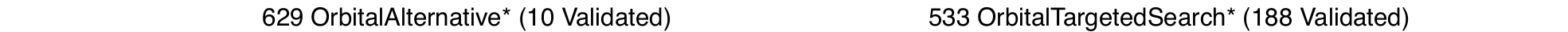} 
  \includegraphics[width=0.49\textwidth]{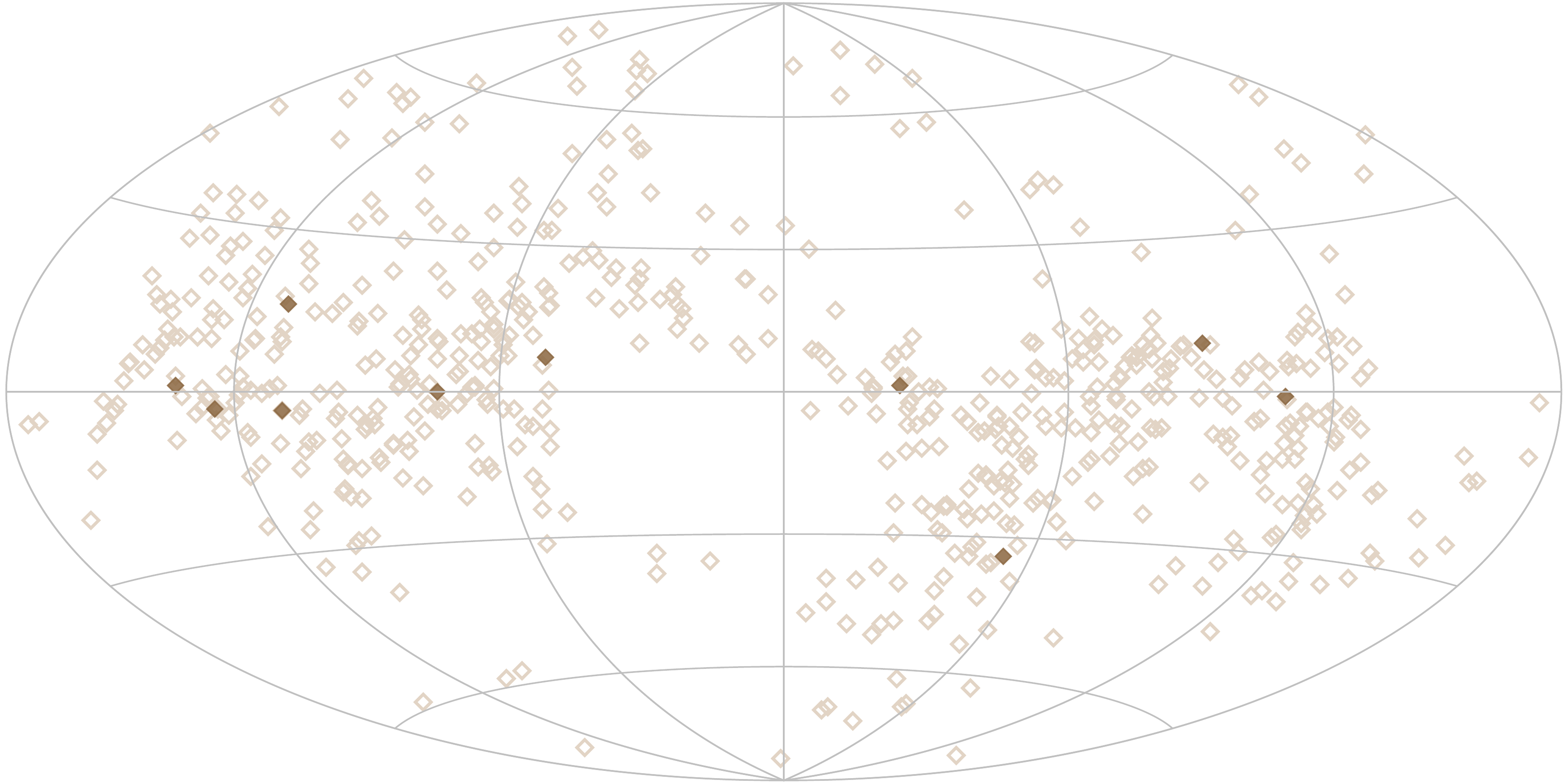} 
  \includegraphics[width=0.49\textwidth]{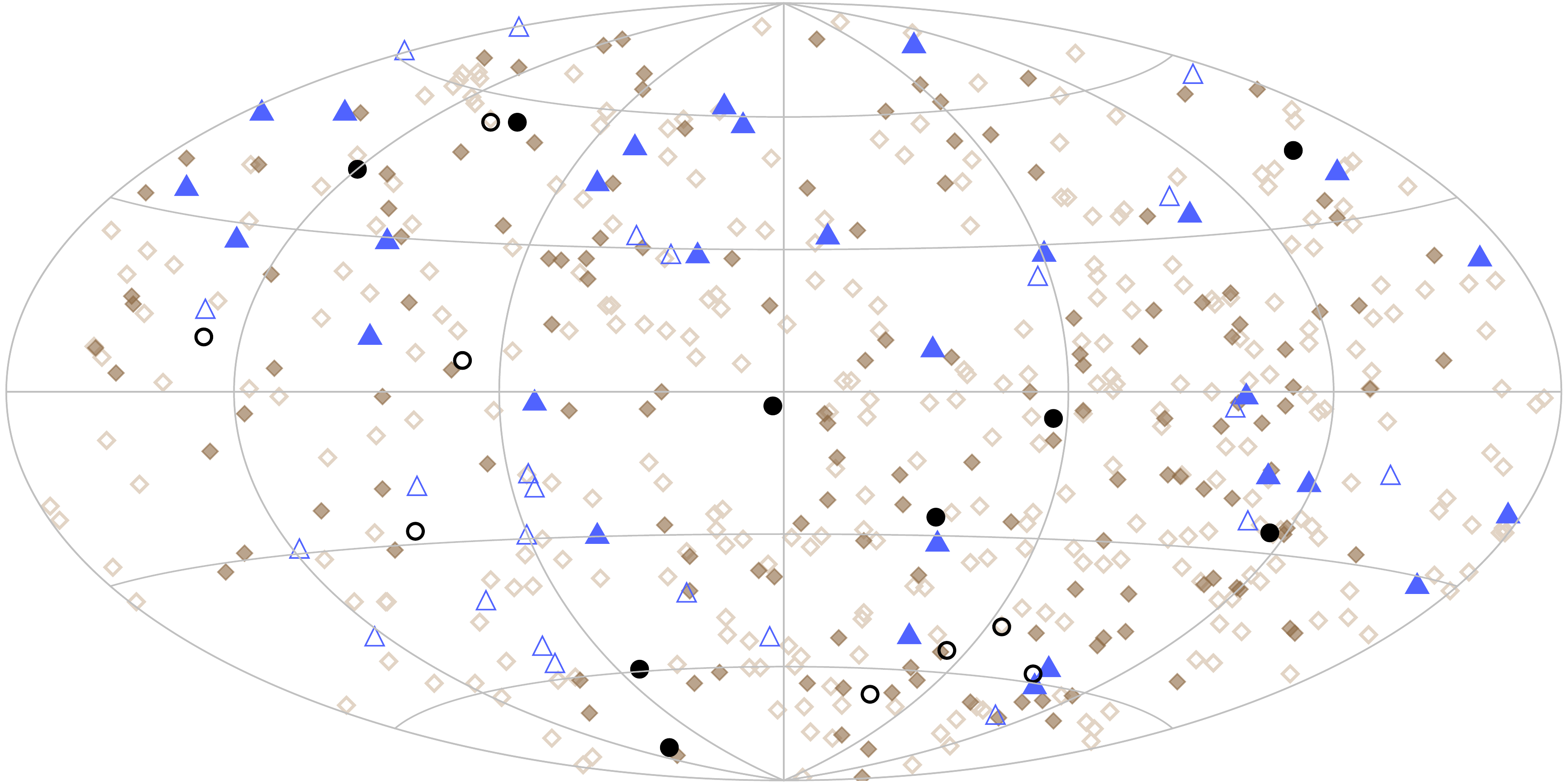}   
  \includegraphics[width=0.49\textwidth]{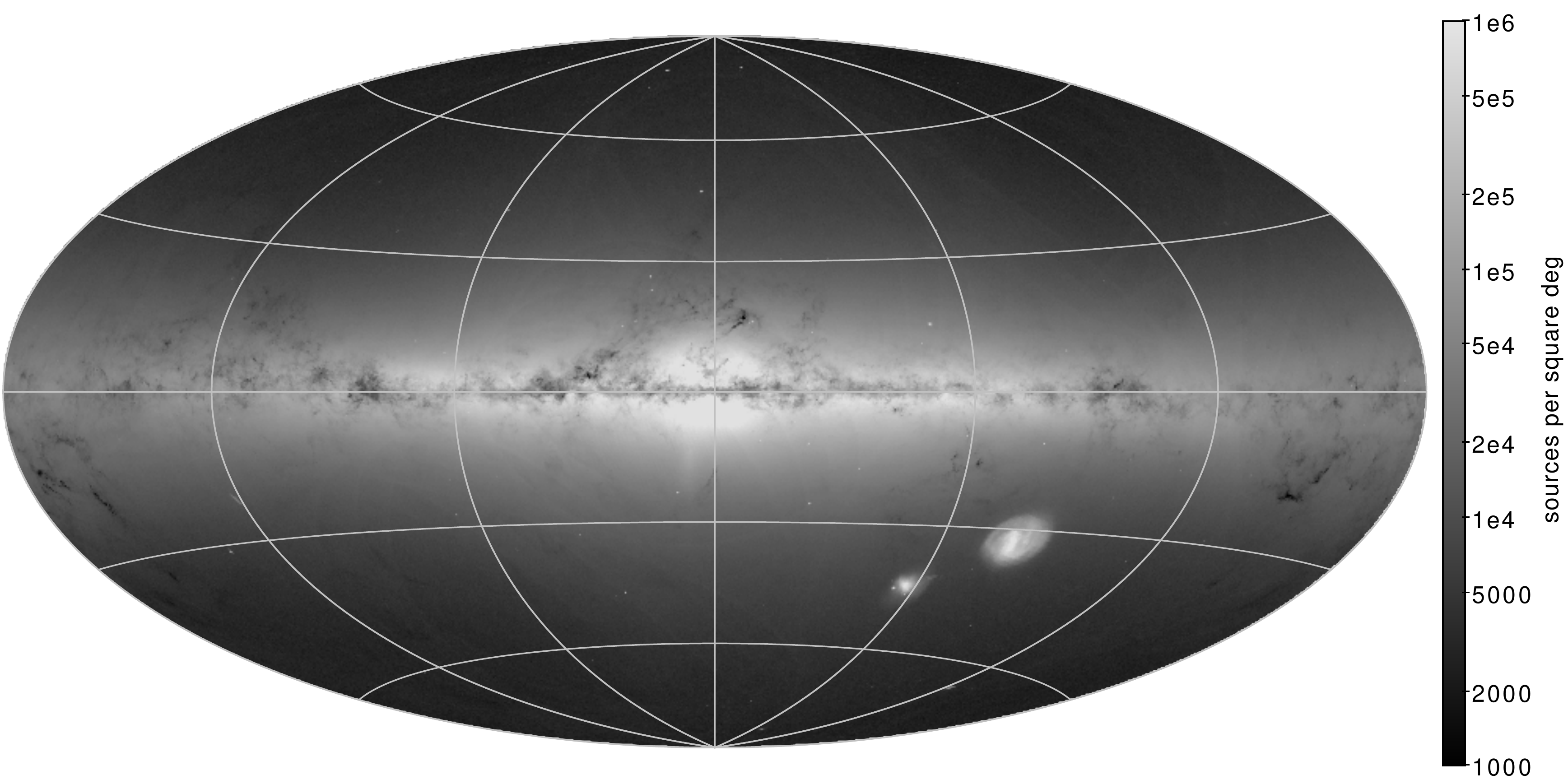} 
  \includegraphics[width=0.49\textwidth]{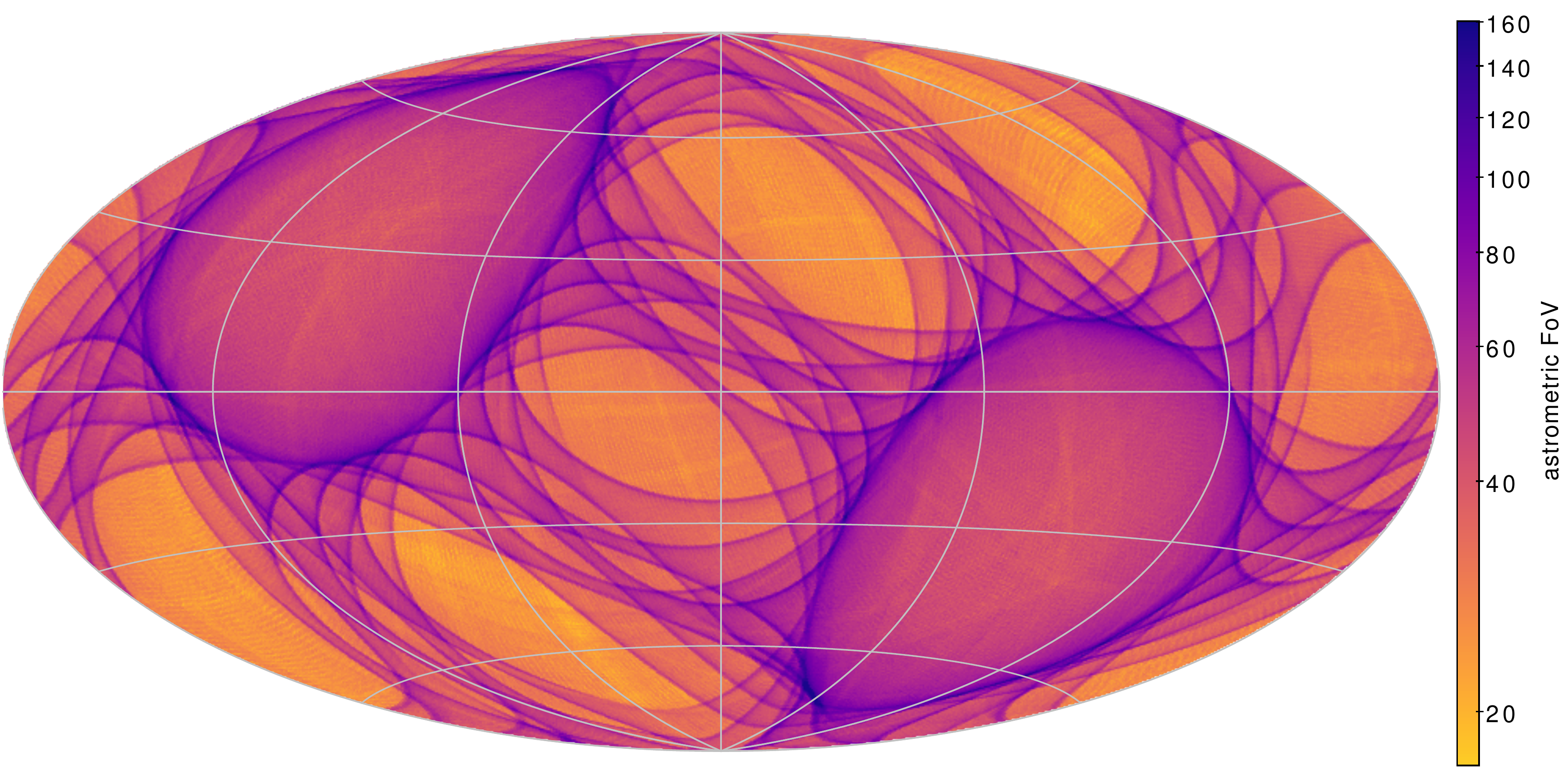}  	
%  \vspace{-0.3cm}
\caption{Galactic sky distribution of our published solutions\MOD{, longitude increasing to the left. We broadly categorise the different mass regimes using the pseudo-companion mass \Mcomp assuming a solar-mass host %(\Mhost=1\Msun) 
and define three solution groups:
   17 (9 validated) solutions with companions in the planetary-mass regime (\Mcomp~<~20~\Mjup, black circles), 52 (29 validated) in the brown dwarf regime (20~\Mjup~$\leq$~\Mcomp $\leq 120$~\Mjup, blue triangles), and 1093 (160 validated) in the low mass stellar companion regime (\Mcomp~>~120~\Mjup, orange diamonds). Validated targets are plotted as (dark) filled symbols, while open symbols are used for the non-validated targets. Same symbols are used in Fig.~\ref{fig:a1Sn} and following figures.
   %All sources in the upper left panels (showing  \typeOrbAltPVal solutions) are in the highest mass category and thus contain only orange diamonds, which would make it near-impossible to distinguish filled symbols. Therefore the validated targets on the left panels are filled with a darker orange colour to make them readily identifiable.
Top left panel: \typeOrbAltStar;
%, all in the highest mass category and thus contain only orange diamonds, which would make it near-impossible to distinguish filled symbols. Therefore the validated targets on the left panels are filled with a darker orange colour to make them readily identifiable.; 
top right panel: \typeOrbTarStar; bottom left panel: \gaia DR3 source sky density; bottom right panel: (maximum) number of DR3 astrometric FoV transits.}}
\label{fig:skyPlots}
\end{figure*}

The sky distribution of our solutions is shown in Fig.~\ref{fig:skyPlots}.
Clearly the filtering on \typeOrbAltStar has selected sources in regions of the sky with sufficiently dense sampling, causing `holes' mainly around low ecliptic latitudes ($|\beta|<45^\circ$) that are generally less well sampled. In contrast, the external input catalog-based \typeOrbTarStar has a much more uniform distribution.

%\paragraph{\textbf{Signal to noise estimate}}

\subsubsection{Signal to noise estimate}

\begin{figure*}[ht!]
  \includegraphics[width=0.98\textwidth]{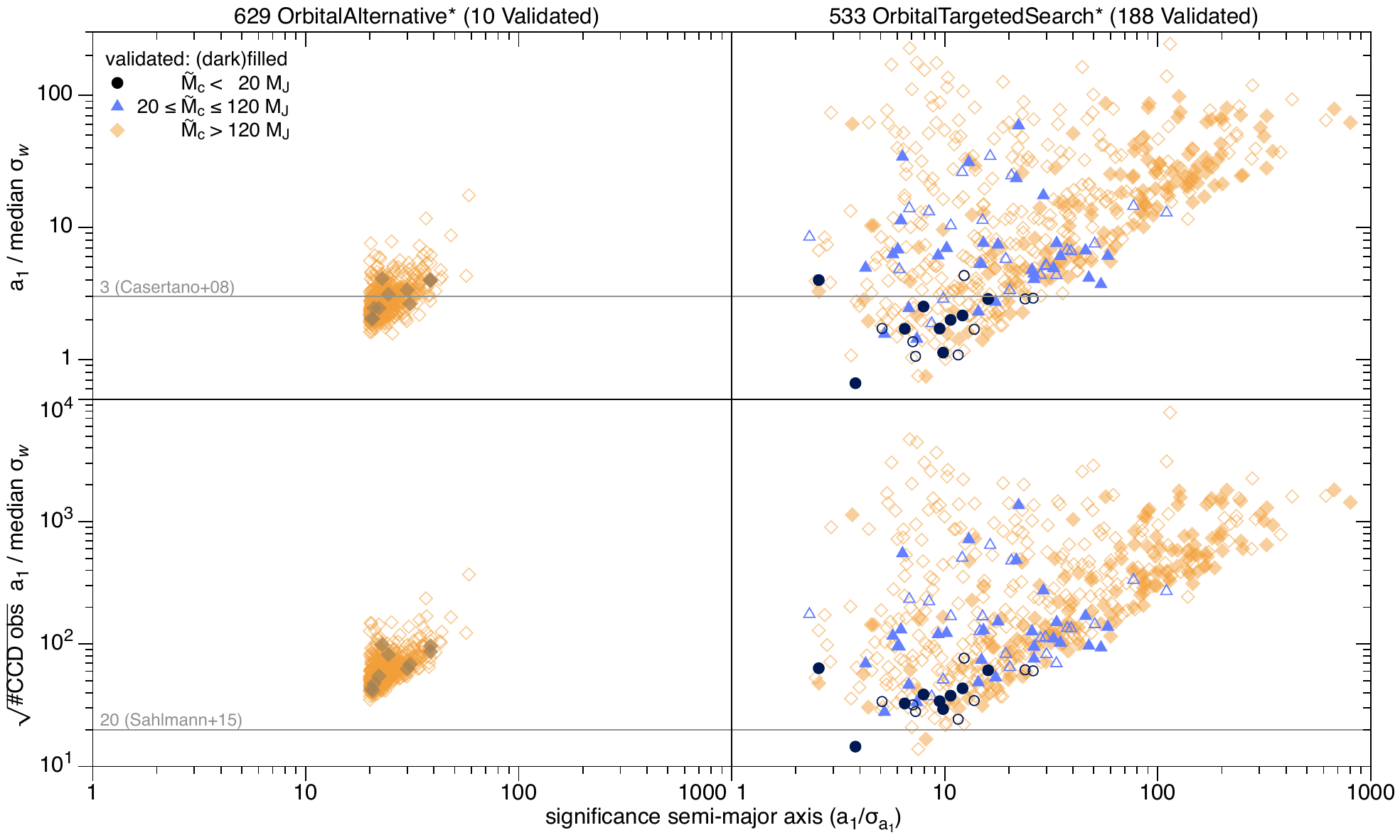}
  \vspace{-0.2cm}
\caption{Signal to noise ratio of $a_1$ with respect to the median abscissa uncertainty. Top panel: statistic used in \cite{2008A&A...482..699C} with its typical proposed threshold of 3, bottom panel: statistic used in \cite{Sahlmann:2015aa} with its proposed threshold of 20.} 
\label{fig:a1Sn}
\end{figure*}

In Fig.~\ref{fig:a1Sn} we explore the approximate signal-to-noise ratio of our results, where the fitted semi-major axis ($a_1$~[mas]) is taken as a proxy for the signal level, and the median abscissa uncertainty as the noise proxy. We compare two commonly used definitions: the top panels show that of \cite{2008A&A...482..699C} with its typical proposed threshold of 3, and the bottom panel shows that of \cite{Sahlmann:2015aa} with its proposed threshold of 20. %\FEEDBACK{JSA: 
% 3/1162   = 0.00258 = 0.3% of all in DR3
% 3/533    = 0.0056. = 0.6% of OrbitalTargetedSeach
% 515/1162 = 0.4432  = 44% of all in DR3
Only three (0.3 \%; two validated) solutions fall below the \cite{Sahlmann:2015aa} threshold, whereas 515 (44 \%; 33 validated) solutions fall below the \cite{2008A&A...482..699C} threshold.
%}

We see that the majority of our sample has reasonable to high signal to noise, though targets having very low signal to noise levels generally are the least massive companions, as expected.

On the ordinate axis we plot the significance of the derived semi-major axis ($a_1/\sigma_{a_1}$), which shows the expected trend that higher signal to noise is associated with better constrained parameter estimates.

\subsubsection{Goodness-of-fit statistics\label{sssec:gof}}
\begin{figure}[ht!]
  \includegraphics[width=0.98\columnwidth]{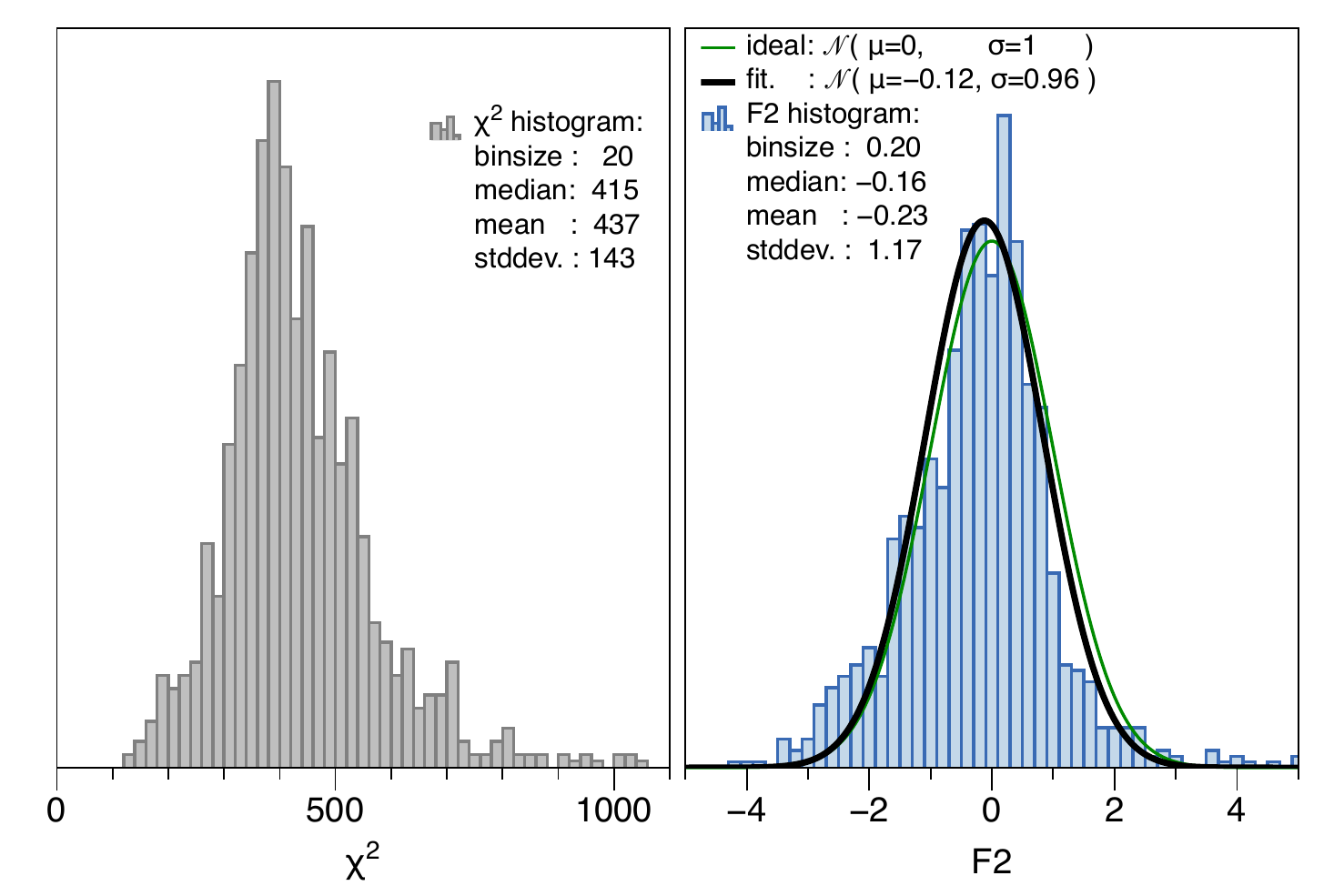}  
  \vspace{-0.2cm}
\caption{Goodness of fit statistics. Left panel: $\chi^2$ (\texttt{obj\_func} in Table~\ref{tab:archiveTabFields}); right panel: F2 (\texttt{goodness\_of\_fit}  in Table~\ref{tab:archiveTabFields}).}
\label{fig:gof}
\end{figure}

In Fig.~\ref{fig:gof} we present the goodness-of-fit statistics that are available in the \gaia data archive (Sect.~\ref{ssec:archiveModelParams}): $\chi^2$ (\texttt{obj\_func}) and F2 (\texttt{goodness\_of\_fit}). While the former is difficult to interpret globally without compensating for the varying number of degrees of freedom (i.e. the $\chi^2_\text{red}$), the latter F2, ‘gaussianized chi-square’, is expected to follows a normal distribution with zero mean and unit standard deviation, as shown with the thin green line. Comparison to the histogram and a fit to it (thick black line) shows that the distribution is not completely symmetric but generally is rather well-behaved. This can however largely be subscribed to the inclusion of the non-linear jitter term $\sigma_\text{jit}$ model parameter (see Sect.~\ref{ssec:mathModelDescr} and Figs.~\ref{fig:a1SumFigHist} and \ref{fig:agisExcessNoise}) which was meant to absorb any unmodelled variance in the data, and thus likely contributed to the `normalisation' of the expected goodness-of-fit statistics. %\WORKSUGGESTION{[BH: Do you agree with this statement?]}

%\paragraph{\textbf{Parameter distributions}}
\subsubsection{Parameter distributions\label{sssec:paramDistr}}

\begin{figure*}%[t]
  \includegraphics[width=0.98\textwidth]{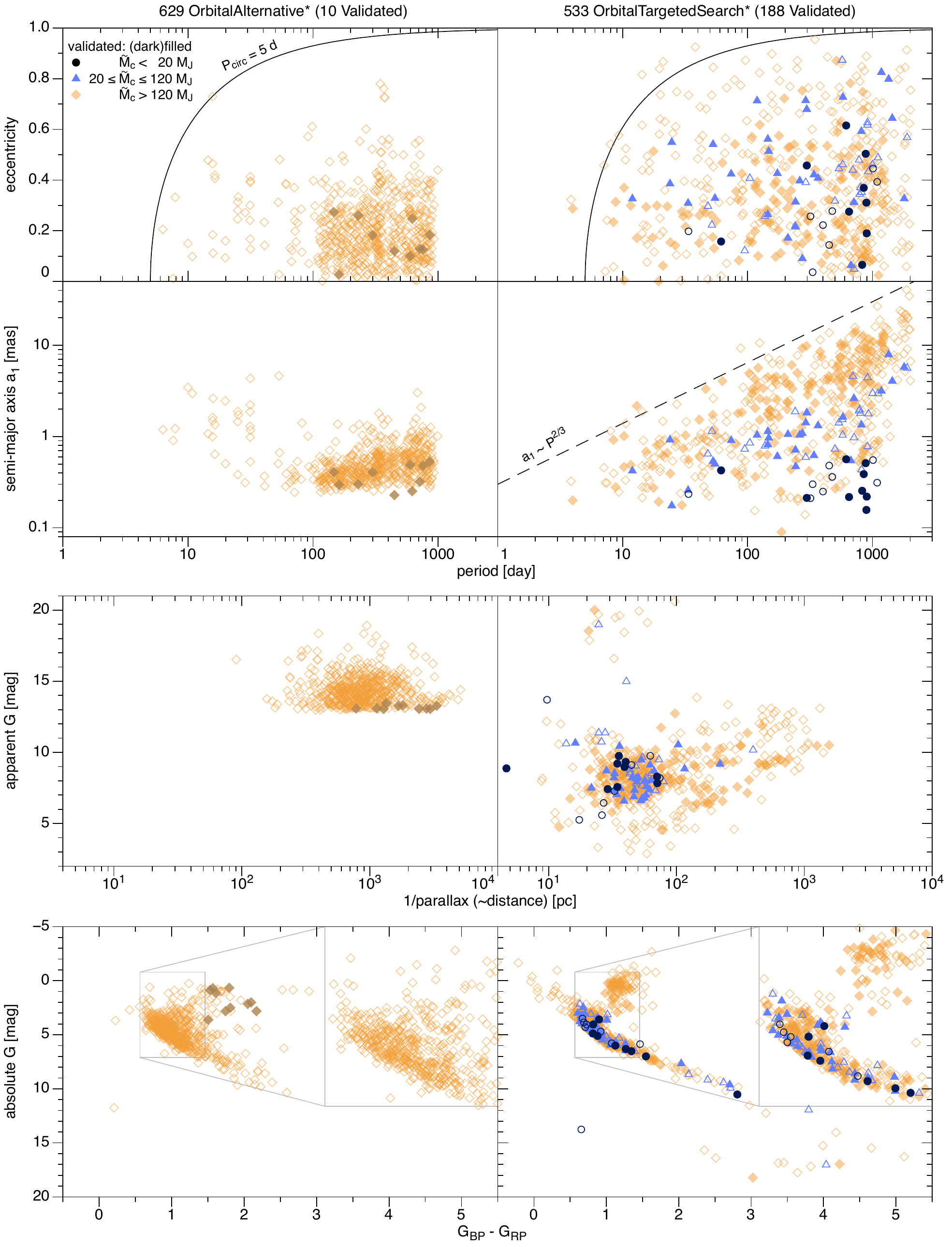}  	
  \vspace{-0.3cm}
\caption{ Parameter distributions of the published solutions: period-eccentricity  and period versus semi-major axis (top panels), apparent magnitude versus inverse parallax (third panel), zero-extinction absolute magnitude versus colour (bottom panel).
See Sect.~\ref{sssec:paramDistr} and \ref{sec:results_verification} for discussion. }
\label{fig:sumFig1}
\end{figure*}

Starting with Fig.~\ref{fig:sumFig1}, we see in the top row panels that the period eccentricity is well constrained by a 5-day circularisation period \citep[formulation of][]{Halbwachs2005} and that our validated targets generally have medium to low eccentricities. Due to the aggressive filtering on the \typeOrbAltStar mainly periods above 100~d and orbits with eccentricities below 0.4 were selected. 

The second row right panel show that the different pseudo-mass groups roughly follow a $a_1 \propto P^{2/3}$ relation with different offsets, which is what one would expect from Kepler's third law for the astrometric signal of systems with similar distance and mass (ratio) companions. Looking at the third row right panel we see that indeed the sources are roughly at a typical distance of about 50~pc. In the left panel we see that the distance distribution is rather widely spread out and thus would not produce a nice relation in the period versus $a_1$ plot above it. Note the excess of high semi-major axis solution for short periods in the left plot: these are likely spurious or incorrect period detections \cite[see][]{DR3-DPACP-100}.

The third row panels also illustrate that the `blind search' \typeOrbAltStar sample (left) is at much fainter magnitudes and larger distances than the \typeOrbTarStar (right), the latter being largely compiled from radial velocity literature sources thus not surprisingly consisting of mostly relatively bright targets.

The fourth row of panels show the zero-extinction absolute magnitude estimate, based on parallax and G-band apparent magnitude, see discussion in Sect.~\ref{sssec:hrDiagram} for more details. %As shown, most of our host-stars lie along the main sequence, though a small fraction are likely (sub-)giants.

\begin{figure*}
  \includegraphics[width=0.98\textwidth]{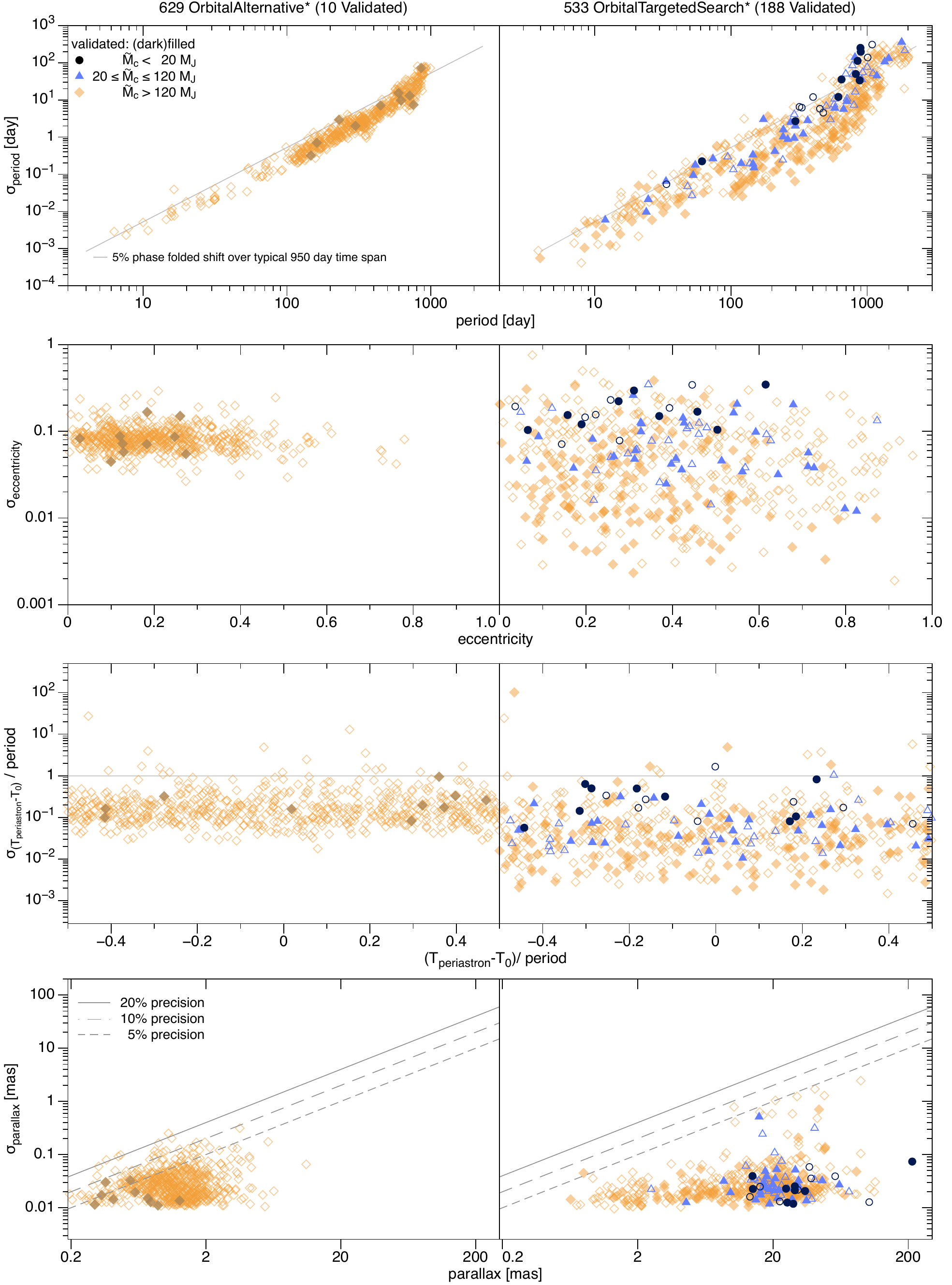} 
  \vspace{-0.3cm}
\caption{Uncertainties for period (top panel), eccentricity (second panel), periastron epoch (third panel), and parallax (fourth panel) as function of the parameter value itself. See Sect.~\ref{sssec:paramUnc} for discussion.}
\label{fig:sigmPETp}
\end{figure*}

Note that both the discussed parallax-based `distance' and absolute magnitude estimates are meaningful given that the uncertainty on the parallax is generally (much) smaller than 20\% of the value, see bottom panel of Fig.~\ref{fig:sigmPETp}. This is further supported % alternatives: strengthened / reinforced
by the relatively `tightness' of the HR diagram in the bottom right panel.

\begin{figure*}%[t]
  \includegraphics[width=0.98\textwidth]{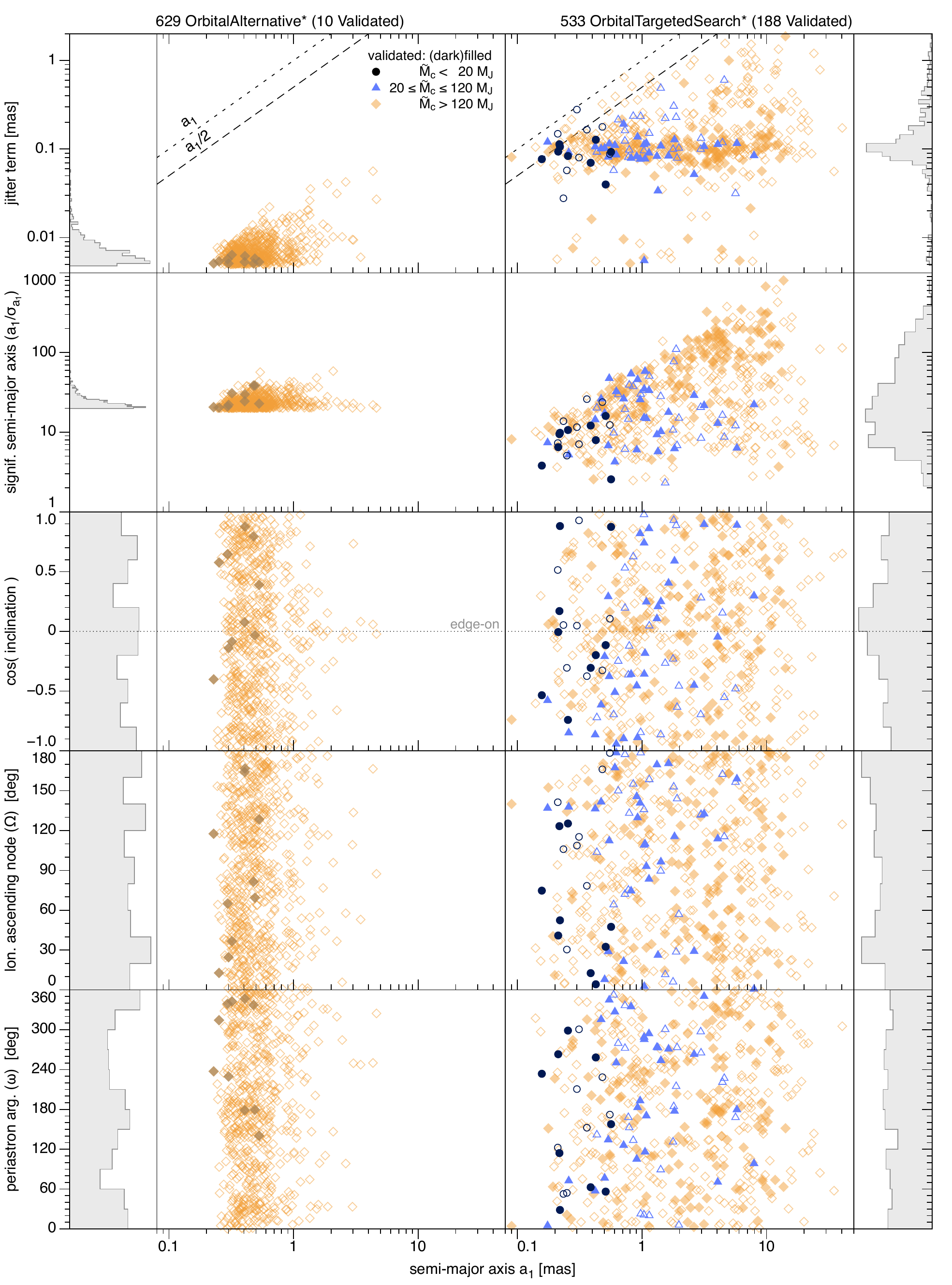}  			
  \vspace{-0.3cm}
\caption{Parameter distributions as function of (photocentre) semi-major axis $\mathrm{a_1}$: jitter term (top panel), semi-major axis (second panel), cosine of the inclination (third panel), longitude of the ascending node (fourth panel), and periodastron argument (bottom panel). See Sect.~\ref{sssec:paramDistr} for discussion.}
\label{fig:a1SumFigHist}
\end{figure*}

Figure~\ref{fig:a1SumFigHist} presents us with several parameters as function of $a_1$. We start with the jitter term, which for most \typeOrbAltStar is around or below an insignificant 0.01~mas. For the \typeOrbTarStar the level is generally around 0.1~mas, and for about a dozen the jitter level is above $a_1/2$, i.e., an (unmodelled) noise of the same order as the orbital solution semi-major axis.

The second row of panels shows the significance of the semi-major axis ($a_1/\sigma_{a_1}$), which is around 20-30 for the \typeOrbAltStar sample and spans several orders of magnitude for the \typeOrbTarStar sample. As expected, in both cases the validated samples generally have a relatively high significance.

The cosine inclination distribution on the third row of panels is rather flat for the \typeOrbAltStar as expected for random oriented systems. For the \typeOrbTarStar we see an excess of edge-on systems, as expected given that the input selection was mainly based on radial velocity targets. 
The bottom two rows show the longitude of the ascending node ($\Omega$) and periastron argument ($\omega$) which both are relatively flat for both samples, as expected for randomly-oriented orbits. 
A general discussion of the expected distributions and observed biases in the geometric orbital elements of astrometric orbits is given in \citet{DR3-DPACP-100}.

\subsubsection{Parameter uncertainties\label{sssec:paramUnc}}
We systematically inspect the parameter uncertainties of all fitted parameters versus their value and do not see any unexpected trends or outliers. As noted in Sect.~\ref{ssec:mathModelDescr}, we exported the formal uncertainties as provided from the covariance matrix of the best-solution least-squares solution, without any scaling. As we know from the astrometric jitter term that there is some level of unmodelled noise left in the data, these values might not always give reliable estimates of the true uncertainties of these parameters. See also Sect.~\ref{ssec:TItoCampbell} related to propagation of uncertainties on derived parameters.

For a few interesting parameters we plot the data shown in Fig.~\ref{fig:sigmPETp}: the top panels show the period uncertainty versus period along with an observation time span (i.e. cycle-normalised) phase shift of 5\%, below which almost all solutions lie, except for the longest periods as expected due to the mild constrains on the period due to the very low cycle count.

The second row of panels show the eccentricity uncertainty versus their value, which is typically between 0.06 -- 0.1 for the \typeOrbAltStar, but varies over a much wider range for the \typeOrbTarStar, though the majority of the latter still remains below a meaningful uncertainty of 0.2. 

The epoch periastron has been distributed between -0.5 -- 0.5 of the period around $T_0$ as shown in the third row of panels, and is rather flatly distributed in this range, as expected. A relative uncertainty above 1 (solid line) clearly is non-informative, which luckily only happens for a dozen of objects in either category.

Finally, we show also the relative parallax uncertainty in the bottom row of panels, which shows that the majority of our sample has relative precision better than 5\%, and almost all better than 10\%. Given that 20\% is an absolute minimum to use parallaxes as distance estimator, we are confident that the distance and  absolute magnitude estimates in the bottom two panel rows of Fig.~\ref{fig:sumFig1} are meaningful.

\subsubsection{Campbell element estimation\label{sssec:campbElEst}}
\begin{figure}[t]
  \includegraphics[width=0.98\columnwidth]{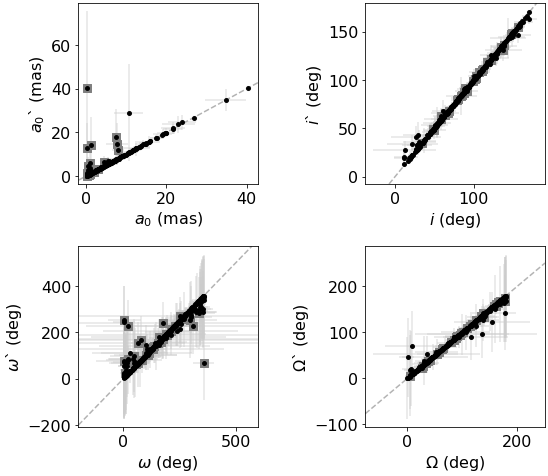}  	
  \vspace{-0.3cm}
\caption{Values and uncertainties for geometric elements. Estimates using linear error propagation and Monte-carlo resampling are shown on the x- and y-axes respectively. In the latter case, the median value is adopted with a symmetric uncertainty computed as the mean of the upper and lower 1-$\sigma$-equivalent confidence interval. The dashed line indicates equality and solutions with $e/\sigma_e<1$ are marked with grey squares. Large discrepancies in $\omega$ at the 360\degr$\xrightarrow{}$0\deg boundary \MOD{(e.g., the point on the bottom right)} are of no major concern.}
\label{fig:geometric_element_uncertainties}
\end{figure}

\begin{figure}[t]
  \includegraphics[width=0.98\columnwidth]{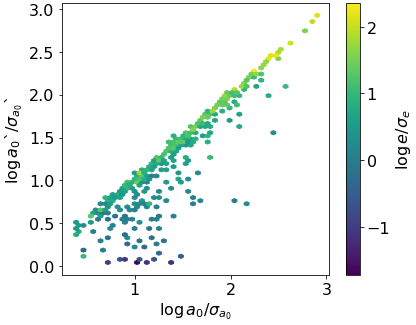}  	
  \vspace{-0.3cm}
\caption{Significance estimates (a over sigma\_a). Density histogram of estimates using linear error propagation and Monte-carlo resampling on the x- and y-axis, respectively. The histogram bins are colour-coded by the average eccentricity significance $e/\sigma_e$.}
\label{fig:significance_mc_versus_significance}
\end{figure}

%In Sect.~\ref{ssec:TItoCampbell} we discussed the subtleties related to the estimation of Campbell elements from the Thiele-Innes elements that we provide in our solution. 
Figures \ref{fig:geometric_element_uncertainties} and \ref{fig:significance_mc_versus_significance} show the differences between  using Monte-Carlo resampling or linear propagation for calculating values and uncertainties of these geometric parameters and their significance. Generally there is good agreement, but  for about ten solutions the Monte-Carlo estimate for the semimajor axis is much larger than the linearly-estimated one. These cases typically correspond to solutions with poorly-constrained eccentricities (i.e.\ $e/\sigma_e<1$) for which Monte Carlo resampling is not recommended because of unrealistic variances of the Thiele-Innes coefficients \citep{DR3-DPACP-127}. This is also reflected in the comparison of the semimajor significance estimators (Fig.\ \ref{fig:significance_mc_versus_significance}), which shows significant discrepancies predominantly for solutions with poorly-constrained eccentricities.

There are four solutions with linearly-propagated $\omega$-uncertainties $\sigma_\omega > 1000$ deg and those also have small or very small eccentricities. Therefore both Monte Carlo resampling and linear propagation can lead to unrealistic results for some almost-circular orbits.

%------------------------------------------------------------------
% Results SUB-SECTION: Verification
%------------------------------------------------------------------
\subsection{Verification \label{sec:results_verification}}
%\TODO{[Rough definition of `Verification' from documentation: \textit{"e.g., verified that results, distributions, time series, etc. match expectations of bona fide objects."}
%]}
In this verification section the focus is on confirming  bona fide companions based on internal consistency checks and expectations.

%\TODO{[Copied from documentation:] Verification included extensive model comparison with alternative (less complex) models to safeguard detectionof bona fide candidated.  The candidates were visual inspected in aggregated one- and two-dimensional diagramssuch as astrometric residual and orbital model plots, as well as placement in the observational Hertzsprung–Russell diagram and agreement with derived astrophysical parameters.}

%\TODO{[Old general text:]
%The candidates were visual inspected in aggregated one- and two-dimensional diagrams such as astrometric residual and orbital-model plots, as well as placement in the observational Hertzsprung--Russell diagram and agreement with derived astrophysical parameters.
%}

\subsubsection{HR-diagram position\label{sssec:hrDiagram}}
%\WORKSUGGESTION{[Move discussion on HR diagram from Sect.~\ref{sssec:paramDistr} to here?]}
The bottom row of panels in Fig.~\ref{fig:sumFig1} shows the zero-extinction absolute magnitude estimate (i.e. $G_{\rm apparent} + 5 (\log10( \varpi/1000) +1 )$, with $\varpi$ in [mas]). Most of our host-stars lie along the main sequence, though a small fraction are likely (sub-)giants. Interestingly, all \typeOrbAltVal sources \MOD{(i.e., the filled dark orange diamonds)} appear to belong to the latter class. The relative tightness of the HR diagram in the right panel reinforces % alternatives: strengthened / reinforced
the correctness of our parallaxes having high relative precision, and that extinction amongst most of these targets is likely low. %No binary sequence is obviously visible
A binary sequence is not clearly identifiable \citep[in contrast to fig. 47 of][]{DR3-DPACP-100}, which further reinforces the expectation that the sample is not significantly polluted by 'impostors' masquerading as systems with small-mass ratios and negligible flux ratios but that are instead binaries with a flux ratio close to the mass ratio.

%\WORKSUGGESTION{[one could potentially see a binary sequence up-right from the MS data.. should we comment on that?]}

%\WORKSUGGESTION{BH->AS: I think you mentioned in earlier text " well as placement in the observational Hertzsprung--Russell diagram and agreement with derived astrophysical parameters.". Since we do not have access to astrophysical params, what can we say more here? Should we refer to DPACP-100?}

%\begin{figure*}[h]
%  \includegraphics[width=0.98\textwidth]{figures/sigm_abfg_full.pdf}  			
%  \vspace{-0.3cm}
%\caption{Full range of uncertainties for the Thiele-Innes parameters as function of the parameter value itself. Note that the axis are adjusted in each plot to show the full range of data.}
%\label{fig:sigmABFGfull}
%\end{figure*} 

%\begin{figure*}[h]
%  \includegraphics[width=0.98\textwidth]{figures/sigm_abfg_limit.pdf}  		
%  \vspace{-0.3cm}
%\caption{Uncertainties for the Thiele-Innes parameters as function of the parameter value itself. Note that the parameter range was the axis are adjusted in each plot to show the full range of data.}
%\label{fig:sigmABFGlimit}
%\end{figure*} 

\subsubsection{Astrometric orbit visualisation \label{sssec:orbit_visualisation}}
We generated graphical representations of all \typeOrbTarStar solutions. These contain the modelled astrometric motion, the post-fit residuals, and auxiliary information describing the properties of the source's data and fit-quality metrics. These figures were used as an empirical tool to assess the quality of the solutions, but not to validate them. The visual discovery of a doubtful solution usually led to the identification or refinement of filter criteria. In other words, this visual inspection was most efficiently used for identifying spurious solutions that should be rejected. 

% They were not used to validate the solutions but they were used when a solution did not match the radial velocities available, and if the astrometric orbit visualisation showed a messy orbit \TODO{[difficult to describe this in a concrete way as this step was very subjective in nature. But because this was mostly used for rejections it should be fine -NU]} then this helped to inform the rejection of a solution.

\begin{figure*}[h]
\centering
  \includegraphics[width=0.8\textwidth]{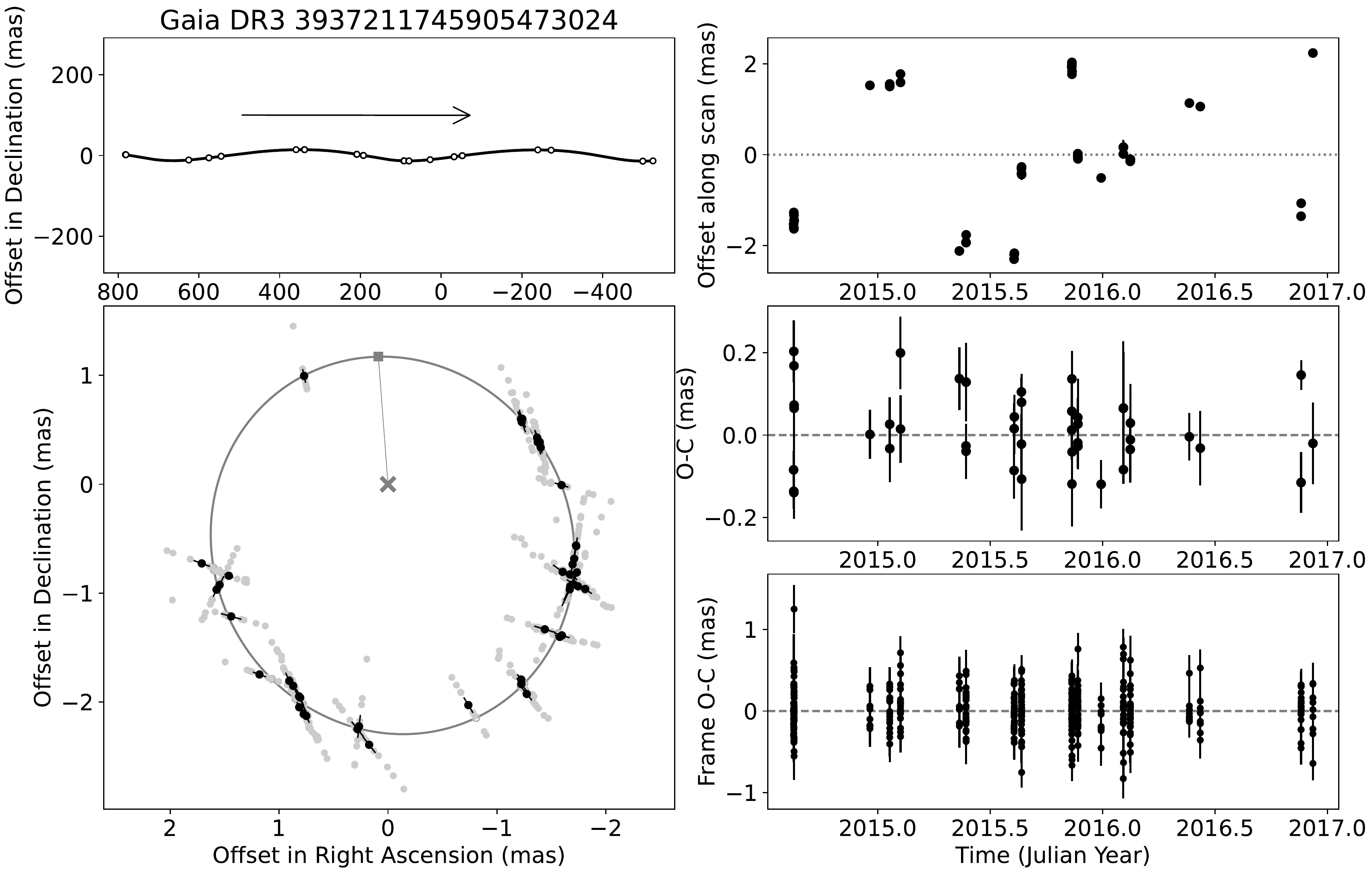}  		
%   \vspace{-0.3cm}
\caption{Astrometric orbit of \object{HD 114762} i.e.\ Gaia DR3 3937211745905473024 (left bottom panel) as determined by Gaia ($G=7.15$\,mag, $P={83.74}\pm{0.12}$\,day, $e={0.32}\pm{0.04}$, $\varpi={25.36}\pm{0.04}$\,mas). North is up and East is left. The sky-projected orbit model about the system barycentre marked with an "x" is shown in grey and astrometric measurements and normal-points after subtraction of parallax and proper motion are shown in grey and black, respectively. Only one-dimensional ("along-scan") astrometry was used, therefore the shown offsets are projected along Gaia's instantaneous scan angle, whose orientation is also indicated by the error-bars. The star's modelled parallax and proper motion is shown in the top-left panel by the solid curve, where open circles indicate the times when the star crossed the Gaia field-of-view. The arrow indicates the direction of motion. The top-right panel shows the normal points after subtraction of the parallax and proper motion as a function of time. 
The middle-right and bottom-right panel shows the post-fit residual normal-points and individual CCD-transit data, respectively. Normal-points are computed at every field-of-view transit of the star from the $\sim$9 individual CCD transits and are only used for visualisation, whereas the data processing uses individual CCD-transit data.}
\label{fig:orbit_1d_summary_DR3_3937211745905473024_companion0}
\end{figure*}

\begin{figure*}[h]
\centering
\includegraphics[width=0.8\textwidth]{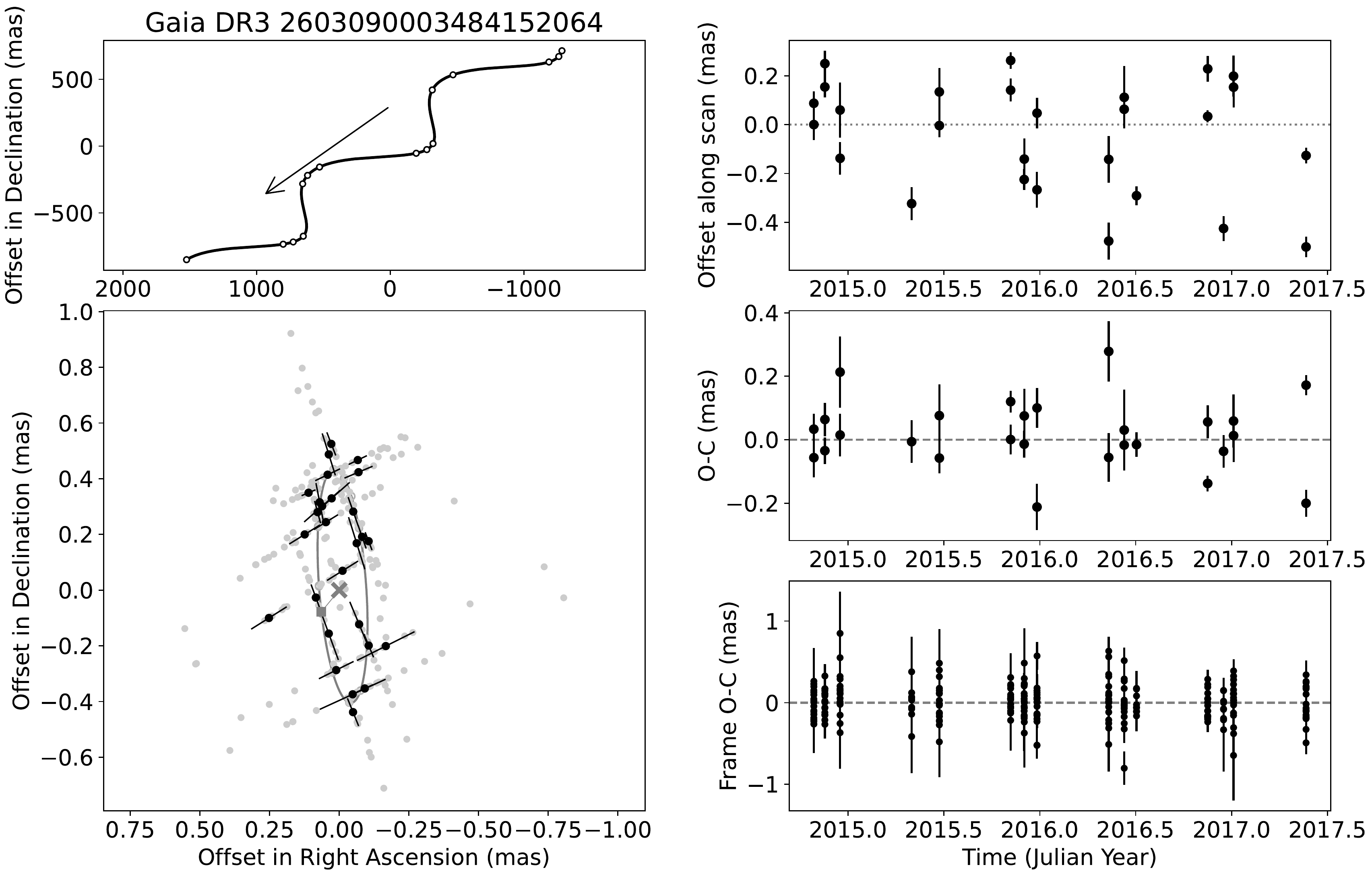}  		
%   \vspace{-0.3cm}
\caption{Same as Fig.\ \ref{fig:orbit_1d_summary_DR3_3937211745905473024_companion0} but for \object{Gl 876} i.e.\ Gaia DR3 2603090003484152064 ($G=8.88$\,mag, $P={61.36}\pm{0.22}$\,day, $e={0.16}\pm{0.15}$, $\varpi={213.79}\pm{0.07}$\,mas).}
\label{fig:orbit_1d_summary_DR3_2603090003484152064_companion0}
\end{figure*}

\begin{figure*}[h]
\centering
  \includegraphics[width=0.8\textwidth]{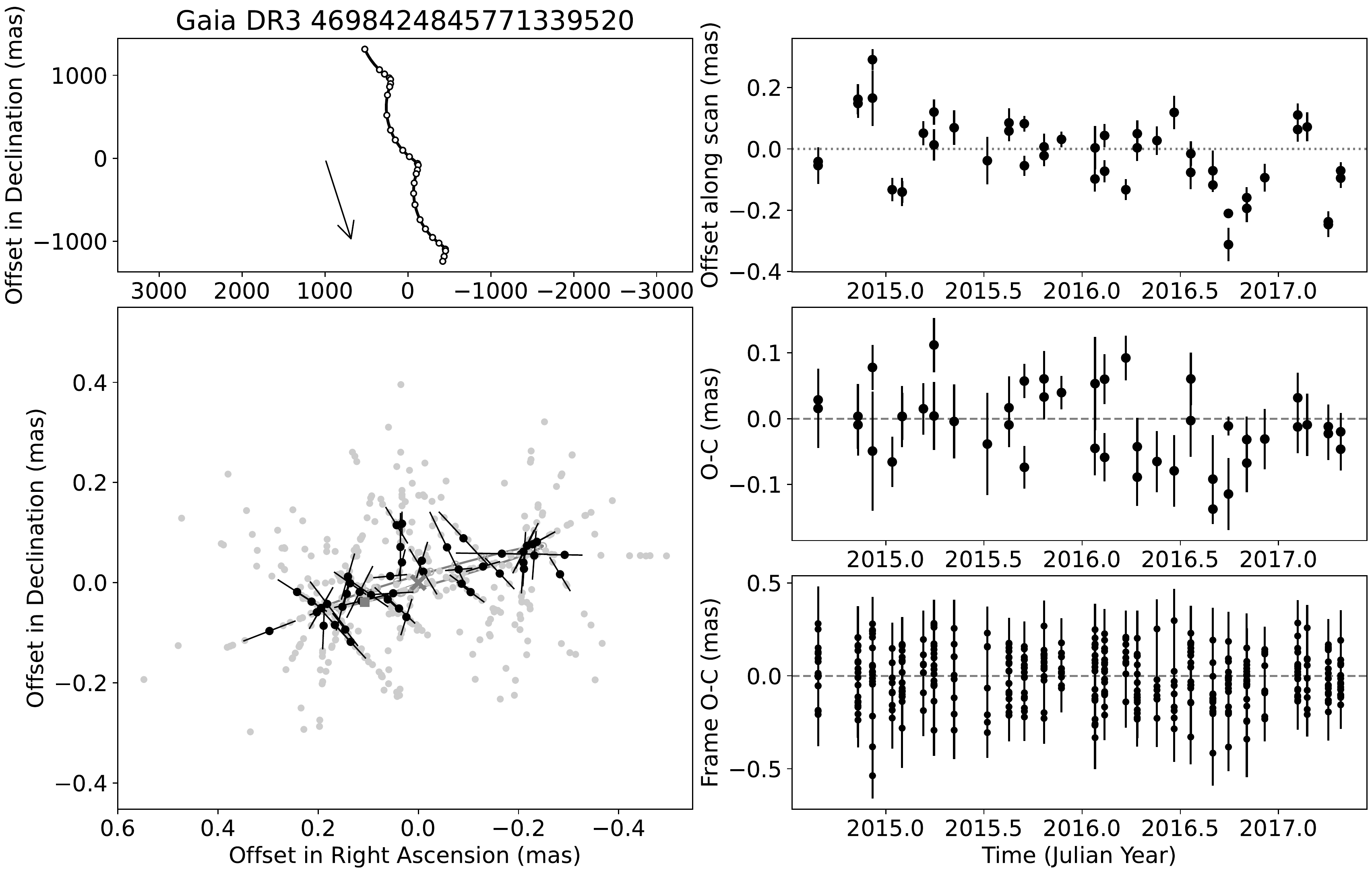}  		
%   \vspace{-0.3cm}
\caption{Same as Fig.\ \ref{fig:orbit_1d_summary_DR3_3937211745905473024_companion0} but for \object{WD0141-279} i.e.\ Gaia DR3 4698424845771339520 ($G=13.70$\,mag, $P={33.65}\pm{0.05}$\,day, $e={0.20}\pm{0.15}$, $\varpi={102.87}\pm{0.01}$\,mas).}
\label{fig:orbit_1d_summary_DR3_4698424845771339520_companion0}
\end{figure*}

\begin{figure*}[h]
\centering
  \includegraphics[width=0.8\textwidth]{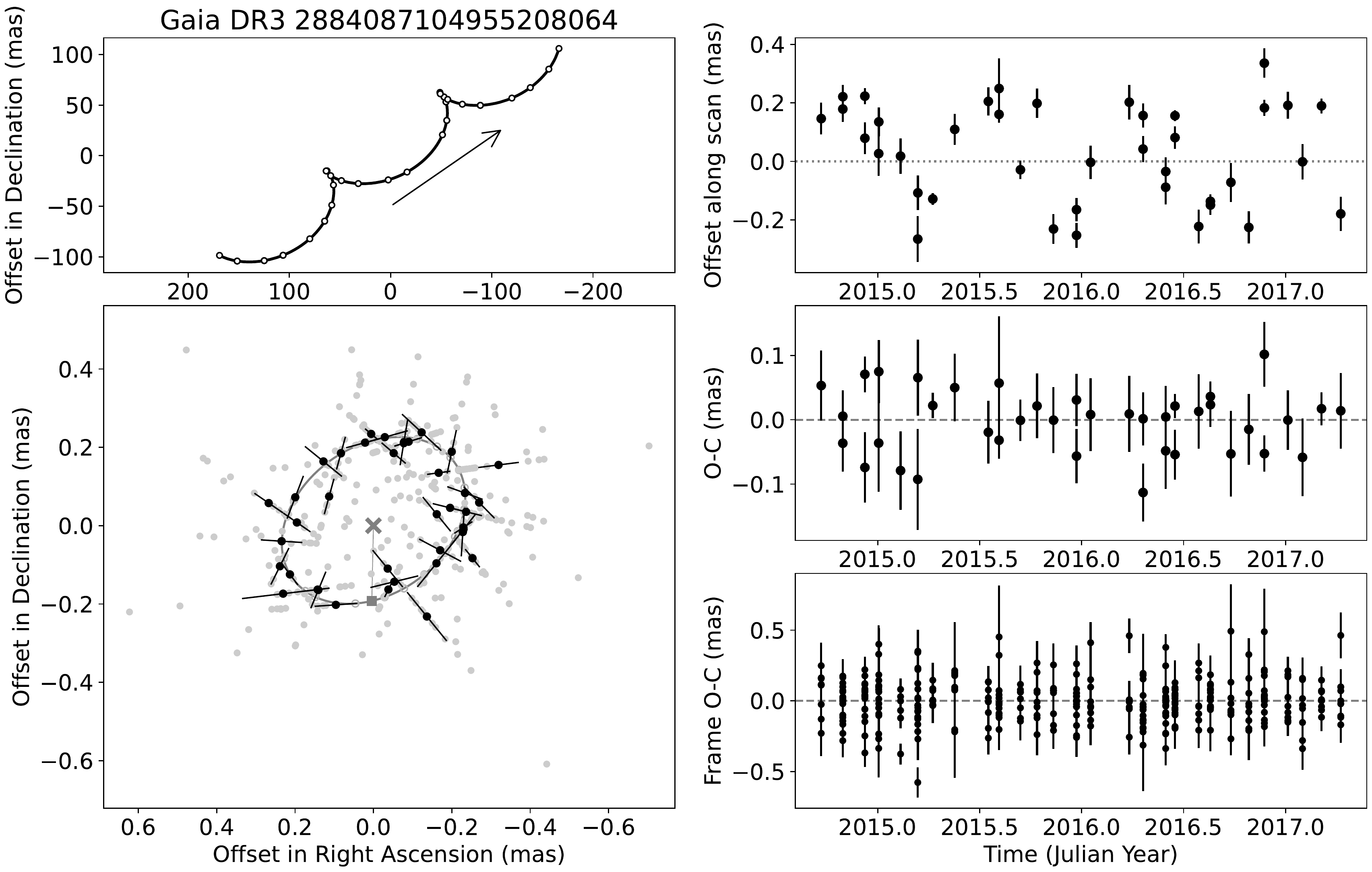}  		
%   \vspace{-0.3cm}
\caption{Same as Fig.\ \ref{fig:orbit_1d_summary_DR3_3937211745905473024_companion0} but for \object{HD 40503} i.e.\ Gaia DR3 2884087104955208064 ($G=8.97$\,mag, $P={826.53}\pm{49.89}$\,day, $e={0.07}\pm{0.10}$, $\varpi={25.49}\pm{0.01}$\,mas).}
\label{fig:orbit_1d_summary_DR3_2884087104955208064_companion0}
\end{figure*}

\begin{figure*}[h]
\centering
  \includegraphics[width=0.8\textwidth]{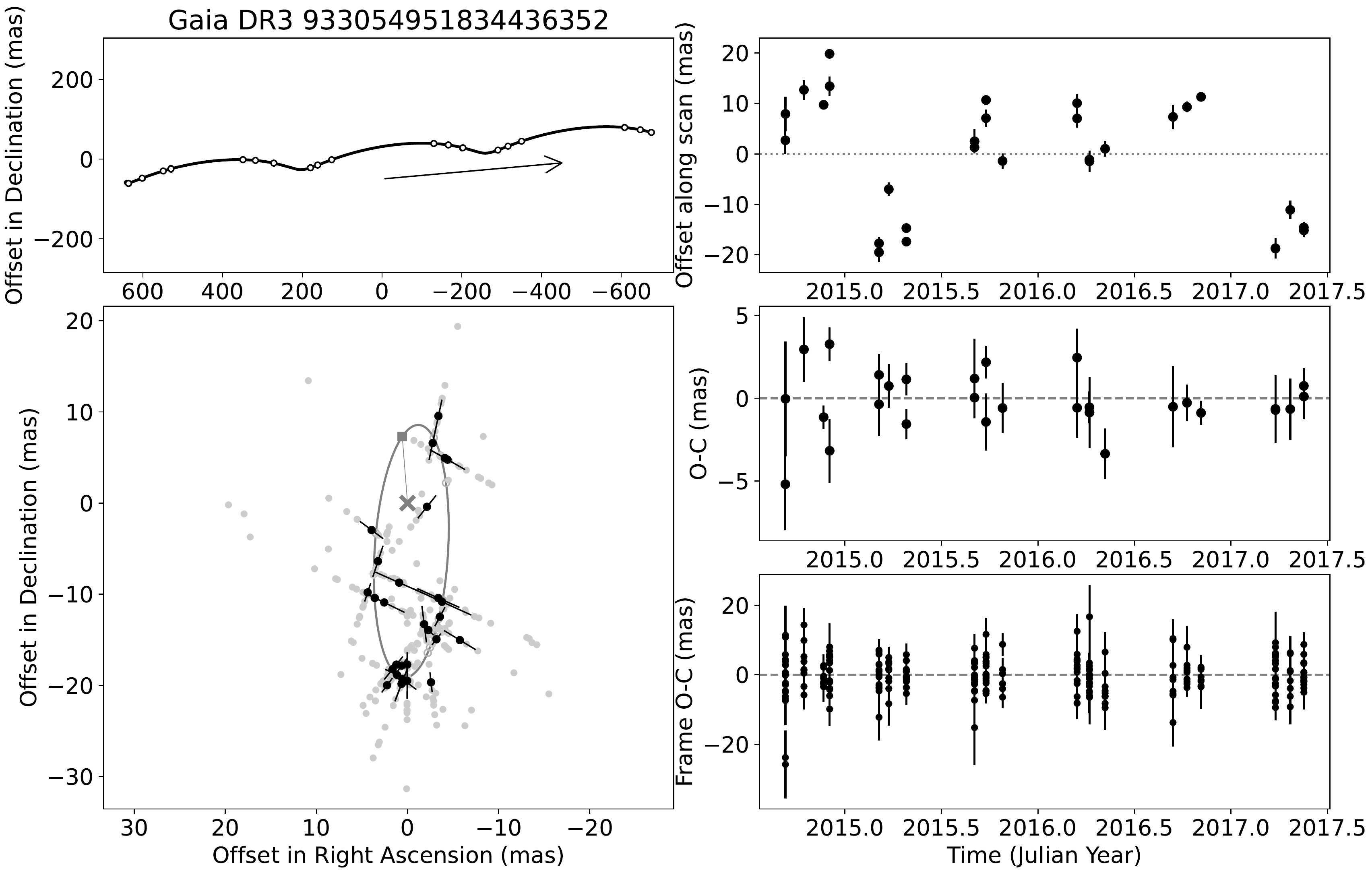}  		
%   \vspace{-0.3cm}
\caption{Same as Fig.\ \ref{fig:orbit_1d_summary_DR3_3937211745905473024_companion0} but for \object{2MASS J08053189+4812330} i.e.\ Gaia DR3 933054951834436352 ($G=20.01$\,mag, $P={735.91}\pm{22.99}$\,day, $e={0.42}\pm{0.23}$, $\varpi={43.77}\pm{0.71}$\,mas).}
\label{fig:orbit_1d_summary_DR3_933054951834436352_companion0}
\end{figure*}

Figures \ref{fig:orbit_1d_summary_DR3_3937211745905473024_companion0} -- \ref{fig:orbit_1d_summary_DR3_933054951834436352_companion0} show examples of orbit visualisations, which were obtained on the basis of the \texttt{pystrometry} package \citep{johannes_sahlmann_2019_3515526}\footnote{\url{https://github.com/Johannes-Sahlmann/pystrometry}}.
An additional example which also demonstrated the solution validation with external RVs is the case of \object{HD 81040}, which was showcased in a Gaia Image of the Week (\url{https://www.cosmos.esa.int/web/gaia/iow_20220131}). 

These figures illustrate the different regimes in terms of sampling, source magnitude, measurement uncertainty, and orbit size in which our algorithms were successful in identifying significant solutions. The scientific implications of the shown orbital solutions are discussed in \citet{DR3-DPACP-100}.

\subsubsection{Comparison with AGIS solution excess noise\label{sssec:results_compAgis}}
\begin{figure*}[th!]
  \includegraphics[width=0.98\textwidth]{figures/topLabel1.pdf}  
  \includegraphics[width=0.98\textwidth]{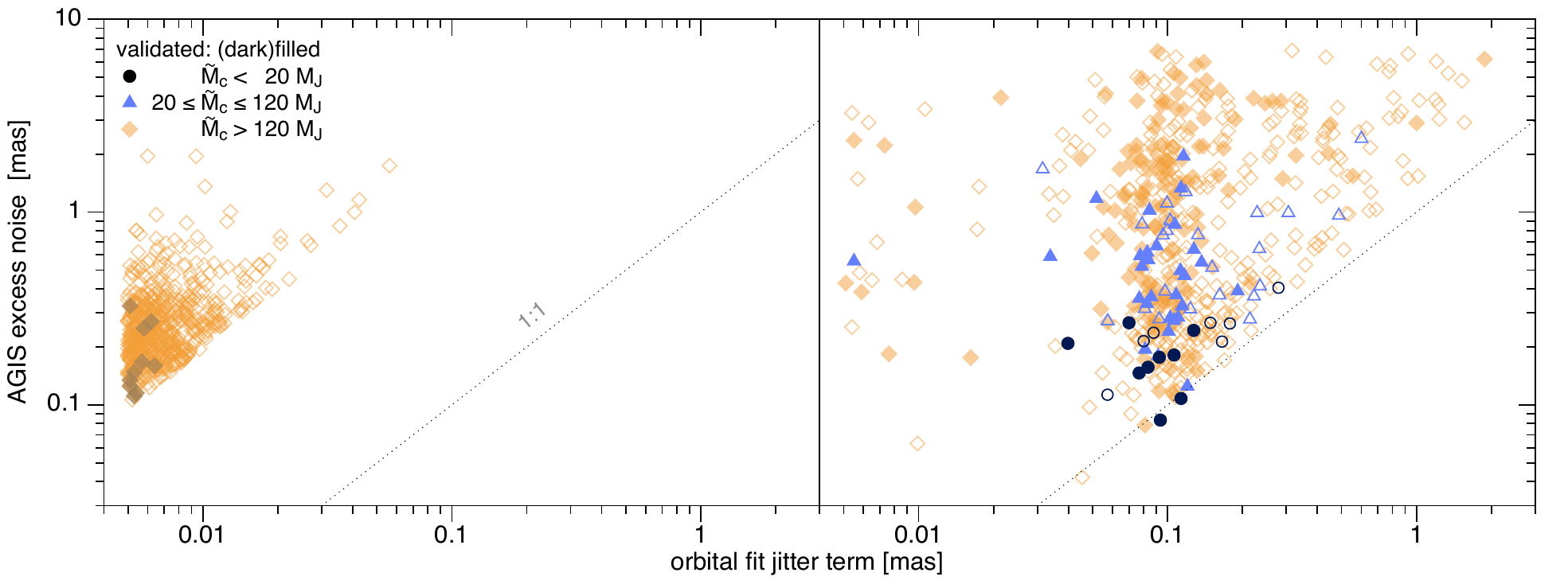}  
  \includegraphics[width=0.98\textwidth]{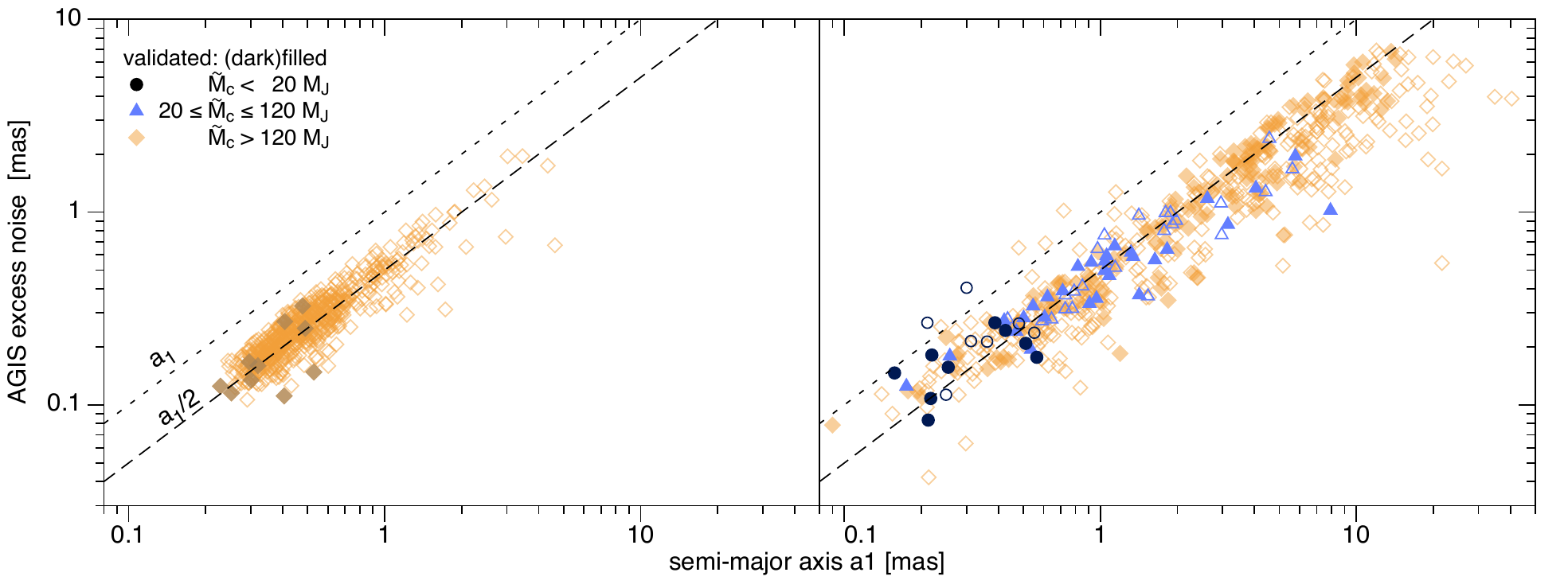}  			
  \vspace{-0.3cm}
\caption{Comparison with the single-star AGIS excess noise.}
\label{fig:agisExcessNoise}
\end{figure*}

A crude, but generally effective indicator of improved modelling is to compare the excess noise level between the 5-parameter AGIS solution and Keplerian orbit solution (called jitter term in the latter). The top panels of Fig.~\ref{fig:agisExcessNoise} show that for the most massive companions we have more than an order of magnitude decrease in the residuals noise level, though for the less massive companions this difference is reduced, as expected.

The second row of panels shows an interesting relation between the AGIS excess noise and fitted semi-major axis: typically the semi-major axis is about half the AGIS excess noise, which holds true for the wide range of masses and semi-major axis in our sample. 
%\WORKSUGGESTION{[say something more?]}
Comparison of other parameters with those from AGIS do not present any unexpected deviations and are not shown.

\subsubsection{Spurious orbits \label{sssec:results_spuriousOrbits}}

\begin{figure}[htp]
\centerline{
   \includegraphics[width=0.5\textwidth]{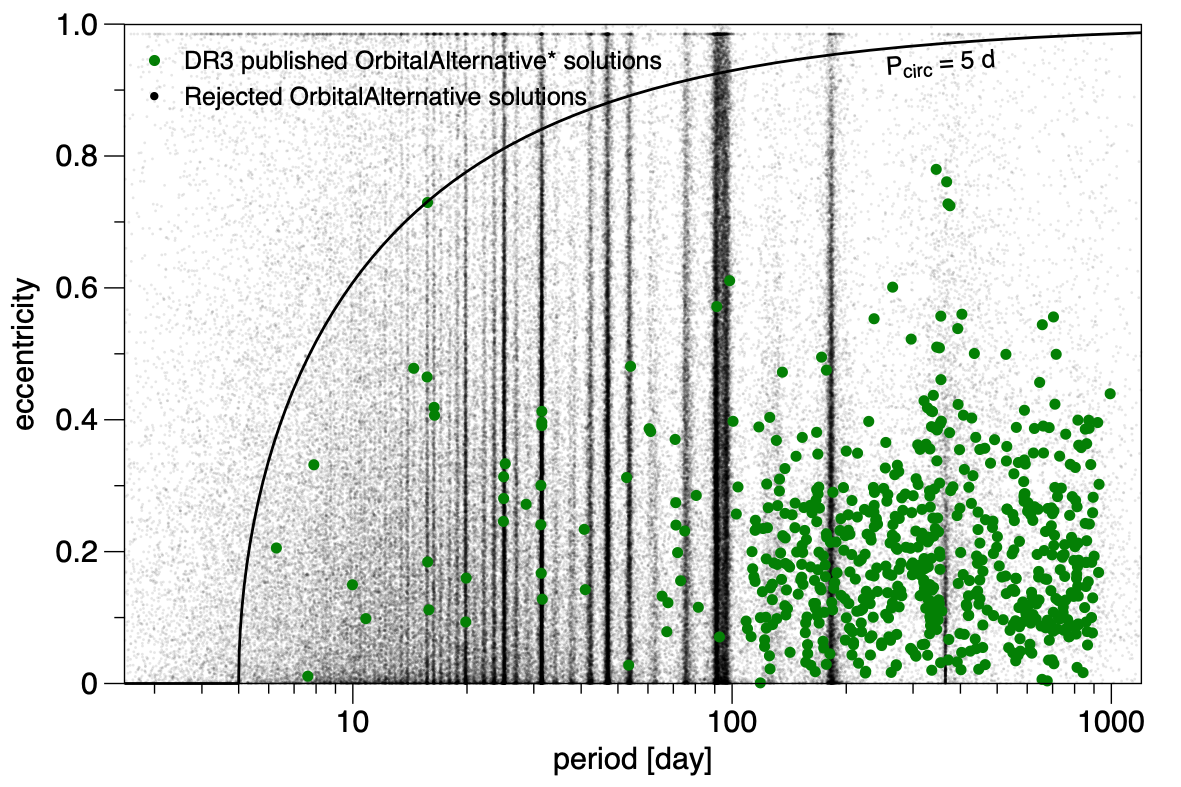}
}\caption{Period-eccentricity diagram for a random sample of sources (black points) with orbits obtained by the DE-MCMC or GA algorithms from the processing of the stochastic solution sample. 
The clear structure of period aliases are symptoms of incorrectly derived orbits and are further discussed in \cite{DR3-DPACP-164}.
%The clear structure of period aliases reproduces analogous patterns seen already in e.g. \TODO{[FIX...]}, which are symptoms of incorrectly derived orbits. 
The green filled dots correspond to $P$ and $e$ values from \typeOrbAlt  solutions. The solid black curve indicates the maximum eccentricity attainable for an orbit unaffected by tides, assuming a cut-off period of 5 days (see e.g. \citet{Halbwachs2005})}
\label{fig:cu4nss_OrbAlt_P_e_diagram}
\end{figure}

%\WORKSUGGESTION{[ASO: add summary of \url{https://gitlab.astro.unige.ch/gaia/du437-validation/-/issues/32}]}
Figure~\ref{fig:cu4nss_OrbAlt_P_e_diagram} shows an example of the period-eccentricity diagram from the unfiltered stochastic solution sample. The clear structure of period aliases
corresponding to e.g. 1 year, 6 months, the 63-days precession period of the satellite, and their harmonics is further discussed in \cite{DR3-DPACP-164}. The green filled dots correspond to $P$ and $e$ values from \typeOrbAlt  solutions equivalent to the data in the top left panel of Fig.~\ref{fig:sumFig1}. Note that these large number of spurious solutions were mostly filtered out by our aggressive filter criteria described in Sect.~\ref{sec:sourceSelection}, at the cost of removing potentially good solution and thus overall low completeness.

The adopted solution filtering procedure did not include constraints on the mass function. A few percent of unrealistically large values of $f(\mathcal{M})$ primarily with short orbital periods ($P\lesssim100$ days) is still present. As further discussed in \citet{DR3-DPACP-100}, these are likely to be spurious, and therefore the level of contamination of the \typeOrbAlt  solutions is probably around 5\%. In \citet{DR3-DPACP-100} (see Sect. 5.1) a recipe is provided for effectively excluding such spurious orbits based on constraints of the parallax significance as a function of the orbital period of the solution.  

The identification of likely spurious solutions in the \typeOrbTar sample, and corresponding estimate of the degree of contamination, was performed as part of the validation analysis, and is described in the following sections. 

%\WORKSUGGESTION{[polish/expand?, regenerate image]}

%s an example, we show in Fig.\ref{fig:cu4nss_OrbAlt_P_e_diagram} the $P$-$e$ diagram of a representative subset of orbits obtained for the stochastic solution sample, compared to the distribution in eccentricity and period of the \typeOrbAlt  sample selected using the acceptance criteria listed in Sec.\ref{sssec:cu4nss_substellar_sf_orbalt}. 

%------------------------------------------------------------------
% Results SUB-SECTION: Validation
%------------------------------------------------------------------
\subsection{Validation \label{sec:results_validation}}
%\TODO{[Rough definition of `Validation' from documentation: \textit{"e.g., comparison of results with the literature, expected completeness and contamination, etc."}
%]}
Validation includes comparison with available literature radial velocity or astrometry data, as well as \gaia radial velocity solutions. These were used to grant certain candidate orbits the status of `Validated', identified by this suffix to their \nssSolutionType.

%\WORKSUGGESTION{Include estimation of (completeness and) contamination!}

\subsubsection{Literature astrometric solutions}
In the \typeOrbTar category, literature astrometric orbits for two targets were available: DE0823$-$49  \cite[\gdr{3} 5514929155583865216,][]{2013A&A...556A.133S} and 2M0805$+$48   \cite[\gdr{3} 933054951834436352,][]{2020MNRAS.495.1136S} both being consistent with the \gaia orbit and thus leading to their `Validated' suffix. They are listed in Appendix~\ref{sec:refSolParams}.

\subsubsection{Literature radial velocity solutions\label{sssec:validationRv}}
%  \WORKSUGGESTION{[Unger et al.]} and \WORKSUGGESTION{[Barbato et al.]}
When available, literature radial velocity data was used to vet the full subset of candidate companions with $\Mcomp<120$~M$_\text{J}$ (i.e., assuming 1~M$_{\odot}$ host) and a subset with $\Mcomp>120$~M$_\text{J}$ in the \typeOrbTar candidate set. When the orbital parameters (typically the period and eccentricity) between the RV solution and the \gaia solution were found to be consistent, it resulted in the `Validated' suffix, all of which are listed in Appendix~\ref{sec:refSolParams}. 
 
For several sources the RV reference parameters are not given, this is to indicate that the RV data alone was not enough to validate the orbit on its own (e.g. there were multiple significant peaks in the RV periodogram). In those cases we validated the target if the RV data was consistent after constraining the period of the keplerian to the \gaia orbital period.

% \FEEDBACK{NU: [Explain JSA comments: Note that for several sources the RV reference parameters are not provided: this is to indicate that the external RV data on itself was insufficient to identify an orbit (e.g. there were multiple significant peaks in the RV periodogram), but that constrained with the \gaia orbital period, the RV data was found to be consistent. ]}

If no literature RV solution was found the source was kept without additional suffix in the name, i.e., it stayed a candidate.
When an astrometric orbit was found to be incompatible with RV data it was removed from our publication list, thus making this validation step part of the filtering process (see Sect. \ref{ssec:sol_filter}).

While this step of the validation procedures was performed quite carefully, it is not entirely free from pitfalls. For example, a small number of sources with good matches between the fitted and literature $P$ values where not flagged as 'Validated', and are still listed in Table \ref{tab:refSolParamsOrbTarNG} with solution type \typeOrbTar. These include two known RV planet hosts, HR 810 (\gdr{3} 4745373133284418816) and HD 142 (\gdr{3} 4976894960284258048), as well as %\WORKSUGGESTION{[LET US ADD THIS TO THE TABLE!]}

\gdr{3} 2133476355197071616 (Kepler-16 AB). The latter source hosts the first circumbinary planet detected by the {\it Kepler} mission, with $P=105$ d \citep{2011Sci...333.1602D,Triaud2022}. In this case \gaia detects a companion with $P=41$ d and an almost edge-on orbit, which is in fact the low-mass stellar companion Kepler 16 B eclipsing Kepler-16 A. In a few cases, inconsistencies between literature RV data and the \gaia solutions were overlooked. Two such examples are those corresponding to \gdr{3} 1748596020745038208 (WASP-2) and \gdr{3} 5656896924435896832 (HATS-26, TOI-574). The two known companions are hot Jupiters with orbital periods of 2.1 d \citep{Cameron2007,Knutson2014} and 3.3~d \citep{Espinoza2016}, respectively. The \gaia solutions have $P=38$ d and $P=193$ d, respectively. No additional RV trends or modulations are detected for WASP-2 and HATS-26, indicating that the \gaia detections might be spurious. 

% DS: ADDED 9 JUNE 2022 (after submission)
Another illustration of the challenges of the validation procedure is the case of GJ~812A (\object{HIP 103393}) for which 9 HARPS measurement were taken between August 2011 and August 2012 that are publicly available on DACE (see Table~\ref{tab:refSolParamsOrbTarNG}). The periodogram of the RV time series does not show any significant peak at the \gaia period ($P=23.926\pm 0.010$~d) but a compatible RV orbit can be found once the period is fixed to the \gaia period. Furthermore, a close inspection of the HARPS' CCFs shows that GJ~812A is a  double-line spectroscopic binary and rules out the presence of a brown dwarf companion to GJ~812A.

\subsubsection{Other literature solutions}

Among literature solutions obtained with techniques other than astrometry and RVs, we report in Table \ref{tab:refSolParamsOrbTarNG} the good agreement between the \gaia period and that obtained by \citet{Murphy2016} for the companion orbiting \gdr{3} 2075978592919858432 (KIC 7917485, Kepler-1648). The primary is a $\delta$ Scuti variable, A-type star, and the companion was identified based on phase modulation of the pulsations of the host. 

Among the sample of Kepler transiting planets that have been statistically validated, Kepler-1697b orbits a K-dwarf primary (\gdr{3} 2102991776844251264) with $P=33.5$ d \citep{Armstrong2021}. The \gaia \typeOrbTar solution has $P=98.5$ d and $i\simeq 65$ deg, which might indicate either a spurious orbit, or the presence of a third body in the system. 

A small number of transiting planet candidates from the TESS mission (TESS Objects of Interest, TOIs) are worth mentioning. TOI-288 (\gdr{3} 6608926350294211328), TOI 289 (\gdr{3} 4919562197762515456), TOI-1104 (\gdr{3} 6345896578791113472) and TOI-2008 (\gdr{3} 5096613016130459136) are indicated as single-transit candidates, and the \gaia solutions of type \typeOrbTarVal have $P=79$ d, $P=223$ d, $P=170$ d, and $P=52$ d, respectively, and close to edge-on configurations, indicating these are likely to be the companions responsible for the eclipses. In the case of TOI-355 (\gdr{3} 5013703860801457280), the transit candidate has $P=1.03$ d, while the \gaia solution is validated with $P=297$ d, indicating the presence of a third body in the system (not necessarily transiting). Similar conclusions could be drawn for  TOI-614 (\gdr{3} 570355367386642611), TOI-746 (\gdr{3} 528033714422882547), TOI-946 (\gdr{3} 5549740136101187840), and TOI-1113 (\gdr{3} 6356417496318028928), all found to be hosting short-period transiting planet candidates and for which \gaia solutions of type \typeOrbTar have much longer $P$ values. With no additional external validation, however, they could also correspond to spurious solutions, as in the case of TOI-574 discussed in Sect \ref{sssec:validationRv}. For TOI-933 (\gdr{3} 5499342375670742784), the period of the transit candidate is almost exactly twice that fitted in the \gaia astrometry. The latter solution is far from edge-on, which might be a symptom of a spurious solution. 

Finally, the case of HD 185501 (\gdr{3} 2047188847334279424) is particularly instructive: the \gaia solution has $P=450$ d and $a_1=0.48$ mas, which at the distance of 32.75 pc (assuming a $1-M_\odot$ primary) would imply detection of a companion well in the substellar regime. However, speckle imaging data have revealed HD 185501 to be an equal-mass binary with $\sim0.04$ arcsec separation and $P=434$ d (\citealt{Horch2020}, and references therein). This is to our knowledge the only clear case in the \typeOrbTar sample of a system incorrectly classified as having a low-mass companion with negligible flux ratio, but rather being one with two components having flux ratio very close to the mass ratio (see \ref{sssec:hrDiagram}).  

%\FEEDBACK{JSA: Very nice section, I like it!}

% \WORKSUGGESTION{[Johannes: please have a look at the details of the 1st section copied below from the documentation)]}

%\WORKSUGGESTION{[Jean-Baptiste have a first look.)]}

%\WORKSUGGESTION{[AS+others: EXPAND: add discussion about targets that were discrepant with published data or sources that we labelled validated but still have some discrepancy]}

%\WORKSUGGESTION{[AS+others: estimate contamination.]}

\subsubsection{Compatibility with independent \gaia solutions \label{sssec:overlapGaiaSols}}

The overlap with alternative solution published by the \gaia NSS pipeline includes a total of 230 cases: 213 in \nssTwoBodyOrbit and 17 in \nssNonLinearSpectro, as tabulated in Tab.~\ref{tab:overlapOtherTabs}. See Appendix~\ref{sec:archiveQueries} for the related queries. Our sample has no overlap with the \nssAccelerationAstro and \nssVimFl solution tables. 

Generally, a given source can have solutions of different \nssSolutionType in \nssTwoBodyOrbit, except for solutions that were obtained from astrometric data only, i.e.\ an \typeOrbTarPVal solution would supersede an \typeOrb solution. For the ten \typeOrbAltVal sources the alternative, and completely independent `SB1' solutions, have compatible periods (to within 10\%) and eccentricities with respect to the orbital solutions from our exoplanet pipeline, and thus were used to provide these sources with the `Validated' suffix. They are listed in Appendix~\ref{sec:refSolParams}. Similarly, out of 147 \typeOrbAltVal sources with alternative \gaia solutions, 142 were found to have matching period to within 10\% with solutions of different \nssSolutionType in \nssTwoBodyOrbit. 

%Only the `SB*' solutions can be considered fully `independent' orbital solutions and thus were allowed to `Validate' any of our astrometric orbits. 
%\WORKSUGGESTION{[Along the same reasoning also `EclipsingSpectro' should be considered independent: check this one for sure!]} 
The identification of discrepant orbital solutions for the same source\_id in \gaia astrometry, spectroscopy and/or photometry does not necessarily mean that either one is incorrect as their respective sensitivities are largely non-overlapping, especially for the lower mass regime, thus they may correspond to effects induced by different components of the system. No sources were thus filtered out based on this \gaia internal comparison. 

Five \typeOrbTarVal sources have orbital periods differing by more than 10\% from those of other \gaia solutions. \gdr{3} 4748772376561143424, \gdr{3} 3309006602007842048, and \gdr{3}  276487905502478720 have `AstroSpectroSB1' and `SB1' (the latter two) long-period solutions, which appear compatible with the fitted values from astrometry. The `SB1' solution for \gdr{3} 3550762648877966336 has $P=6.8$ d, while the astrometric orbit has $P=1213$ d. Based on criteria of significance of the semi-major axis and RV semi-amplitude of the astrometric and spectroscopic orbit and period ratios such as the ones listed in \citet{DR3-DPACP-100}, the system is robustly classified as a hierarchical triple. As for \gdr{3} 2370173652144123008, this is discussed further in the following section.  

A total of 57 sources have \typeOrbTar solutions as well as another orbital solution of type: `SB1', `SB2', `SB2C', `EclipsingSpectro', or `AstroSpectroSB1'. Of these, 48 have period discrepancies in excess of 10\%. For 22 of them both periods are $>750$ d, so they can be considered compatible on the grounds of of the expectedly larger uncertainties as $P$ is similar to, or exceeds the DR3 timespan of the observations. Seven sources with astrometric orbits with $P>330$~d have additional short-period solutions (1 `EclipsingSpectro': \gdr{3} 5556931152602576000; 3 `SB1': \gdr{3} 3537815929524205568, \gdr{3} 4975005999307397376, \gdr{3} 3775615766054509568; 3 `SB2': \gdr{3} 5963704000561744768, \gdr{3} 3304340751399786752, \gdr{3} 5549740136101187840). These could still be interpreted as triple systems. The remainder of the sources have two solutions with both periods $\lesssim120$ d. In a few cases (\gdr{3} 5294105439288288256, \gdr{3} 2729543564483732608, \gdr{3} 6531037981670835584) the period ratio is approximately 2, but for the majority of the sources the period values do not appear compatible, indicating a fraction of the solutions might be spurious. 

For 15 sources with \typeOrbTar orbits (and one \typeOrbTarVal orbit) the additional \gaia solution is of 'FirstDegreeTrendSB1' or ‘SecondDegreeTrendSB1’ type. With one exception, all astrometric orbits have $P>820$ d, likely indicating that the fitted astrometric orbits have significantly under-estimated the true periods of the companions. This is an expected feature of the orbit fitting process in the limit of periods longer than the timespan of the \gaia observations (e.g., \citealt{2008A&A...482..699C}). For \gdr{3} 5140730507877918592, the $P\simeq53$ d orbit might possibly be spurious. 

Overall, based on the above consideration we roughly estimate a degree of contamination from spurious solutions of $\sim10\%$ in the \typeOrbTar sample.

\subsubsection{Solutions of particular interest \label{sssec:solPartInterest}}

Three \typeOrbTarPVal solutions imply the discovery of previously-unknown planetary-mass companions
if a diluted binary scenario can be discarded.
%\WORKSUGGESTION{AS: I am not sure we should describe the companion to HD68638A as a planetary companion. Its inferred mass is closer to 30 MJUP, given the primary mass, with an orbit size 3 times larger than those of the other two objects. I understand the interest to support our cautiousness, but we should then clearer on the perspective with which we discuss the three objects.}
%\FEEDBACK{[JSA], if a diluted binary scenarion can be discarded.}. 
These refer to 
\object{HIP~66074} %\WORKSUGGESTION{[\url{https://gitlab.astro.unige.ch/gaia/du437-validation/-/issues/34}]} 
,
\object{HD~40503} and  \object{HD~68638A} %\WORKSUGGESTION{[\url{https://gitlab.astro.unige.ch/gaia/du437-validation/-/issues/28}]}
 which properties are discussed in \citet{DR3-DPACP-100}.
 
HIP~66074 has 11 radial velocity measurements taken with the HIRES instrument \citep{2017AJ....153..208B} which show a significant dispersion (11 m/s) compared to the measurement uncertainties (1.5 m/s). Fitting the RV data with a constant model and a jitter term ($\sigma_{RV_{Jit}}$) results in a log({\rm likelihood})=-37.82 and $\sigma_{RV_{Jit}}=10.5\pm2.5$~m/s. A thorough frequency analysis of the radial velocity time series is not possible due to the limited number of measurements, but we can nevertheless assess whether the radial velocities are compatible with the \gaia orbit solution: To do so, we fitted a single Keplerian model to the radial velocity with an additional jitter term while keeping the period, the eccentricity, and the argument of periastron fixed at the \gaia-derived values (i.e.\ $P$=297.6~d; $e$=0.46 and $\omega=263.11$ deg.). The adjustment converges to $log({\rm likelihood})=-18.45$, with  $\sigma_V=0.4\pm1.2$~m/s and $T'_0=57\,420.044\pm0.86$ MJD which is compatible within 1~$\sigma$ with the \gaia-derived value of 
%$T_0=57\,389.0\pm 31.4$ MJD. -> the former is the reference epoch in JD at 2016 TCB!!! Fixed now, thanks JBD for spotting this error.
\MOD{$T_0=57\,443.7$}$\pm 31.4$ MJD.
The comparison between the two RV models based on $\Delta$BIC favours the single Keplerian model with a highly-significant value of $\Delta \mathrm{BIC}=-34$. \MOD{Assuming the exoplanet scenario is correct, \citet{DR3-DPACP-100} report a value of $7.3\pm1.1$ M$_\mathrm{Jup}$ for the companion}. 

% not really enough data to say something about the periodigra, In the periodogram there is a peak at 294 days (although not very significant with a Baluev False Alarm Probability of 28\%) 
%\FEEDBACK{[AS], which way is the FAP computed? With GLS I obtain a theoretical FAP of 2\%.} \WORKSUGGESTION{[NU]: The 28\% is from DACE, which is calculated following Baluev 2008}. 

%When fitting a Keplerian to this radial velocity peak we obtain $P=304.2 \pm 7.7$ days and an eccentricity of $0.55 \pm 0.34$. The solution from Gaia gives a period of $297.6 \pm 2.7$ days and an eccentricity of $0.46 \pm 0.17$, so this target was validated based on the agreement of these two orbital parameters between the radial velocities and the Gaia solution.

% BH: update 9 June 2022 (after submmission) reference to refSolParamsOrbTarNG instead of refSolParamsOrbTarG1
HD~40503 (\object{HIP 28193}) has 4 HARPS and 13 CORALIE radial velocity measurements taken between Dec 2003 and Nov 2021 that are publicly available on DACE (see Table~\ref{tab:refSolParamsOrbTarNG}). With a chromospheric activity index of $\log{R^{'}_{\rm HK}}=4.55\pm0.02$ derived from the HARPS spectra, the K2 dwarf is considered as active which makes the analysis of the few publicly available RV measurements challenging. The periodogram of the RV time series is also impacted by the sparsity of the data and the different instrumental offsets.  It can nevertheless be used to validate the \gaia-orbit. Two dominant peaks at 748~d and 917~d are within 2 $\sigma$ of the \gaia-derived period ($P=826\pm50$~d). In addition, the corresponding Keplerian solutions both lead to a significant improvement of the $\Delta \mathrm{BIC}=-52$ compared to the constant model with an additional RV jitter term. However, due to the limited number of  high-resolution spectra and RV measurements, we are not able to rule out the scenario of a blended double-line spectroscopic binary which would require a better phase coverage of the orbital period.

% BH: update 9 June 2022 (after submmission) reference to refSolParamsOrbTarNG instead of refSolParamsOrbTarG1
HD~68638A (\object{HIP 40497}) has 27 ELODIE  radial velocity measurements taken between Nov 1997 and Nov 2003 that are publicly available on DACE (see Table~\ref{tab:refSolParamsOrbTarNG}). The periodogram of the RV time series shows a significant peak at 243 days with a FAP lower than 0.01. The corresponding Keplerian solution derived using DACE leads to a significant improvement of the $\Delta \mathrm{BIC}=-85$ compared to the constant model with an additional RV jitter term and a period $P=240.85\pm0.38$~d that is compatible with the \gaia-derived period ($ 241.6\pm1.0$~d). 
However,  \cite{2007A&A...466.1089B} describe the target as a double-line spectroscopic binary - which is confirmed by an inspection of the ELODIE's CCFs available on the "Observatoire de Haute-Provence" archive - ruling out the substellar nature of the companion.  % http://atlas.obs-hp.fr/elodie/fE.cgi?n=e501&c=o&a=hexp&z=d&fql=[datenuit%20=%2719980111%27],[imanum%20=%270035%27] 

%-153.89  sigJit 193.86 +- 28.86
%-103.72. sigJit 14.18+-4.89  DeltaBIC -84.67; 240.854+-0.380; ecc=0.57+- 0.04; 270.3+- 3.9  / 241.6+-1.0; 0.31+-0.06;126+-22

%However, the star is active with a $\log(R'hk)=-4.XX$ whicgused to search for SB2 with a negative result. However, a correlation between the CCF-FHWM and the velocity as well as 
%may be present leaallow us taken over a period of 300 days allow taken with the HIRES instrument  was validated on the basis of ... \WORKSUGGESTION{[NU/JBD/DS: discuss how these were validated]}
%20.00/32 fixed offsets
%-73.12, sigmaJit=18.89+- 8.43
%-50.75,  sigmaJit 0+-1.40, DeltaBic=-52.75 which is highly significant. P=917+-6 which is compatible within 2$\sigma$
%-51.02 , sigmaJit 0+-1.26, DeltaBic=-52.21  which is highly significant.P=748+-8d
%\FEEDBACK{[BH added this as general comment to this section in relation to above two sources and to indicate there are others like that?. Or should it be moved somewhere else? or removed?]}
%\FEEDBACK{[Remove paragraph?]} As mentioned in Sect.~\ref{sssec:validationRv}, for some sources the literature radial velocity by itself did not provide a constrained orbital solution, but sources were still labelled as `Validated' when the RV data was consistent with our astrometric orbital period. However, even when an independent radial velocity orbit is available with compatible orbital period, there are various instances where there exists a tension between other derived parameters like semi-major amplitude and eccentricity, some of which are also discussed in  \citet{DR3-DPACP-100}.

\subsubsection{Compatibility with reference data\label{sssec:compatibilityRefData}}

For 194 sources (of which 188 validated) we have a reference period and eccentricity available, as listed in Appendix~\ref{sec:refSolParams}. Here we plot the period and eccentricity compatibility in Fig.~\ref{fig:pVsPref} and \ref{fig:eVsEref}, respectively.
In the left panels (\typeOrbAltVal), the reference data is exclusively from the \gaia \typeSBone solutions.
%The error bars of the periods are within the symbols and thus are only plotted for the eccentricity data. 

\begin{figure}[t!]
  \includegraphics[width=0.98\columnwidth]{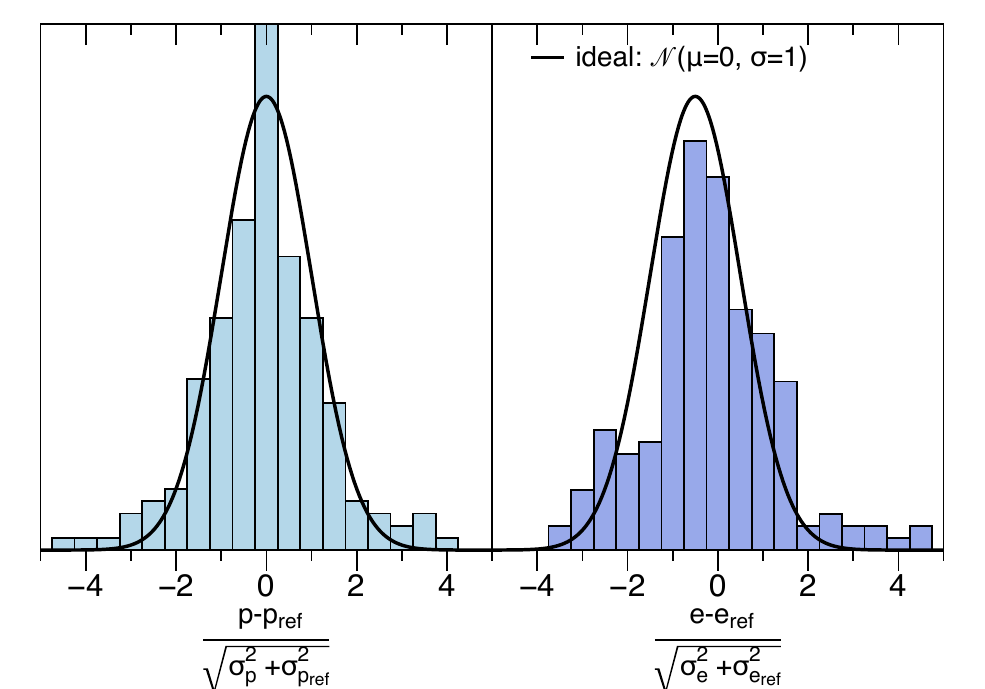}  
  \vspace{-0.2cm}
\caption{Distribution of normalised reference value offsets for the 184~solutions in the \typeOrbTarPVal sample.}
\label{fig:pAndERefHist}
\end{figure}

\begin{figure*}[t]
  \includegraphics[width=0.99\textwidth]{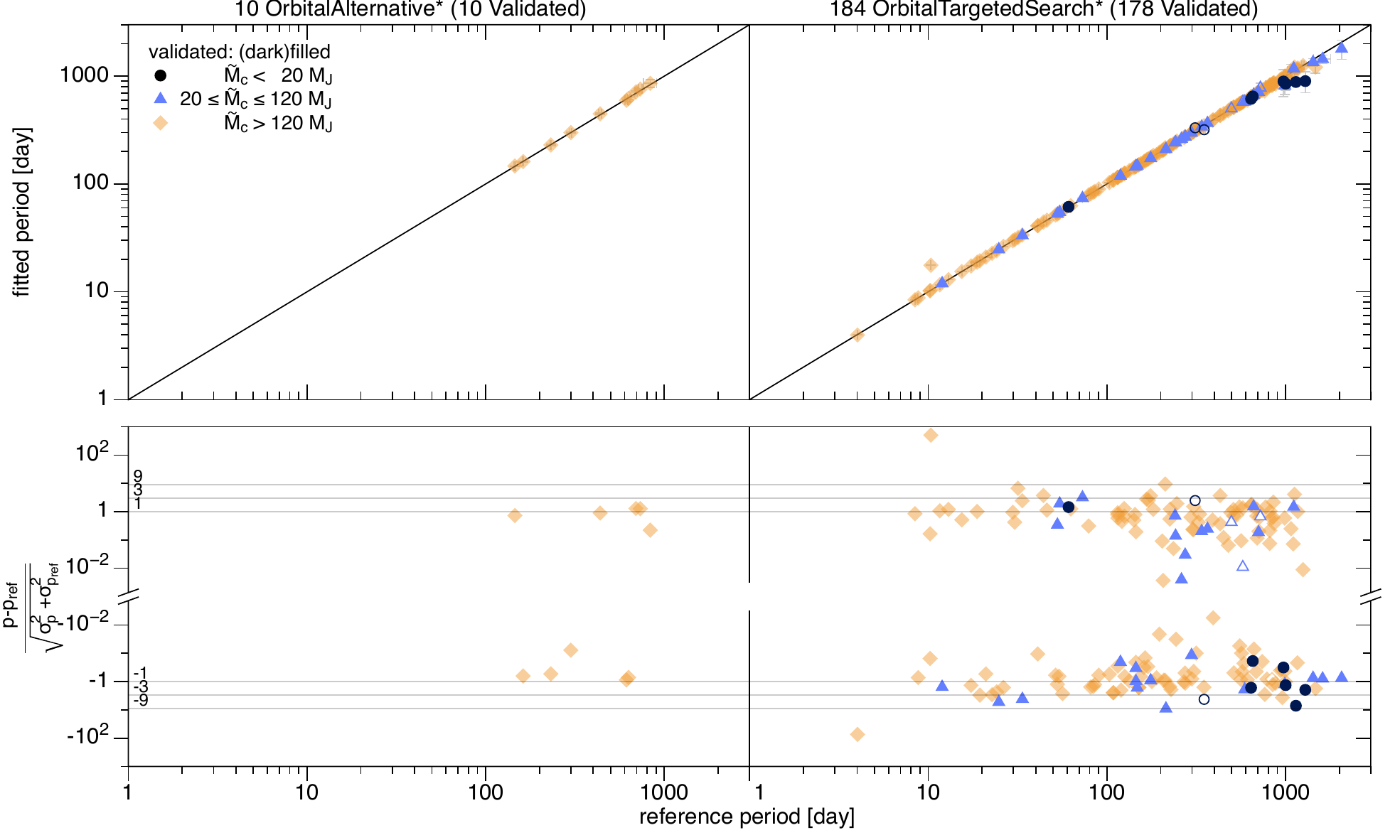}  			
  \vspace{-0.3cm}
\caption{Period versus reference period of sources with reference data available. %Only \texttt{Validated} sources are shown.
}
\label{fig:pVsPref}
\end{figure*}

\begin{figure*}[h!]
  \includegraphics[width=0.99\textwidth]{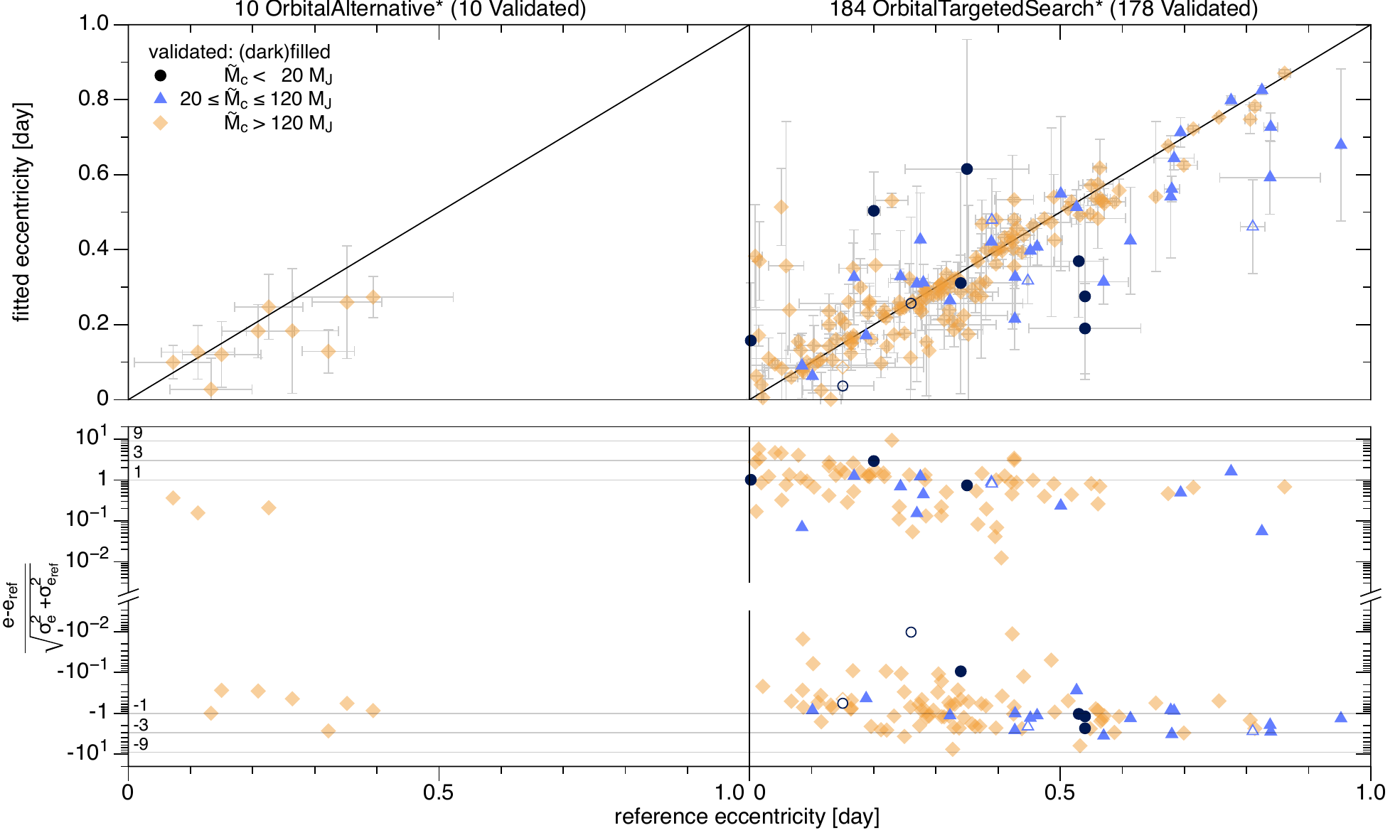}  			
  \vspace{-0.3cm}
\caption{Eccentricity versus reference eccentricity of sources with reference data available. %Only \texttt{Validated} sources are shown.
}
\label{fig:eVsEref}
\end{figure*}

In the bottom panels of each figure we plot the uncertainly normalised absolute differences %\FEEDBACK{
(assuming independent solutions and normal distributed uncertainties), so-as to get an idea of the X-sigma offset between the the reference and fitted values. As shown, the vast majority lies within $\pm$3 sigma, with only a dozen or so beyond it. The most notable `outliers' are the periods of HD21703  (5086152743542494592, $p_\text{ref}=4.0199$~d) and BD-18113 (2370173652144123008, $p_\text{ref}=10.3$~d) which are different by 0.04 and 7.3~d, respectively. Together with very small uncertainties in both astrometric and literature RV period this causes the large divergence which we accepted in this case. 
In Fig.~\ref{fig:pAndERefHist} we see that for the \typeOrbTarPVal samples the normalised $p - p_\text{ref}$ distributions is reasonably well represented by the expected normal distribution centred around 0 with standard deviation 1, while the normalized $e - e_\text{ref}$ distribution has a rather non-Gaussian shape with excesses and deficiencies on both positive and negative side, as also clearly visible in Fig.~\ref{fig:eVsEref}. The median offset -0.22 indicates that overall we tend to fit smaller eccentricities.
%}

%\WORKSUGGESTION{[BH-> AS further discussion here on the distributions of e.g. the eccentricities?]}

\begin{table*}
\caption{\label{tab:overlapOtherTabs} Overview of 213 \texttt{source\_id}'s with an alternative solution in either \nssTwoBodyOrbit or \nssNonLinearSpectro.}
\centering                                      % used for centering table
%\begin{tiny}
\begin{tabular}{llrl}
\hline\hline
Exoplanet pipeline \nssSolutionType  & Alternative \nssSolutionType  & Counts  & Alternative table name \\
\hline
\typeOrbAlt	    &	`SecondDegreeTrendSB1'  	&	1	&	\nssNonLinearSpectro	\\
\hline
\typeOrbAltVal	&	`SB1'	                    &	10	&	\nssTwoBodyOrbit	\\
\hline
\typeOrbTar	    &	`SB1'	                    &	40	&	\nssTwoBodyOrbit	\\
\typeOrbTar	    &	`SecondDegreeTrendSB1'	    &	11	&	\nssNonLinearSpectro	\\
\typeOrbTar	    &	`SB2'	                    &	9	&	\nssTwoBodyOrbit	\\
\typeOrbTar	    &	`AstroSpectroSB1'	        &	6	&	\nssTwoBodyOrbit	\\
\typeOrbTar	    &	`FirstDegreeTrendSB1'	    &	4	&	\nssNonLinearSpectro	\\
\typeOrbTar	    &	`EclipsingSpectro'	        &	1	&	\nssTwoBodyOrbit	\\
\typeOrbTar	    &	`SB2C'	                    &	1	&	\nssTwoBodyOrbit	\\
\hline
\typeOrbTarVal	&	`AstroSpectroSB1'	        &	92	&	\nssTwoBodyOrbit	\\
\typeOrbTarVal	&	`SB1'	                    &	46	&	\nssTwoBodyOrbit	\\
\typeOrbTarVal	&	`SB2'	                    &	8	&	\nssTwoBodyOrbit	\\
\typeOrbTarVal	&	`SecondDegreeTrendSB1'	    &	1	&	\nssNonLinearSpectro	\\
\hline
\end{tabular}
%\tablebib{
%n2bo = \nssTwoBodyOrbit; nnls = \nssNonLinearSpectro. 
%}
%\end{tiny}
\end{table*}

\section{Conclusions \label{sec:conclusion}}
We present a sample of 1162 orbital solutions of \gaia astrometric data produced by the exoplanet pipeline, spanning the planetary mass regime up to low-mass stellar companions and probing a low astrometric signal-to-noise regime. The host-star distribution is dominated by main-sequence stars along with a small fraction of (sub-)giant stars spanning the apparent magnitude range of $G\sim 3-20$. The vast majority of least massive (potentially brown dwarf and planetary mass) companions are confined to $G\lesssim 11$. Semi-major axes range between $a_1\sim 0.1-10$~mas with the majority of periods found between $100-2000$~d and a noticeable tail down to $\sim8$~d. The least massive companions are detected out to $\sim100$~pc, with (sub-)stellar  mass companions up to (several) kiloparsecs. The sky distribution is rather uniform for the (nearby) least massive companions and more confined to the Galactic plane for the more massive (further out) ones.

The sample orbital solutions is subdivided in the \gdr{3} \nssTwoBodyOrbit archive table into four \nssSolutionType: 629~\typeOrbAltPVal of which 10 validated and 533 \typeOrbTarPVal of which 188 validated (Sect.~\ref{sec:sourceSelection}).

%\WORKSUGGESTION{CONVEY NOTION OF LOW ASTROMETRIC SNR REGIME} . 
Due to the rather short available time span of 34~months of data, limited number of observations, and still to be improved error model (in particular calibrations) in the bright-star regime, we had to adopt a complex, inhomogeneous filtering procedure, and a large variety of verification and validation steps (often on source-per-source basis) in order to present a sample with high reliability (i.e. low contamination from spurious solutions and incorrectly classified objects) at the cost of very low completeness and very uneven selection function. 
%\FEEDBACK{AS: ADD A BRIEF SUMMARY OF OTHER RELEVANT NUMBERS, INCLUDING POSSIBLE FRACTION OF "CONTAMINANTS". WE CAN ALSO BRIEFLY FUTURE SOFTWARE IMPROVEMENTS?}.
We estimate the level of spurious/incorrect solutions in our sample to be of the order of $\sim5\%$ and $\sim10\%$ in the \typeOrbAlt and \typeOrbTar sample, respectively.

Given the above mentioned difficulties, it is therefore no surprise that the \gdr{3} sample of (known and new) exoplanets and brown dwarfs orbiting bright F-G-K dwarfs with reliable astrometric orbits is small\footnote{See \citet{DR3-DPACP-100} for a different perspective on the sample of astrometrically detected substellar companions to low-mass M dwarfs.}. Over the past two decades, estimates of the \gaia harvest of exoplanets and brown dwarfs \citep{Lattanzi2000,Sozzetti2001,2008A&A...482..699C,Sozzetti2014b,Sozzetti2014,Perryman2014,Sahlmann:2015aa,Holl2021} have converged on ballpark numbers of (tens of) thousands of new detections. These studies have always provided end-of-mission (nominal or extend) figures, which cannot therefore be directly compared with the DR3-level sensitivity of the \gaia survey. Indeed, \gdr{3} provides the first-ever full orbital solutions for a number of known exoplanets and brown dwarfs, and allows to identify a few previously unknown planetary-mass companions based on astrometric data alone. This should be regarded by no means a small feat, but rather a fundamental stepping stone for the expected improvements in future data releases. 

%\WORKSUGGESTION{We do not reference anywhere papers with predicted Gaia planet and BD counts; I believe we should, though I am not sure where to do it. Here perhaps, or in the introduction. We should at least comment that there are many more predicted and that this is just a first (rather safe) sample.

%BH+AS: add some references.}

% SHOULD NOT COMMENT ON WHAT COMES NEXT (as things change anyways)
%Future data releases will each time double the amount and time span of the data which, along with calibration improvements, will likely bring us closer to the predicted \gaia harvest of (tens of) thousands of planets and brown dwarfs \TODO{[references planet and BD estimates]}. Additionally these will also expose the underlying astrometric time series, which will allow anyone to re-analyse the data and combine it with any other available data. \WORKSUGGESTION{[A bit of a sad ending, probably should end with a positive note...]}

%\WORKSUGGESTION{[TODO]}

%   \begin{enumerate}
%      \item ...
%   \end{enumerate}

%%%%%%%%%%%%%%%%%%%%%%%%%%%%%%%%%%%%%%%%%%%%%%%%%%%%%%%%%%%%%%%
% Acknowledgements
\begin{acknowledgements}
\MOD{We thank the anonymous referee for helpful feedback and suggestions that improved the quality of this paper.}
This work has, in part, been carried out within the framework of the National Centre for Competence in Research PlanetS supported by SNSF.
%\WORKSUGGESTION{[ADD YOUR REQUIRED FUNDING ACKNOWLEDGEMENTS HERE]}
A.S., P.G., M.G.L., and R.M. gratefully acknowledge financial support of the Italian Space Agency (ASI) under contracts 2018-24-HH.0 and 2018-24-HH.1-2022 in support of the Italian participation to the Gaia mission.

This work presents results from the European Space Agency (ESA) space mission \gaia. \gaia\ data are being processed by the \gaia\ Data Processing and Analysis Consortium (DPAC). Funding for the DPAC is provided by national institutions, in particular the institutions participating in the \gaia\ MultiLateral Agreement (MLA). The \gaia\ mission website is \url{https://www.cosmos.esa.int/gaia}. The \gaia\ archive website is \url{https://archives.esac.esa.int/gaia}.
Acknowledgements are given in Appendix~\ref{ssec:appendixA}

%This work has made use of data from the European Space Agency (ESA) mission
% {\it Gaia} (\url{https://www.cosmos.esa.int/gaia}), processed by the {\it Gaia}
% Data Processing and Analysis Consortium (DPAC,
% \url{https://www.cosmos.esa.int/web/gaia/dpac/consortium}). Funding for the DPAC
% has been provided by national institutions, in particular the institutions
%participating in the {\it Gaia} Multilateral Agreement.

This publication makes use of the Data \& Analysis Center for Exoplanets (DACE), which is a facility based at the University of Geneva (CH) dedicated to extrasolar planets data visualisation, exchange and analysis. DACE is a platform of the Swiss National Centre of Competence in Research (NCCR) PlanetS, federating the Swiss expertise in Exoplanet research. The DACE platform is available at \url{https://dace.unige.ch}.

This research has made use of the NASA Exoplanet Archive, which is operated by the California Institute of Technology, under contract with the National Aeronautics and Space Administration under the Exoplanet Exploration Program.

This research has made use of the SIMBAD database, operated at CDS, Strasbourg, France.

The authors made use of 
%\FEEDBACK[JSA]{This is a commercial product, right? If so, I suggest not to list it here.} 
\href{https://www.visualdatatools.com/DataGraph/}{\textsc{DataGraph}}, 
\textsc{TOPCAT} \citep{2005ASPC..347...29T},  \textsc{ASTROPY} (a community-developed core Python package for Astronomy \citep{Astropy-Collaboration:2013aa}), \textsc{SCIPY} \citep{Jones:2001aa}, \textsc{NUMPY} \citep{Oliphant2007}, \textsc{IPYTHON} \citep{Perez2007}, \textsc{PANDAS} \citep{jeff_reback_2022_6408044} and \textsc{MATPLOTLIB} \citep{hunter2007}.

\end{acknowledgements}


\bibliographystyle{aa}
\raggedbottom
\bibliography{local}

\begin{appendix}

\section{Reference solution parameters} \label{sec:refSolParams}
Table~\ref{tab:refSolParamsOrbAlt}, \ref{tab:refSolParamsOrbTarNG}, \ref{tab:refSolParamsOrbTarG1}, and \ref{tab:refSolParamsOrbTarG2} contain an overview of all sources we published into \gdr{3} table \nssTwoBodyOrbit for which we identified reference data. The precision in the numbers in the printed table is occasionally truncated, but has been preserved in the online table version. Asymmetric uncertainties available for some of the literature sources have been set to the largest absolute uncertainty. \MOD{The last column contains an index relating to our pseudo-companion mass estimate \Mcomp assuming a solar-mass host, with 0 = \{\Mcomp~<~20~\Mjup\}, 1 = \{20~\Mjup~$\leq$~\Mcomp $\leq 120$~\Mjup\}, and 2 = \{\Mcomp~>~120~\Mjup\}.}

Note that for several sources the non-\gaia RV reference parameters are not provided: this corresponds to sources mentioned in Sect.~\ref{sssec:validationRv} for which the literature radial velocity by itself did not provide a constrained orbital solution, but the data is consistent with astrometric orbital period.
%(e.g. there were multiple significant peaks in the RV periodogram), but that constrained with the \gaia orbital period, the RV data was found to be consistent. 
Three validated solutions marked with "$^\dagger$"  were validated based on internal \gaia reference data, which was eventually not published in \gdr{3}, therefore no reference values are provided.
All sources with available period and eccentricity were used in the analyses of Sect.~\ref{sssec:compatibilityRefData}.

References to data that is publicly available on the \href{https://dace.unige.ch/dashboard/}{Data and Analysis Center for Exoplanets} (DACE) platform are marked with "*" and can be directly queried online as indicated in the note of Table~\ref{tab:refSolParamsOrbTarNG}.

The HARPS and CORALIE radial velocity light curve for HD40503 has been published as part of this paper in the online material.
%\FEEDBACK{[BH: Mention we publish reference RV data in Vizier for source HD40503.]}

%\FEEDBACK{[BH: Provide electronic file version of the tables online (need to check if it needs to be ADS or can be with A\&A).]}

%\WORKSUGGESTION{[BH: finalise references list once \url{https://gitlab.astro.unige.ch/gaia/du437-validation/-/issues/30} completed. Can still be done during DPAC review.]}

\begin{table*}[b]
\caption{\label{tab:refSolParamsOrbAlt} Reference solution parameters for \typeOrbAltPVal solutions. See Appendix~\ref{sec:refSolParams} for details.  }
\centering                                      % used for centering table
\begin{tiny}
\begin{tabular}{
    % see https://texblog.org/2019/06/03/control-the-width-of-table-columns-tabular-in-latex/
    m{0.15\textwidth}
    >{\raggedleft}m{0.16\textwidth}
    m{0.03\textwidth}
    >{\raggedleft}m{0.03\textwidth}
    |rr|rr|c}

\hline
\hline
\nssSolutionType	&	\gdr{3} \texttt{source\_id}	&	Name	&	Ref.		&	\multicolumn{1}{|l}{Ref. $P$ \MOD{[d]}}			&	\multicolumn{1}{l}{Our $P$ \MOD{[d]}}		&	\multicolumn{1}{|l}{Ref. $e$}			&	\multicolumn{1}{l|}{Our $e$}	&  \MOD{\Mcomp} 	\\
\hline																							
\typeOrbAltVal	&	4517375515957545216	&		&	\ \ 	1a		&	231.7	$\pm$ 	\ \ 1.5	&	229.9	$\pm$ 	\ \ 2.9	&	0.35	$\pm$ 	0.06	&	0.26	$\pm$ 	0.15	&	2	\\
\typeOrbAltVal	&	2009052252148821632	&		&	\ \ 	1a		&	630.6	$\pm$ 	\ \ 9.4	&	620.9	$\pm$ 	\ \ 9.8	&	0.23	$\pm$ 	0.06	&	0.25	$\pm$ 	0.09	&	2	\\
\typeOrbAltVal	&	6350499649858805120	&		&	\ \ 	1a		&	162.1	$\pm$ 	\ \ 1.0	&	161.3	$\pm$ 	\ \ 0.7	&	0.13	$\pm$ 	0.07	&	0.03	$\pm$ 	0.08	&	2	\\
\typeOrbAltVal	&	426186585428243840	&		&	\ \ 	1a		&	693.3	$\pm$ 	\ \ 9.2	&	714.0	$\pm$ 	13.0	&	0.32	$\pm$ 	0.04	&	0.13	$\pm$ 	0.06	&	2	\\
\typeOrbAltVal	&	5545870301853637504	&		&	\ \ 	1a		&	437.1	$\pm$ 	\ \ 8.9	&	447.3	$\pm$ 	\ \ 7.1	&	0.15	$\pm$ 	0.06	&	0.12	$\pm$ 	0.09	&	2	\\
\typeOrbAltVal	&	5431358403498310656	&		&	\ \ 	1a		&	733.3	$\pm$ 	17.6	&	757.8	$\pm$ 	\ \ 7.4	&	0.11	$\pm$ 	0.06	&	0.13	$\pm$ 	0.07	&	2	\\
\typeOrbAltVal	&	5941647007018169728	&		&	\ \ 	1a		&	834.2	$\pm$ 	70.8	&	856.7	$\pm$ 	72.5	&	0.26	$\pm$ 	0.07	&	0.18	$\pm$ 	0.17	&	2	\\
\typeOrbAltVal	&	2274022837765746304	&		&	\ \ 	1a		&	145.7	$\pm$ 	\ \ 1.2	&	146.6	$\pm$ 	\ \ 0.3	&	0.39	$\pm$ 	0.13	&	0.27	$\pm$ 	0.05	&	2	\\
\typeOrbAltVal	&	2055801936074059264	&		&	\ \ 	1a		&	613.6	$\pm$ 	11.7	&	596.7	$\pm$ 	15.4	&	0.07	$\pm$ 	0.06	&	0.10	$\pm$ 	0.04	&	2	\\
\typeOrbAltVal	&	513567110946268544	&		&	\ \ 	1a		&	299.6	$\pm$ 	\ \ 3.3	&	299.3	$\pm$ 	\ \ 2.0	&	0.21	$\pm$ 	0.06	&	0.18	$\pm$ 	0.07	&	2	\\
\end{tabular}																							
\tablebib{																						
(1a) \gdr{3} table \nssTwoBodyOrbit with \nssSolutionType = \texttt{SB1};
}																																																		\end{tiny}
\end{table*}

\begin{table*}
\caption{\label{tab:refSolParamsOrbTarNG} Non-\gaia reference solution parameters for \typeOrbTarPVal solutions. See Appendix~\ref{sec:refSolParams} for details. }
\centering                                      % used for centering table
\begin{tiny}
\begin{tabular}{%lllr|rr|rr|r}
    % see https://texblog.org/2019/06/03/control-the-width-of-table-columns-tabular-in-latex/
    m{0.18\textwidth}
    >{\raggedleft}m{0.155\textwidth}
    p{0.089\textwidth}
    >{\raggedright}m{0.023\textwidth}
    |rr|rr|c}
\hline
\hline
\nssSolutionType	&	\gdr{3} \texttt{source\_id}	&	Name	&	Ref.		&	\multicolumn{1}{|l}{Ref. $P$ \MOD{[d]}}			&	\multicolumn{1}{l}{Our $P$ \MOD{[d]}}		&	\multicolumn{1}{|l}{Ref. $e$}			&	\multicolumn{1}{l|}{Our $e$}	&  \MOD{\Mcomp}	\\
\hline
\typeOrbTarVal	&	1318110830190386048	&	HD148284	&	\ \ 	4		&	339.3	$\pm$ 	2e-2	&	339.6	$\pm$ 	\ \ 1.2	&	0.39	$\pm$ 	9e-4	&	0.42	$\pm$ 	0.04	&	1	\\
\typeOrbTarVal	&	409909484005053440	&	HD8054	&	\ \ 	5		&	847.7	$\pm$ 	\ \ 3.8	&	835.4	$\pm$ 	\ \ 8.8	&	0.81	$\pm$ 	0.01	&	0.75	$\pm$ 	0.04	&	2	\\
\typeOrbTarVal	&	4062446910648807168	&	HD164604	&	\ \ 	6		&	641.5	$\pm$ 	10.1	&	615.5	$\pm$ 	12.0	&	0.35	$\pm$ 	0.10	&	0.62	$\pm$ 	0.35	&	0	\\
\typeOrbTarVal	&	6374231714992810752	&	EBLM \ J2011-71	&	\ \ 	7		&	663.0	$\pm$ 	\ \ 0.3	&	671.5	$\pm$ 	\ \ 5.6	&	0.10	$\pm$ 	3e-3	&	0.06	$\pm$ 	0.05	&	1	\\
\typeOrbTarVal	&	2370173652144123008	&	BD-18113	&		31\href{https://dace.unige.ch/radialVelocities/?pattern=BD-18113}{*}	&	10.3	$\pm$ 	1e-6	&	17.6	$\pm$ 	1e-2	&	0.05	$\pm$ 	6e-5	&	0.51	$\pm$ 	0.10	&	2	\\
\typeOrbTarVal	&	873616860770228352	&	BD+291539	&	\ \ 	8		&	175.9	$\pm$ 	1e-2	&	173.2	$\pm$ 	\ \ 3.0	&	0.27	$\pm$ 	1e-3	&	0.43	$\pm$ 	0.12	&	1	\\
\typeOrbTarVal	&	1142214430312151424	&	HD48679	&	\ \ 	8		&	1111.6	$\pm$ 	\ \ 0.3	&	1177.8	$\pm$ 	45.0	&	0.82	$\pm$ 	5e-4	&	0.83	$\pm$ 	0.01	&	1	\\
\typeOrbTarVal	&	3550762648877966336	&	HD94340	&	\ \ 	3		&	1122.7	$\pm$ 	\ \ 0.4	&	1213.8	$\pm$ 	22.0	&	0.30	$\pm$ 	3e-3	&	0.30	$\pm$ 	0.01	&	2	\\
\typeOrbTarVal	&	3626268998574790656	&	HD112758	&	\ \ 	9		&	103.3	$\pm$ 	3e-2	&	103.2	$\pm$ 	3e-2	&	0.14	$\pm$ 	0.01	&	0.16	$\pm$ 	0.01	&	2	\\
\typeOrbTarVal	&	6264881882000588672	&	HD137812	&	\ \ 	3		&	73.0	$\pm$ 	3e-4	&	73.9	$\pm$ 	\ \ 0.3	&	0.61	$\pm$ 	6e-4	&	0.42	$\pm$ 	0.14	&	1	\\
\typeOrbTarVal	&	2603090003484152064	&	Gl876	&		10		&	61.0	$\pm$ 	8e-4	&	61.4	$\pm$ 	\ \ 0.2	&	2e-3	$\pm$ 	2e-3	&	0.16	$\pm$ 	0.15	&	0	\\
\typeOrbTarVal	&	637329067477530368	&	HD81040	&		11		&	1001.7	$\pm$ 	\ \ 7.0	&	850.8	$\pm$ 	~113	&	0.53	$\pm$ 	0.04	&	0.37	$\pm$ 	0.15	&	0	\\
\typeOrbTarVal	&	1035000055055287680	&	HD68638A	&		33\href{https://dace.unige.ch/radialVelocities/?pattern=HD68638A}{*}	&	240.9	$\pm$ 	\ \ 0.4	&	241.6	$\pm$ 	\ \ 1.0	&	0.57	$\pm$ 	0.04	&	0.31	$\pm$ 	0.06	&	1	\\
\typeOrbTarVal	&	1594127865540229888	&	HD132406	&		12		&	974.0	$\pm$ 	39.0	&	893.2	$\pm$ 	~251	&	0.34	$\pm$ 	0.09	&	0.31	$\pm$ 	0.30	&	0	\\
\typeOrbTarVal	&	6421118739093252224	&	HD175167	&		13		&	1290.0	$\pm$ 	22.0	&	898.7	$\pm$ 	~198	&	0.54	$\pm$ 	0.09	&	0.19	$\pm$ 	0.12	&	0	\\
\typeOrbTarVal	&	5957920668132624256	&	HD162020	&		14		&	8.4	$\pm$ 	6e-5	&	8.4	$\pm$ 	1e-3	&	0.28	$\pm$ 	2e-3	&	0.23	$\pm$ 	0.05	&	2	\\
\typeOrbTarVal	&	3921176983720146560	&	HD106888	&	\ \ 	8		&	365.6	$\pm$ 	\ \ 0.1	&	366.3	$\pm$ 	\ \ 2.8	&	0.46	$\pm$ 	4e-3	&	0.41	$\pm$ 	0.05	&	1	\\
\typeOrbTarVal	&	4385502286022654464	&	HD151465	&		15		&	54.4	$\pm$ 	2e-3	&	54.8	$\pm$ 	\ \ 0.2	&	0.24	$\pm$ 	0.01	&	0.33	$\pm$ 	0.12	&	1	\\
\typeOrbTarVal	&	5808612830236138368	&	HD147584	&		29		&	13.0	$\pm$ 	1e-5	&	13.0	$\pm$ 	1e-3	&	0.01	$\pm$ 	4e-4	&	0.17	$\pm$ 	0.03	&	2	\\
\typeOrbTarVal	&	5086152743542494592	&	HD21703	&		30		&	4.0	$\pm$ 	1e-4	&	4.0	$\pm$ 	6e-4	&	0.34	$\pm$ 	0.06	&	0.29	$\pm$ 	0.10	&	2	\\
\typeOrbTarVal	&	685029558383335168	&	HD77065	&		16		&	119.1	$\pm$ 	3e-3	&	119.1	$\pm$ 	\ \ 0.2	&	0.69	$\pm$ 	4e-4	&	0.71	$\pm$ 	0.04	&	1	\\
\typeOrbTarVal	&	3751763647996317056	&	HD89707	&	\ \ 	9		&	297.7	$\pm$ 	6e-3	&	297.5	$\pm$ 	\ \ 2.0	&	0.95	$\pm$ 	1e-3	&	0.68	$\pm$ 	0.20	&	1	\\
\typeOrbTarVal	&	933054951834436352	&	J0805+4812	&		17		&	740.4	$\pm$ 	\ \ 1.6	&	735.9	$\pm$ 	23.0	&	0.42	$\pm$ 	0.02	&	0.42	$\pm$ 	0.23	&	2	\\
\typeOrbTarVal	&	824461960796102528	&	HD82460	&	\ \ 	8		&	590.9	$\pm$ 	\ \  0.2	&	579.3	$\pm$ 	\ \ 6.1	&	0.84	$\pm$ 	0.01	&	0.73	$\pm$ 	0.04	&	1	\\
\typeOrbTarVal	&	2367734656180397952	&	BD-170063	&		18		&	655.6	$\pm$ 	\ \ 0.6	&	648.9	$\pm$ 	36.0	&	0.54	$\pm$ 	0.01	&	0.28	$\pm$ 	0.22	&	0	\\
\typeOrbTarVal	&	4994200964065634432	&	HD3277	&		19		&	46.2	$\pm$ 	2e-4	&	46.2	$\pm$ 	9e-3	&	0.29	$\pm$ 	1e-3	&	0.27	$\pm$ 	0.01	&	2	\\
\typeOrbTarVal	&	5514929155583865216	&	J0823-4912	&		20		&	246.4	$\pm$ 	\ \ 1.4	&	250.0	$\pm$ 	\ \ 1.2	&	0.35	$\pm$ 	0.07	&	0.22	$\pm$ 	0.06	&	2	\\
\typeOrbTarVal	&	4753355209745022208	&	HD17155	&	\ \ 	3		&	1426.1	$\pm$ 	\ \ 0.7	&	1346.1	$\pm$ 	~107	&	0.78	$\pm$ 	6e-3	&	0.80	$\pm$ 	0.01	&	1	\\
\typeOrbTarVal	&	3309006602007842048	&	HD30246	&		16		&	990.7	$\pm$ 	\ \ 5.6	&	814.7	$\pm$ 	~141	&	0.84	$\pm$ 	0.08	&	0.59	$\pm$ 	0.10	&	1	\\
\typeOrbTarVal	&	6608926350294211328	&	TOI-288	&		32\href{https://dace.unige.ch/radialVelocities/?pattern=TOI-288}{*}	&	79.1	$\pm$ 	\ \ 0.1	&	79.3	$\pm$ 	\ \ 0.5	&	0.65	$\pm$ 	3e-3	&	0.54	$\pm$ 	0.20	&	2	\\
\typeOrbTarVal	&	4942195301023352320	&	HIP9095	&	\ \ 	3\href{https://dace.unige.ch/radialVelocities/?pattern=HIP9095}{*}	&	960.0	$\pm$ 	\ \ 1.4	&	919.6	$\pm$ 	10.9	&	0.33	$\pm$ 	1e-3	&	0.21	$\pm$ 	0.02	&	2	\\
\typeOrbTarVal	&	5902262122552686848	&	HD134237	&	\ \ 	3		&	774.3	$\pm$ 	\ \ 3.4	&	780.4	$\pm$ 	\ \ 2.4	&	0.43	$\pm$ 	0.03	&	0.53	$\pm$ 	0.01	&	2	\\
\typeOrbTarVal	&	2161507648230817792	&	HD166356	&		15		&	261.5	$\pm$ 	\ \ 0.1	&	261.5	$\pm$ 	\ \ 0.9	&	0.45	$\pm$ 	4e-3	&	0.40	$\pm$ 	0.04	&	1	\\
\typeOrbTarVal	&	6647630950597964544	&	HD164427A	&		21		&	108.6	$\pm$ 	4e-2	&	108.4	$\pm$ 	4e-2	&	0.55	$\pm$ 	0.02	&	0.57	$\pm$ 	0.02	&	2	\\
\typeOrbTarVal	&	5563001178343925376	&	HD52756	&		19		&	52.9	$\pm$ 	1e-4	&	52.9	$\pm$ 	\ \ 0.1	&	0.68	$\pm$ 	3e-4	&	0.54	$\pm$ 	0.16	&	1	\\
\typeOrbTarVal	&	5294069567720883968	&	HD63581	&	\ \ 	3		&	1472.6	$\pm$ 	\ \ 0.2	&	1210.5	$\pm$ 	~150	&	0.59	$\pm$ 	1e-3	&	0.56	$\pm$ 	0.03	&	2	\\
\typeOrbTarVal	&	855523714036230016	&	HD92320	&		16		&	145.4	$\pm$ 	1e-2	&	145.1	$\pm$ 	\ \ 0.3	&	0.32	$\pm$ 	1e-3	&	0.26	$\pm$ 	0.05	&	1	\\
\typeOrbTarVal	&	4724313637321332864	&	HD17289	&		19		&	562.1	$\pm$ 	\ \ 0.4	&	561.7	$\pm$ 	\ \ 0.3	&	0.53	$\pm$ 	4e-3	&	0.49	$\pm$ 	5e-3	&	2	\\
\typeOrbTarVal	&	1224551770875466496	&	HD140913	&	\ \ 	9		&	147.9	$\pm$ 	3e-2	&	147.6	$\pm$ 	\ \ 0.2	&	0.53	$\pm$ 	0.01	&	0.51	$\pm$ 	0.05	&	1	\\
\typeOrbTarVal	&	5855730584310531200	&	HD111232	&		22		&	1143.0	$\pm$ 	14.0	&	882.1	$\pm$ 	34.0	&	0.20	$\pm$ 	0.01	&	0.50	$\pm$ 	0.10	&	0	\\
\typeOrbTarVal	&	6334716469679728000	&	HD134251	&	\ \ 	3		&	54.0	$\pm$ 	1e-4	&	53.9	$\pm$ 	\ \ 0.1	&	0.29	$\pm$ 	3e-4	&	0.13	$\pm$ 	0.14	&	2	\\
\typeOrbTarVal	&	1236764218322666880	&	HD130396	&	\ \ 	8		&	2060.6	$\pm$ 	\ \ 7.3	&	1792.6	$\pm$ 	~359	&	0.43	$\pm$ 	4e-3	&	0.33	$\pm$ 	0.10	&	1	\\
\typeOrbTarVal	&	3937211745905473024	&	HD114762	&		23		&	83.9	$\pm$ 	3e-3	&	83.7	$\pm$ 	\ \ 0.1	&	0.34	$\pm$ 	5e-3	&	0.32	$\pm$ 	0.04	&	2	\\
\typeOrbTarVal	&	1181993180456516864	&	HD132032	&		16		&	274.3	$\pm$ 	\ \ 0.2	&	274.4	$\pm$ 	\ \ 2.5	&	0.08	$\pm$ 	2e-3	&	0.09	$\pm$ 	0.09	&	1	\\
\typeOrbTarVal	&	2133476355197071616	&	Kepler-16 AB	&		24		&	41.1	$\pm$ 	8e-5	&	41.1	$\pm$ 	2e-2	&	0.16	$\pm$ 	6e-4	&	0.26	$\pm$ 	0.05	&	2	\\
\typeOrbTarVal	&	1712614124767394816	&	HIP66074	&		25		&				&	297.6	$\pm$ 	\ \ 2.7	&				&	0.46	$\pm$ 	0.17	&	0	\\
\typeOrbTarVal	&	2884087104955208064	&	HD40503	&	\ \ 	2\href{https://dace.unige.ch/radialVelocities/?pattern=HD40503}{*}	&				&	826.5	$\pm$ 	49.9	&				&	0.07	$\pm$ 	0.10	&	0	\\
\typeOrbTarVal	&	276487905502478720	&	HD26596	&	\ \ 	\ \ 		 \href{https://dace.unige.ch/radialVelocities/?pattern=HD26596}{*}	&				&	900.4	$\pm$ 	\ \ 7.6	&				&	0.45	$\pm$ 	0.01	&	2	\\
\typeOrbTarVal	&	725469767850488064	&	HD89010	&	\ \ 	\ \ 		 \href{https://dace.unige.ch/radialVelocities/?pattern=HD89010}{*}	&				&	523.8	$\pm$ 	\ \ 6.2	&				&	0.59	$\pm$ 	0.08	&	2	\\
\typeOrbTarVal	&	6913810483612308480	&	GJ812A	&	\ \ 	\ \ 		 \href{https://dace.unige.ch/radialVelocities/?pattern=GJ812A}{*}	&				&	23.9	$\pm$ 	1e-2	&				&	0.39	$\pm$ 	0.02	&	1	\\
\typeOrbTarVal	&	6399966162596931712	&	GJ9732	&	\ \ 	\ \ 		 \href{https://dace.unige.ch/radialVelocities/?pattern=GJ9732}{*}	&				&	381.2	$\pm$ 	\ \ 0.1	&				&	0.40	$\pm$ 	0.01	&	2	\\
\typeOrbTarVal	&	4296383402592198016	&	HD183162	& \ \ 	\ \ 		 \href{https://dace.unige.ch/radialVelocities/?pattern=HD183162}{*}	&				&	923.5	$\pm$ 	11.0	&				&	0.12	$\pm$ 	0.02	&	2	\\
\typeOrbTarVal	&	5999024986946599808	&	1SWASPJ152523.\ 04-463833.9 / \ CD-46 10046	&	\ \ 	\ \ 		 \href{https://dace.unige.ch/radialVelocities/?pattern=1SWASPJ152523.04-463833.9}{*}	&	242.5	$\pm$ 	\ \ 0.4	&	242.7	$\pm$ 	\ \ 1.6	&	0.43	$\pm$ 	0.01	&	0.22	$\pm$ 	0.08	&	1	\\
\hline
\typeOrbTar	&	2075978592919858432	&	KIC 7917485	&		26		&	840.0	$\pm$ 	22.0	&	810.5	$\pm$ 	28.0	&	0.15	$\pm$ 	0.13	&	0.09	$\pm$ 	0.06	&	2	\\
\typeOrbTar	&	4745373133284418816	&	HR 810	&		27		&	312.0	$\pm$ 	\ \ 5.0	&	331.7	$\pm$ 	\ \ 6.2	&	0.15	$\pm$ 	0.05	&	0.04	$\pm$ 	0.19	&	0	\\
\typeOrbTar	&	2778298280881817984	&	HD5433	&		15		&	576.6	$\pm$ 	\ \ 1.6	&	576.7	$\pm$ 	11.0	&	0.81	$\pm$ 	0.02	&	0.46	$\pm$ 	0.12	&	1	\\
\typeOrbTar	&	3750881083756656128	&	HD91669	&		16		&	497.5	$\pm$ 	\ \ 0.6	&	500.5	$\pm$ 	\ \ 6.9	&	0.45	$\pm$ 	2e-3	&	0.32	$\pm$ 	0.06	&	1	\\
\typeOrbTar	&	4976894960284258048	&	HD142	&		28		&	350.3	$\pm$ 	\ \ 3.6	&	318.6	$\pm$ 	\ \ 6.5	&	0.26	$\pm$ 	0.18	&	0.26	$\pm$ 	0.23	&	0	\\
\typeOrbTar	&	2651390587219807744	&	BD-004475	&		15		&	723.2	$\pm$ 	\ \ 0.7	&	780.0	$\pm$ 	84.0	&	0.39	$\pm$ 	0.01	&	0.48	$\pm$ 	0.11	&	1	\\
\end{tabular}
\tablebib{
%(1) \gdr{3} table \nssTwoBodyOrbit with \nssSolutionType = (a) \texttt{SB1}, (b) \texttt{AstroSpectroSB1}, (c) \texttt{SB2};
(2)~RV data published with this work;
(3)~\cite{Barbato2022}; 
(4)~\cite{2018AJ....156..213M};
(5)~\cite{2002AJ....124.1144L};
(6)~\cite{2019ApJS..242...25F};
(7)~\cite{2019A&A...624A..68M};
(8)~\cite{2019A&A...631A.125K};
(9)~\cite{2000A&A...355..581H};
(10)~\cite{2021ApJS..255....8R};
(11)~\cite{2006A&A...449..417S};
(12)~\cite{2007A&A...473..323D};
(13)~\cite{2010ApJ...711.1229A};
(14)~\cite{2002A&A...390..267U};
(15)~\cite{2021A&A...651A..11D};
(16)~\cite{2016A&A...588A.144W};
(17)~\cite{2020MNRAS.495.1136S}; 
(18)~\cite{2009A&A...496..513M};
(19)~\cite{2011A&A...525A..95S};
(20)~\cite{2013A&A...556A.133S}; 
(21)~\cite{2001ApJ...551..507T};
(22)~\cite{2009ApJ...693.1424M};
(23)~\cite{2011ApJ...735L..41K};
(24)~\cite{2011Sci...333.1602D};
(25)~\cite{2017AJ....153..208B};
(26)~\cite{2016ApJ...827L..17M};
(27)~\cite{2001ApJ...555..410B};
(28)~\cite{2006ApJ...646..505B};
(29)~\cite{2004MNRAS.352..975S};
(30)~\cite{2014A&A...567A..64H};
(31)~\cite{2017A&A...608A.129T};
(32)~\cite{2022A&A...664A..94P};
% BH: CHANGED 9 JUNE 2022 (after submission)
(33)~Reference RV orbital parameters derived from analyses on DACE platform using public data.%~\cite{2007A&A...466.1089B};
}
\tablefoot{
* : data available on: \url{https://dace.unige.ch/radialVelocities/?pattern=Name} with `Name' the value from that column.
}																							
\end{tiny}
\end{table*}

\begin{table*}
\caption{\label{tab:refSolParamsOrbTarG1} \gaia reference solution parameters for \typeOrbTarVal solutions, 1 of 2. See Appendix~\ref{sec:refSolParams} for details. }
\centering                                      % used for centering table
\begin{tiny}
\begin{tabular}{%lllr|rr|rr|r}
    % see https://texblog.org/2019/06/03/control-the-width-of-table-columns-tabular-in-latex/
    m{0.18\textwidth}
    >{\raggedleft}m{0.155\textwidth}
    m{0.06\textwidth}
    >{\raggedright}m{0.02\textwidth}
    |rr|rr|c}
\hline
\hline
\nssSolutionType	&	\gdr{3} \texttt{source\_id}	&	Name	&	Ref.		&	\multicolumn{1}{l}{Ref. $P$ \MOD{[d]}}			&	\multicolumn{1}{l}{Our $P$ \MOD{[d]}}		&	\multicolumn{1}{|l}{Ref. $e$}			&	\multicolumn{1}{l|}{Our $e$}	&  \MOD{\Mcomp}	\\
\hline	
\typeOrbTarVal	&	4919562197762515456	&	TOI-289	&	1a      &	224.7	$\pm$ 	\ \ 0.9	&	222.8	$\pm$ 	\ \ 0.9	&	0.20	$\pm$ 	0.04	&	0.36	$\pm$ 	0.08	&	2	\\
\typeOrbTarVal	&	3280086388180617600	&	HD27642	&	1a		&	706.1	$\pm$ 	16.6	&	709.7	$\pm$ 	\ \ 9.1	&	0.28	$\pm$ 	0.06	&	0.31	$\pm$ 	0.05	&	1	\\
\typeOrbTarVal	&	6873251629971269632	&		&		1a		&	126.0	$\pm$ 	\ \ 0.2	&	125.9	$\pm$ 	\ \ 0.2	&	0.17	$\pm$ 	0.02	&	0.35	$\pm$ 	0.07	&	2	\\
\typeOrbTarVal	&	3176470817561155200	&		&		1a		&	17.4	$\pm$ 	4e-3	&	17.4	$\pm$ 	7e-3	&	0.23	$\pm$ 	0.02	&	0.17	$\pm$ 	0.05	&	2	\\
\typeOrbTarVal	&	4552227182675443584	&		&		1a		&	10.2	$\pm$ 	2e-4	&	10.2	$\pm$ 	2e-3	&	0.02	$\pm$ 	2e-3	&	0.01	$\pm$ 	0.07	&	2	\\
\typeOrbTarVal	&	6448639343335797888	&		&		1a		&	29.7	$\pm$ 	7e-3	&	29.8	$\pm$ 	5e-2	&	0.37	$\pm$ 	0.02	&	0.28	$\pm$ 	0.17	&	2	\\
\typeOrbTarVal	&	5657709399107097600	&		&		1a		&	243.9	$\pm$ 	\ \ 0.9	&	243.8	$\pm$ 	\ \ 0.8	&	0.10	$\pm$ 	0.03	&	0.14	$\pm$ 	0.05	&	2	\\
\typeOrbTarVal	&	4597154602175212672	&		&		1a		&	181.9	$\pm$ 	\ \ 0.3	&	183.9	$\pm$ 	\ \ 1.6	&	0.26	$\pm$ 	0.02	&	0.11	$\pm$ 	0.21	&	2	\\
\typeOrbTarVal	&	1905308073023457920	&		&		1a		&	8.8	$\pm$ 	1e-4	&	8.8	$\pm$ 	5e-3	&	0.38	$\pm$ 	4e-3	&	0.31	$\pm$ 	0.10	&	2	\\
\typeOrbTarVal	&	5547864407933928320	&		&		1a		&	993.0	$\pm$ 	29.7	&	975.5	$\pm$ 	28.3	&	0.43	$\pm$ 	0.03	&	0.46	$\pm$ 	0.02	&	2	\\
\typeOrbTarVal	&	5961294592641279104	&		&		1a		&	765.1	$\pm$ 	\ \ 7.4	&	728.9	$\pm$ 	10.3	&	0.08	$\pm$ 	0.02	&	0.13	$\pm$ 	0.04	&	2	\\
\typeOrbTarVal	&	5220689364275188224	&		&		1a		&	226.2	$\pm$ 	\ \ 0.1	&	228.0	$\pm$ 	\ \ 1.4	&	0.38	$\pm$ 	3e-3	&	0.40	$\pm$ 	0.10	&	2	\\
\typeOrbTarVal	&	797106798693569536	&		&		1a		&	611.8	$\pm$ 	14.8	&	599.2	$\pm$ 	\ \ 0.4	&	0.34	$\pm$ 	0.06	&	0.20	$\pm$ 	4e-3	&	2	\\
\typeOrbTarVal	&	6130370305216737408	&		&		1a		&	19.5	$\pm$ 	3e-3	&	19.4	$\pm$ 	2e-2	&	0.13	$\pm$ 	0.02	&	8e-4	$\pm$ 	0.21	&	2	\\
\typeOrbTarVal	&	670216834655936128	&		&		1a		&	161.3	$\pm$ 	\ \ 0.4	&	161.1	$\pm$ 	\ \ 0.2	&	0.26	$\pm$ 	0.03	&	0.32	$\pm$ 	0.04	&	2	\\
\typeOrbTarVal	&	1401541841225053696	&		&		1a		&	325.2	$\pm$ 	\ \ 1.2	&	325.7	$\pm$ 	\ \ 0.4	&	0.28	$\pm$ 	0.01	&	0.30	$\pm$ 	0.01	&	2	\\
\typeOrbTarVal	&	5096613016130459136	&		&		1a		&	51.6	$\pm$ 	4e-2	&	51.6	$\pm$ 	\ \ 0.1	&	0.24	$\pm$ 	0.01	&	0.26	$\pm$ 	0.09	&	2	\\
\typeOrbTarVal	&	5503370333440752384	&		&		1a		&	815.4	$\pm$ 	41.6	&	827.5	$\pm$ 	37.3	&	0.10	$\pm$ 	0.12	&	0.09	$\pm$ 	0.13	&	2	\\
\typeOrbTarVal	&	3703975672901804160	&		&		1a		&	56.5	$\pm$ 	3e-2	&	56.2	$\pm$ 	\ \ 0.1	&	0.32	$\pm$ 	0.01	&	0.24	$\pm$ 	0.07	&	2	\\
\typeOrbTarVal	&	596878550087942528	&		&		1a		&	10.2	$\pm$ 	2e-4	&	10.2	$\pm$ 	5e-3	&	0.02	$\pm$ 	2e-3	&	0.37	$\pm$ 	0.10	&	2	\\
\typeOrbTarVal	&	4525943082344640256	&		&		1a		&	26.4	$\pm$ 	8e-4	&	26.4	$\pm$ 	6e-3	&	0.22	$\pm$ 	3e-3	&	0.14	$\pm$ 	0.03	&	2	\\
\typeOrbTarVal	&	2431188610386566784	&		&		1a		&	879.3	$\pm$ 	22.8	&	844.6	$\pm$ 	19.9	&	0.16	$\pm$ 	0.02	&	0.12	$\pm$ 	0.04	&	2	\\
\typeOrbTarVal	&	4253017049075877120	&		&		1a		&	21.0	$\pm$ 	6e-3	&	21.0	$\pm$ 	5e-2	&	0.01	$\pm$ 	0.01	&	0.06	$\pm$ 	0.31	&	2	\\
\typeOrbTarVal	&	4921427313081081600	&		&		1a		&	1073.8	$\pm$ 	50.3	&	1117.5	$\pm$ 	~168	&	0.12	$\pm$ 	0.03	&	0.02	$\pm$ 	0.05	&	2	\\
\typeOrbTarVal	&	4254091856092504192	&		&		1a		&	52.8	$\pm$ 	2e-2	&	52.6	$\pm$ 	\ \ 0.2	&	0.35	$\pm$ 	0.02	&	0.17	$\pm$ 	0.34	&	2	\\
\typeOrbTarVal	&	603701328976020864	&		&		1a		&	44.1	$\pm$ 	3e-2	&	44.3	$\pm$ 	3e-2	&	0.30	$\pm$ 	0.01	&	0.28	$\pm$ 	0.07	&	2	\\
\typeOrbTarVal	&	5065735499806985856	&		&		1a		&	125.0	$\pm$ 	\ \ 0.1	&	125.1	$\pm$ 	\ \ 0.1	&	0.56	$\pm$ 	0.01	&	0.62	$\pm$ 	0.08	&	2	\\
\typeOrbTarVal	&	4556905020537693696	&		&		1a		&	641.1	$\pm$ 	\ \ 2.0	&	644.8	$\pm$ 	\ \ 0.9	&	0.25	$\pm$ 	0.01	&	0.25	$\pm$ 	0.01	&	2	\\
\typeOrbTarVal	&	5361232754474635776	&		&		1a		&	815.6	$\pm$ 	\ \ 4.9	&	818.8	$\pm$ 	42.0	&	0.31	$\pm$ 	0.02	&	0.21	$\pm$ 	0.08	&	2	\\
\typeOrbTarVal	&	4544260465016425344	&		&		1a		&	235.7	$\pm$ 	\ \ 0.7	&	235.7	$\pm$ 	\ \ 0.6	&	0.49	$\pm$ 	0.02	&	0.54	$\pm$ 	0.06	&	2	\\
\typeOrbTarVal	&	2633003866585490944	&		&		1a		&	494.3	$\pm$ 	\ \ 4.5	&	503.7	$\pm$ 	\ \ 9.1	&	0.06	$\pm$ 	0.07	&	0.24	$\pm$ 	0.11	&	2	\\
\typeOrbTarVal	&	5997977839551049216	&		&		1a		&	225.2	$\pm$ 	\ \ 0.5	&	225.5	$\pm$ 	\ \ 0.2	&	0.32	$\pm$ 	0.01	&	0.31	$\pm$ 	0.01	&	2	\\
\typeOrbTarVal	&	4902020348734646912	&		&		1a		&	189.7	$\pm$ 	\ \ 0.5	&	187.0	$\pm$ 	\ \ 3.6	&	0.06	$\pm$ 	0.03	&	0.36	$\pm$ 	0.38	&	2	\\
\typeOrbTarVal	&	4824625656537163008	&		&		1a		&	81.1	$\pm$ 	\ \ 0.1	&	80.9	$\pm$ 	\ \ 0.1	&	0.43	$\pm$ 	0.09	&	0.36	$\pm$ 	0.05	&	2	\\
\typeOrbTarVal	&	6525140510537736448	&		&		1a		&	11.6	$\pm$ 	4e-4	&	11.6	$\pm$ 	4e-4	&	0.18	$\pm$ 	4e-3	&	0.30	$\pm$ 	0.08	&	2	\\
\typeOrbTarVal	&	4568198963458261632	&		&		1a		&	430.6	$\pm$ 	\ \ 0.7	&	433.7	$\pm$ 	\ \ 0.5	&	0.57	$\pm$ 	0.02	&	0.54	$\pm$ 	0.01	&	2	\\
\typeOrbTarVal	&	3914521949074389632	&		&		1a		&	24.2	$\pm$ 	2e-3	&	24.1	$\pm$ 	4e-2	&	0.28	$\pm$ 	0.01	&	0.15	$\pm$ 	0.14	&	2	\\
\typeOrbTarVal	&	4666680536328166528	&		&		1a		&	511.2	$\pm$ 	\ \ 2.2	&	515.8	$\pm$ 	\ \ 3.2	&	0.03	$\pm$ 	0.03	&	0.11	$\pm$ 	0.05	&	2	\\
\typeOrbTarVal	&	3683727208499815936	&		&		1a		&	114.8	$\pm$ 	\ \ 0.1	&	115.3	$\pm$ 	\ \ 0.6	&	0.49	$\pm$ 	0.03	&	0.47	$\pm$ 	0.25	&	2	\\
\typeOrbTarVal	&	283446302278774528	&		&		1a		&	22.8	$\pm$ 	3e-3	&	22.8	$\pm$ 	8e-3	&	0.32	$\pm$ 	0.01	&	0.30	$\pm$ 	0.04	&	2	\\
\typeOrbTarVal	&	3064145014608187392	&		&		1a		&	115.1	$\pm$ 	\ \ 0.1	&	115.0	$\pm$ 	\ \ 0.1	&	0.36	$\pm$ 	0.02	&	0.38	$\pm$ 	0.02	&	2	\\
\hline																											
\typeOrbTarVal	&	2185171578009765632	&	HD193468	&		1b$^\dagger$	&				&	291.8	$\pm$ 	\ \ 0.9	&				&	0.71	$\pm$ 	0.06	&	1	\\
\typeOrbTarVal	&	6678530491511225856	&		&		1b$^\dagger$	&				&	911.3	$\pm$ 	\ \ 2.9	&				&	0.49	$\pm$ 	3e-3	&	2	\\
\typeOrbTarVal	&	6778413151435607680	&		&		1b$^\dagger$	&				&	1010.2	$\pm$ 	10.4	&				&	0.35	$\pm$ 	0.01	&	2	\\
\typeOrbTarVal	&	4748772376561143424	&	HIP13685	&		1b		&	1613.2	$\pm$ 	~178	&	1438.9	$\pm$ 	~132	&	0.68	$\pm$ 	0.03	&	0.64	$\pm$ 	0.03	&	1	\\
\typeOrbTarVal	&	5865296850880144256	&		&		1b		&	989.1	$\pm$ 	11.9	&	999.0	$\pm$ 	11.8	&	0.09	$\pm$ 	0.01	&	0.09	$\pm$ 	0.01	&	2	\\
\typeOrbTarVal	&	6511657371244871680	&		&		1b		&	1000.4	$\pm$ 	\ \ 7.0	&	990.6	$\pm$ 	\ \ 6.7	&	0.51	$\pm$ 	4e-3	&	0.51	$\pm$ 	4e-3	&	2	\\
\typeOrbTarVal	&	1568219729458240128	&		&		1b		&	273.9	$\pm$ 	\ \ 0.1	&	273.8	$\pm$ 	\ \ 0.1	&	0.76	$\pm$ 	4e-3	&	0.75	$\pm$ 	3e-3	&	2	\\
\typeOrbTarVal	&	4354357901908595456	&		&		1b		&	133.5	$\pm$ 	\ \ 0.1	&	133.4	$\pm$ 	\ \ 0.1	&	0.27	$\pm$ 	0.01	&	0.24	$\pm$ 	0.01	&	2	\\
\typeOrbTarVal	&	1436561046051723648	&		&		1b		&	316.2	$\pm$ 	\ \ 0.6	&	316.1	$\pm$ 	\ \ 0.5	&	0.56	$\pm$ 	0.02	&	0.54	$\pm$ 	0.02	&	2	\\
\typeOrbTarVal	&	2162250505790084480	&		&		1b		&	585.0	$\pm$ 	\ \ 1.0	&	584.3	$\pm$ 	\ \ 2.9	&	0.27	$\pm$ 	0.01	&	0.25	$\pm$ 	0.03	&	2	\\
\typeOrbTarVal	&	1215279005201870336	&		&		1b		&	151.1	$\pm$ 	\ \ 0.2	&	150.5	$\pm$ 	\ \ 0.3	&	0.57	$\pm$ 	0.01	&	0.52	$\pm$ 	0.04	&	2	\\
\typeOrbTarVal	&	5041087953805187584	&		&		1b		&	665.8	$\pm$ 	\ \ 2.1	&	665.6	$\pm$ 	\ \ 2.4	&	0.81	$\pm$ 	0.01	&	0.78	$\pm$ 	0.01	&	2	\\
\typeOrbTarVal	&	5045853099761604096	&		&		1b		&	114.9	$\pm$ 	\ \ 0.1	&	115.0	$\pm$ 	\ \ 0.1	&	0.13	$\pm$ 	0.01	&	0.20	$\pm$ 	0.02	&	2	\\
\typeOrbTarVal	&	1918953867019478144	&		&		1b		&	349.4	$\pm$ 	\ \ 1.0	&	347.3	$\pm$ 	\ \ 1.0	&	0.42	$\pm$ 	0.03	&	0.44	$\pm$ 	0.03	&	2	\\
\typeOrbTarVal	&	787672026858725632	&		&		1b		&	330.6	$\pm$ 	\ \ 0.4	&	331.0	$\pm$ 	\ \ 0.2	&	0.59	$\pm$ 	0.02	&	0.53	$\pm$ 	0.01	&	2	\\
\typeOrbTarVal	&	2843794470563210496	&		&		1b		&	145.3	$\pm$ 	\ \ 0.1	&	145.2	$\pm$ 	\ \ 0.1	&	0.47	$\pm$ 	0.02	&	0.48	$\pm$ 	0.01	&	2	\\
\typeOrbTarVal	&	6075391292173461760	&		&		1b		&	786.9	$\pm$ 	\ \ 3.9	&	777.6	$\pm$ 	10.5	&	0.11	$\pm$ 	0.01	&	0.11	$\pm$ 	0.02	&	2	\\
\typeOrbTarVal	&	5289793360842855808	&		&		1b		&	690.8	$\pm$ 	\ \ 1.5	&	692.3	$\pm$ 	\ \ 1.5	&	0.07	$\pm$ 	0.01	&	0.06	$\pm$ 	0.01	&	2	\\
\typeOrbTarVal	&	3028470161558479232	&		&		1b		&	121.4	$\pm$ 	3e-2	&	121.4	$\pm$ 	3e-2	&	0.09	$\pm$ 	0.01	&	0.10	$\pm$ 	0.01	&	2	\\
\typeOrbTarVal	&	6708063202046601344	&		&		1b		&	120.0	$\pm$ 	4e-2	&	119.9	$\pm$ 	\ \ 0.1	&	0.44	$\pm$ 	0.01	&	0.39	$\pm$ 	0.02	&	2	\\
\typeOrbTarVal	&	2208365672715184256	&		&		1b		&	691.5	$\pm$ 	\ \ 2.7	&	696.5	$\pm$ 	\ \ 3.1	&	0.41	$\pm$ 	0.01	&	0.43	$\pm$ 	0.01	&	2	\\
\typeOrbTarVal	&	5013703860801457280	&		&		1b		&	299.0	$\pm$ 	\ \ 1.2	&	297.8	$\pm$ 	\ \ 1.0	&	0.33	$\pm$ 	0.05	&	0.19	$\pm$ 	0.05	&	2	\\
\typeOrbTarVal	&	2983397469077381632	&		&		1b		&	854.7	$\pm$ 	\ \ 3.3	&	865.6	$\pm$ 	\ \ 3.8	&	0.04	$\pm$ 	4e-3	&	0.09	$\pm$ 	0.01	&	2	\\
\typeOrbTarVal	&	316256588242414592	&		&		1b		&	166.7	$\pm$ 	\ \ 0.2	&	167.7	$\pm$ 	\ \ 0.3	&	0.21	$\pm$ 	0.02	&	0.10	$\pm$ 	0.04	&	2	\\
\typeOrbTarVal	&	6827751124390424320	&		&		1b		&	306.3	$\pm$ 	\ \ 0.3	&	306.2	$\pm$ 	\ \ 0.2	&	0.22	$\pm$ 	0.01	&	0.24	$\pm$ 	0.01	&	2	\\
\typeOrbTarVal	&	3603119331007599232	&		&		1b		&	145.4	$\pm$ 	\ \ 0.1	&	145.5	$\pm$ 	\ \ 0.1	&	0.30	$\pm$ 	0.01	&	0.28	$\pm$ 	0.02	&	2	\\
\typeOrbTarVal	&	6685418107226825216	&		&		1b		&	554.1	$\pm$ 	\ \ 0.9	&	554.0	$\pm$ 	\ \ 0.9	&	0.43	$\pm$ 	0.02	&	0.40	$\pm$ 	0.02	&	2	\\
\typeOrbTarVal	&	5710296119685549440	&		&		1b		&	41.0	$\pm$ 	1e-2	&	41.0	$\pm$ 	2e-2	&	0.16	$\pm$ 	0.01	&	0.15	$\pm$ 	0.02	&	2	\\
\typeOrbTarVal	&	4829273360906556672	&		&		1b		&	305.8	$\pm$ 	\ \ 1.9	&	310.7	$\pm$ 	\ \ 2.6	&	0.56	$\pm$ 	0.04	&	0.48	$\pm$ 	0.08	&	2	\\
\typeOrbTarVal	&	4830490966955576960	&		&		1b		&	693.7	$\pm$ 	\ \ 2.7	&	694.0	$\pm$ 	\ \ 1.7	&	0.57	$\pm$ 	0.02	&	0.53	$\pm$ 	0.02	&	2	\\
\typeOrbTarVal	&	4832672020067562624	&		&		1b		&	204.1	$\pm$ 	\ \ 0.1	&	204.1	$\pm$ 	\ \ 0.1	&	0.20	$\pm$ 	0.01	&	0.16	$\pm$ 	0.01	&	2	\\
\end{tabular}
\tablebib{
(1) \gdr{3} table \nssTwoBodyOrbit with \nssSolutionType = (a) \texttt{SB1}, (b) \texttt{AstroSpectroSB1}, (c) \texttt{SB2};
}
\tablefoot{
$^\dagger$ : \gaia reference data used for validation was eventually not published in \gdr{3}, therefore no reference values are provided.}																							
\end{tiny}
\end{table*}

\begin{table*}
\caption{\label{tab:refSolParamsOrbTarG2} \gaia reference solution parameters for \typeOrbTarVal solutions, 2 of 2. See Appendix~\ref{sec:refSolParams} for details. }
\centering                                      % used for centering table
\begin{tiny}
\begin{tabular}{%lllr|rr|rr|r}
    % see https://texblog.org/2019/06/03/control-the-width-of-table-columns-tabular-in-latex/
    m{0.18\textwidth}
    >{\raggedleft}m{0.155\textwidth}
    m{0.06\textwidth}
    >{\raggedright}m{0.02\textwidth}
    |rr|rr|c}
\hline
\hline
\nssSolutionType	&	\gdr{3} \texttt{source\_id}	&	Name	&	Ref.		&	\multicolumn{1}{l}{Ref. $P$  \MOD{[d]}}			&	\multicolumn{1}{l}{Our $P$  \MOD{[d]}}		&	\multicolumn{1}{|l}{Ref. $e$}			&	\multicolumn{1}{l|}{Our $e$}	&  \MOD{\Mcomp}	\\
\hline
\typeOrbTarVal	&	5841850590016284544	&		&		1b		&	1170.8	$\pm$ 	14.1	&	1189.9	$\pm$ 	12.6	&	0.42	$\pm$ 	0.01	&	0.43	$\pm$ 	0.01	&	2	\\
\typeOrbTarVal	&	5306416671004618240	&		&		1b		&	513.7	$\pm$ 	\ \ 0.7	&	513.2	$\pm$ 	\ \ 0.6	&	0.40	$\pm$ 	0.01	&	0.36	$\pm$ 	0.01	&	2	\\
\typeOrbTarVal	&	6345896578791113472	&		&		1b		&	170.8	$\pm$ 	\ \ 0.1	&	171.2	$\pm$ 	\ \ 0.1	&	0.37	$\pm$ 	0.01	&	0.31	$\pm$ 	0.03	&	2	\\
\typeOrbTarVal	&	5645712490304616320	&		&		1b		&	864.3	$\pm$ 	\ \ 5.1	&	867.3	$\pm$ 	\ \ 6.0	&	0.11	$\pm$ 	0.01	&	0.10	$\pm$ 	0.01	&	2	\\
\typeOrbTarVal	&	5523408692342861696	&		&		1b		&	685.4	$\pm$ 	\ \ 1.5	&	684.2	$\pm$ 	\ \ 2.3	&	0.02	$\pm$ 	0.01	&	0.04	$\pm$ 	0.02	&	2	\\
\typeOrbTarVal	&	4949308041743808384	&		&		1b		&	821.1	$\pm$ 	\ \ 1.9	&	823.9	$\pm$ 	\ \ 2.0	&	0.28	$\pm$ 	4e-3	&	0.29	$\pm$ 	4e-3	&	2	\\
\typeOrbTarVal	&	4800235533695856256	&		&		1b		&	430.6	$\pm$ 	\ \ 1.7	&	431.5	$\pm$ 	\ \ 1.6	&	0.13	$\pm$ 	0.03	&	0.11	$\pm$ 	0.03	&	2	\\
\typeOrbTarVal	&	4950186203640554880	&		&		1b		&	305.5	$\pm$ 	\ \ 0.3	&	305.6	$\pm$ 	\ \ 0.3	&	0.17	$\pm$ 	0.01	&	0.17	$\pm$ 	5e-3	&	2	\\
\typeOrbTarVal	&	6076629204819105664	&		&		1b		&	143.6	$\pm$ 	\ \ 0.1	&	143.9	$\pm$ 	\ \ 0.2	&	0.13	$\pm$ 	0.01	&	0.24	$\pm$ 	0.05	&	2	\\
\typeOrbTarVal	&	2299519550340109696	&		&		1b		&	141.7	$\pm$ 	4e-2	&	141.7	$\pm$ 	\ \ 0.1	&	0.31	$\pm$ 	0.01	&	0.28	$\pm$ 	0.02	&	2	\\
\typeOrbTarVal	&	5629502218573747584	&		&		1b		&	108.8	$\pm$ 	2e-2	&	108.7	$\pm$ 	5e-2	&	0.08	$\pm$ 	0.01	&	0.15	$\pm$ 	0.02	&	2	\\
\typeOrbTarVal	&	5459518648630576768	&		&		1b		&	274.6	$\pm$ 	\ \ 0.2	&	274.2	$\pm$ 	\ \ 0.3	&	0.16	$\pm$ 	0.01	&	0.15	$\pm$ 	0.01	&	2	\\
\typeOrbTarVal	&	2888649013058936448	&		&		1b		&	1105.5	$\pm$ 	22.0	&	1108.9	$\pm$ 	41.9	&	0.41	$\pm$ 	0.01	&	0.40	$\pm$ 	0.02	&	2	\\
\typeOrbTarVal	&	2889380085212270080	&		&		1b		&	1165.6	$\pm$ 	30.4	&	1157.3	$\pm$ 	23.1	&	0.24	$\pm$ 	0.01	&	0.24	$\pm$ 	0.01	&	2	\\
\typeOrbTarVal	&	1527631807474248448	&		&		1b		&	709.9	$\pm$ 	\ \ 0.6	&	710.1	$\pm$ 	\ \ 0.6	&	0.31	$\pm$ 	2e-3	&	0.31	$\pm$ 	2e-3	&	2	\\
\typeOrbTarVal	&	1529439851267809024	&		&		1b		&	196.6	$\pm$ 	\ \ 0.1	&	196.6	$\pm$ 	5e-2	&	0.71	$\pm$ 	0.01	&	0.72	$\pm$ 	0.01	&	2	\\
\typeOrbTarVal	&	3638745672411053952	&		&		1b		&	206.7	$\pm$ 	\ \ 0.2	&	206.7	$\pm$ 	\ \ 0.7	&	0.32	$\pm$ 	0.01	&	0.33	$\pm$ 	0.03	&	2	\\
\typeOrbTarVal	&	6631710606341412096	&		&		1b		&	937.0	$\pm$ 	\ \ 6.4	&	933.3	$\pm$ 	\ \ 6.7	&	0.31	$\pm$ 	0.01	&	0.31	$\pm$ 	0.01	&	2	\\
\typeOrbTarVal	&	5490419250399767680	&		&		1b		&	228.7	$\pm$ 	\ \ 0.2	&	228.2	$\pm$ 	\ \ 0.2	&	0.43	$\pm$ 	0.01	&	0.48	$\pm$ 	0.01	&	2	\\
\typeOrbTarVal	&	5391879235910368768	&		&		1b		&	645.0	$\pm$ 	\ \ 0.9	&	648.7	$\pm$ 	\ \ 1.8	&	0.55	$\pm$ 	5e-3	&	0.49	$\pm$ 	0.02	&	2	\\
\typeOrbTarVal	&	3116065985196889216	&		&		1b		&	852.4	$\pm$ 	\ \ 7.8	&	859.4	$\pm$ 	\ \ 7.8	&	0.46	$\pm$ 	0.01	&	0.46	$\pm$ 	0.01	&	2	\\
\typeOrbTarVal	&	1294000704856807168	&		&		1b		&	125.3	$\pm$ 	\ \ 0.1	&	125.4	$\pm$ 	\ \ 0.1	&	0.28	$\pm$ 	0.01	&	0.27	$\pm$ 	0.01	&	2	\\
\typeOrbTarVal	&	3569106488558337792	&		&		1b		&	212.3	$\pm$ 	\ \ 0.5	&	217.3	$\pm$ 	\ \ 0.1	&	0.23	$\pm$ 	0.03	&	0.53	$\pm$ 	0.02	&	2	\\
\typeOrbTarVal	&	3579784327012046336	&		&		1b		&	90.1	$\pm$ 	5e-2	&	90.0	$\pm$ 	\ \ 0.2	&	0.49	$\pm$ 	0.01	&	0.43	$\pm$ 	0.07	&	2	\\
\typeOrbTarVal	&	861776842822330368	&		&		1b		&	33.6	$\pm$ 	9e-3	&	33.6	$\pm$ 	1e-2	&	0.19	$\pm$ 	0.03	&	0.26	$\pm$ 	0.04	&	2	\\
\typeOrbTarVal	&	5545537291566200320	&		&		1b		&	835.8	$\pm$ 	\ \ 4.9	&	840.2	$\pm$ 	\ \ 5.6	&	0.67	$\pm$ 	5e-3	&	0.68	$\pm$ 	5e-3	&	2	\\
\typeOrbTarVal	&	4810832695483445760	&		&		1b		&	270.3	$\pm$ 	\ \ 0.2	&	269.6	$\pm$ 	\ \ 0.7	&	0.36	$\pm$ 	5e-3	&	0.27	$\pm$ 	0.04	&	2	\\
\typeOrbTarVal	&	4823703990915066880	&		&		1b		&	85.5	$\pm$ 	\ \ 0.1	&	85.2	$\pm$ 	\ \ 0.2	&	0.05	$\pm$ 	0.03	&	0.08	$\pm$ 	0.09	&	2	\\
\typeOrbTarVal	&	6542137929509574144	&		&		1b		&	573.1	$\pm$ 	\ \ 1.1	&	572.4	$\pm$ 	\ \ 0.9	&	0.22	$\pm$ 	0.01	&	0.22	$\pm$ 	0.01	&	2	\\
\typeOrbTarVal	&	1014542369909518976	&		&		1b		&	154.4	$\pm$ 	\ \ 0.2	&	154.1	$\pm$ 	\ \ 0.3	&	0.40	$\pm$ 	0.03	&	0.48	$\pm$ 	0.08	&	2	\\
\typeOrbTarVal	&	1648950576156773760	&		&		1b		&	564.3	$\pm$ 	\ \ 1.2	&	564.5	$\pm$ 	\ \ 0.9	&	0.16	$\pm$ 	0.02	&	0.16	$\pm$ 	0.01	&	2	\\
\typeOrbTarVal	&	107511930591409280	&		&		1b		&	208.1	$\pm$ 	\ \ 0.5	&	207.4	$\pm$ 	\ \ 0.5	&	0.28	$\pm$ 	0.02	&	0.29	$\pm$ 	0.03	&	2	\\
\typeOrbTarVal	&	1484569984328261632	&		&		1b		&	577.0	$\pm$ 	\ \ 0.8	&	579.1	$\pm$ 	\ \ 1.3	&	0.41	$\pm$ 	0.01	&	0.41	$\pm$ 	0.01	&	2	\\
\typeOrbTarVal	&	1695571110421334144	&		&		1b		&	450.2	$\pm$ 	\ \ 0.6	&	450.3	$\pm$ 	\ \ 0.6	&	0.33	$\pm$ 	0.01	&	0.32	$\pm$ 	0.01	&	2	\\
\typeOrbTarVal	&	266076492460606592	&		&		1b		&	567.1	$\pm$ 	\ \ 1.7	&	566.9	$\pm$ 	\ \ 1.3	&	0.44	$\pm$ 	0.02	&	0.44	$\pm$ 	0.01	&	2	\\
\typeOrbTarVal	&	1305309456826990464	&		&		1b		&	533.6	$\pm$ 	\ \ 0.5	&	534.2	$\pm$ 	\ \ 0.5	&	0.22	$\pm$ 	4e-3	&	0.22	$\pm$ 	4e-3	&	2	\\
\typeOrbTarVal	&	5564490982239584768	&		&		1b		&	716.1	$\pm$ 	\ \ 4.3	&	720.5	$\pm$ 	\ \ 5.2	&	0.17	$\pm$ 	0.01	&	0.16	$\pm$ 	0.01	&	2	\\
\typeOrbTarVal	&	2396173975404592512	&		&		1b		&	478.2	$\pm$ 	\ \ 2.1	&	478.4	$\pm$ 	\ \ 1.9	&	0.13	$\pm$ 	0.03	&	0.14	$\pm$ 	0.02	&	2	\\
\typeOrbTarVal	&	6661911579416783616	&		&		1b		&	62.8	$\pm$ 	2e-2	&	62.9	$\pm$ 	3e-2	&	0.37	$\pm$ 	0.01	&	0.47	$\pm$ 	0.06	&	2	\\
\typeOrbTarVal	&	4650689681594278272	&		&		1b		&	951.7	$\pm$ 	11.0	&	945.0	$\pm$ 	14.0	&	0.09	$\pm$ 	0.01	&	0.08	$\pm$ 	0.02	&	2	\\
\typeOrbTarVal	&	1813985523437968256	&		&		1b		&	302.5	$\pm$ 	\ \ 0.5	&	302.7	$\pm$ 	\ \ 0.4	&	0.24	$\pm$ 	0.01	&	0.24	$\pm$ 	0.01	&	2	\\
\typeOrbTarVal	&	6807019282894453504	&		&		1b		&	150.8	$\pm$ 	\ \ 0.1	&	150.6	$\pm$ 	\ \ 0.1	&	0.19	$\pm$ 	0.02	&	0.23	$\pm$ 	0.02	&	2	\\
\typeOrbTarVal	&	4774328222246003840	&		&		1b		&	163.7	$\pm$ 	\ \ 0.2	&	163.5	$\pm$ 	\ \ 1.3	&	0.01	$\pm$ 	0.01	&	0.38	$\pm$ 	0.14	&	2	\\
\typeOrbTarVal	&	422886126401496064	&		&		1b		&	295.5	$\pm$ 	\ \ 0.1	&	295.5	$\pm$ 	\ \ 0.1	&	0.26	$\pm$ 	3e-3	&	0.26	$\pm$ 	3e-3	&	2	\\
\typeOrbTarVal	&	1615450866336763904	&		&		1b		&	390.3	$\pm$ 	\ \ 0.2	&	390.5	$\pm$ 	\ \ 0.3	&	0.09	$\pm$ 	0.01	&	0.07	$\pm$ 	0.02	&	2	\\
\typeOrbTarVal	&	286293109680203008	&		&		1b		&	393.7	$\pm$ 	\ \ 0.5	&	393.7	$\pm$ 	\ \ 0.4	&	0.25	$\pm$ 	0.02	&	0.18	$\pm$ 	0.01	&	2	\\
\typeOrbTarVal	&	4963661822446609792	&		&		1b		&	512.6	$\pm$ 	\ \ 0.5	&	513.1	$\pm$ 	\ \ 0.5	&	0.86	$\pm$ 	0.01	&	0.87	$\pm$ 	0.01	&	2	\\
\typeOrbTarVal	&	4133650458966620672	&		&		1b		&	211.6	$\pm$ 	\ \ 0.4	&	211.1	$\pm$ 	\ \ 0.4	&	0.52	$\pm$ 	0.01	&	0.53	$\pm$ 	0.02	&	2	\\
\typeOrbTarVal	&	4972788662311608960	&		&		1b		&	115.2	$\pm$ 	1e-2	&	115.3	$\pm$ 	4e-2	&	0.37	$\pm$ 	4e-3	&	0.37	$\pm$ 	0.01	&	2	\\
\typeOrbTarVal	&	5236430419447179776	&		&		1b		&	169.6	$\pm$ 	\ \ 0.1	&	169.6	$\pm$ 	\ \ 0.1	&	0.40	$\pm$ 	0.01	&	0.40	$\pm$ 	0.01	&	2	\\
\typeOrbTarVal	&	4594158089392172928	&		&		1b		&	178.7	$\pm$ 	\ \ 0.3	&	178.3	$\pm$ 	\ \ 0.3	&	0.56	$\pm$ 	0.02	&	0.57	$\pm$ 	0.05	&	2	\\
\typeOrbTarVal	&	1348238101626698368	&		&		1b		&	1253.5	$\pm$ 	18.4	&	1253.7	$\pm$ 	20.7	&	0.40	$\pm$ 	0.01	&	0.40	$\pm$ 	0.01	&	2	\\
\typeOrbTarVal	&	5631222984331124864	&		&		1b		&	15.4	$\pm$ 	2e-3	&	15.4	$\pm$ 	2e-3	&	0.36	$\pm$ 	0.02	&	0.29	$\pm$ 	0.03	&	2	\\
\typeOrbTarVal	&	6257172793656011520	&		&		1b		&	552.2	$\pm$ 	\ \ 1.9	&	555.5	$\pm$ 	\ \ 1.1	&	0.70	$\pm$ 	0.02	&	0.63	$\pm$ 	0.01	&	2	\\
\hline																											
\typeOrbTarVal	&	5526720593166247680	&	HD72834	&		1c		&	145.3	$\pm$ 	\ \ 0.1	&	145.3	$\pm$ 	\ \ 0.2	&	0.68	$\pm$ 	0.01	&	0.56	$\pm$ 	0.03	&	1	\\
\typeOrbTarVal	&	5583755078792642304	&	HD47391	&		1c		&	214.1	$\pm$ 	\ \ 0.1	&	210.3	$\pm$ 	\ \ 0.4	&	0.19	$\pm$ 	4e-3	&	0.17	$\pm$ 	0.04	&	1	\\
\typeOrbTarVal	&	4321775627212956672	&	HD184962	&		1c		&	33.7	$\pm$ 	4e-3	&	33.4	$\pm$ 	\ \ 0.1	&	0.27	$\pm$ 	6e-3	&	0.31	$\pm$ 	0.26	&	1	\\
\typeOrbTarVal	&	6490284754986167552	&	GJ4331	&		1c		&	24.8	$\pm$ 	3e-3	&	24.7	$\pm$ 	2e-2	&	0.50	$\pm$ 	7e-3	&	0.55	$\pm$ 	0.21	&	1	\\
\typeOrbTarVal	&	6158160019228728448	&	HD109524	&		1c		&	12.0	$\pm$ 	1e-4	&	12.0	$\pm$ 	6e-3	&	0.17	$\pm$ 	2e-3	&	0.33	$\pm$ 	0.13	&	1	\\
\typeOrbTarVal	&	2035577729682322176	&		&		1c		&	18.8	$\pm$ 	4e-4	&	18.8	$\pm$ 	5e-3	&	0.31	$\pm$ 	2e-3	&	0.30	$\pm$ 	0.06	&	2	\\
\typeOrbTarVal	&	1067685718250692352	&		&		1c		&	30.5	$\pm$ 	6e-4	&	30.5	$\pm$ 	1e-2	&	0.19	$\pm$ 	1e-3	&	0.26	$\pm$ 	0.05	&	2	\\
\typeOrbTarVal	&	6017724140678769024	&		&		1c		&	31.8	$\pm$ 	1e-3	&	31.9	$\pm$ 	9e-3	&	0.15	$\pm$ 	3e-3	&	0.22	$\pm$ 	0.04	&	2	\\
\typeOrbTarVal	&	5534280594604579584	&		&		1c		&	175.5	$\pm$ 	\ \ 0.1	&	176.7	$\pm$ 	\ \ 0.3	&	0.15	$\pm$ 	0.01	&	0.20	$\pm$ 	0.03	&	2	\\
\end{tabular}
\tablebib{
(1) \gdr{3} table \nssTwoBodyOrbit with \nssSolutionType = (a) \texttt{SB1}, (b) \texttt{AstroSpectroSB1}, (c) \texttt{SB2};
}																						
\end{tiny}
\end{table*}

\clearpage
\onecolumn

% archive queries
\section{Examples of \gaia archive queries \label{sec:archiveQueries}}
This section describes Gaia archive queries in the ADQL format that return the various selections and tables presented in this paper. These queries can be made online in the \textit{Gaia} archive at \href{https://gea.esac.esa.int/archive/}{\tt https://gea.esac.esa.int/archive/}:\\

\subsection{\MOD{Sky source density and number of observations}}

      \MOD{Input data for HEALPix \citep{2002ASPC..281..107G} plots of the source sky density and number of FoV observations in the lower two panels of Fig.~\ref{fig:skyPlots}}:
      \begin{tiny}\begin{verbatim}
select 
  gaia_healpix_index(8, source_id) AS hpx8,
  round( count(*) / 0.052441, 0) as sources_per_sq_deg,
  round( avg(astrometric_matched_transits), 1) as mean_agis_fov,
  round( max(astrometric_matched_transits), 0) as max_agis_fov
from gaiaedr3.gaia_source
group by hpx8
      \end{verbatim} \end{tiny}

\subsection{Flag counts}

      Counts of \texttt{flags} in \nssTwoBodyOrbit for our \nssSolutionType's discussed in Sect.~\ref{ssec:archiveModelParams}:
      \begin{tiny}\begin{verbatim}
select
flags,
nss_solution_type,
count(source_id) as counts
from gaiadr3.nss_two_body_orbit 
where nss_solution_type in
    ('OrbitalTargetedSearch','OrbitalAlternative','OrbitalTargetedSearchValidated','OrbitalAlternativeValidated')
group by nss_solution_type, flags
order by flags, counts
      \end{verbatim} \end{tiny}
%lags	nss_solution_type	counts
%0	OrbitalTargetedSearchValidated	2
%0	OrbitalTargetedSearch	14
%0	OrbitalAlternative	148
%64	OrbitalAlternativeValidated	10
%64	OrbitalTargetedSearchValidated	24
%64	OrbitalTargetedSearch	70
%64	OrbitalAlternative	467
%192	OrbitalAlternative	4
%192	OrbitalTargetedSearchValidated	162
%192	OrbitalTargetedSearch	261

% efficiency is zero
%select
%CASE_CONDITION(-1.0, efficiency=0, 1.0, efficiency>0, 0.0) as efficiency_is_zero,
%nss_solution_type,
%count(source_id) as counts
%from 	
%user_dr3int6.nss_two_body_orbit 
%where nss_solution_type in ('OrbitalTargetedSearch','OrbitalAlternative','OrbitalTargetedSearchValidated','OrbitalAlternativeValidated')
%group by nss_solution_type, efficiency_is_zero
%order by efficiency_is_zero, counts
%
%efficiency_is_zero	nss_solution_type	counts
%0	OrbitalAlternativeValidated	1
%0	OrbitalTargetedSearch	42
%0	OrbitalTargetedSearchValidated	44
%0	OrbitalAlternative	134
%1	OrbitalAlternativeValidated	9
%1	OrbitalTargetedSearchValidated	144
%1	OrbitalTargetedSearch	303
%1	OrbitalAlternative	485

\subsection{Table~\ref{tab:overlapOtherTabs} counts}

      Counts of \texttt{source\_id}'s with alternative solutions in table \nssTwoBodyOrbit:
      \begin{tiny}\begin{verbatim}
select 
nss_solution_type_exopl, 
nonexopl.nss_solution_type as nss_solution_type_other, 
count(source_id) as counts from 
(   select source_id,  cnt.num_solutions, orb.nss_solution_type as nss_solution_type_exopl, period as period_exopl
    from gaiadr3.nss_two_body_orbit orb join 
    ( select source_id, count(source_id) as num_solutions from gaiadr3.nss_two_body_orbit group by source_id
    ) as cnt using (source_id)	  
    where nss_solution_type in 
        ('OrbitalTargetedSearch','OrbitalAlternative','OrbitalTargetedSearchValidated','OrbitalAlternativeValidated')
    and cnt.num_solutions = 2 -- subset of source_id with two entries (there are none with >2)
) as exopl
join gaiadr3.nss_two_body_orbit nonexopl using (source_id) where nonexopl.nss_solution_type not in 	
    ('OrbitalTargetedSearch','OrbitalAlternative','OrbitalTargetedSearchValidated','OrbitalAlternativeValidated')
group by nss_solution_type_exopl,nss_solution_type
order by nss_solution_type_exopl asc, counts desc
      \end{verbatim} \end{tiny}

\noindent  Counts of \texttt{source\_id}'s with an alternative solution in table \nssNonLinearSpectro:
      \begin{tiny}\begin{verbatim}
select 
exopl.nss_solution_type as nss_solution_type_exopl, 
nnls.nss_solution_type as nss_solution_type_other, 
count(source_id) as counts 
from gaiadr3.nss_non_linear_spectro nnls
join gaiadr3.nss_two_body_orbit exopl using (source_id) where exopl.nss_solution_type in 
   ('OrbitalTargetedSearch','OrbitalAlternative','OrbitalTargetedSearchValidated','OrbitalAlternativeValidated')
group by exopl.nss_solution_type, nnls.nss_solution_type 
order by exopl.nss_solution_type asc, counts desc
\end{verbatim} \end{tiny}

\clearpage
\twocolumn

%
%+++++++++++++++++++++++++++++++++++++++++++++++++++++++++++++++++++++++++++
\section{Acronyms}\label{sec:acronyms}
{\small
\begin{tabular}{ll}
\hline\hline
\textbf{Acronym}  &  \textbf{Description}  \\ 
%2MASS & Two-Micron All Sky Survey \\
ADQL&Astronomical Data Query Language \\
%AGB&Asymptotic Giant Branch (star) \\
AL & ALong scan (direction) \\
AC & ACross scan (direction) \\
%AP & Astrophysical Parameters \\
%BH & Black hole \\
%BP & Gaia Blue Photometer \\
%CMD & Colour Magnitude Diagram \\
%DoF&Degree(s) of Freedom \\
DACE & \href{https://dace.unige.ch/dashboard/}{Data and Analysis Center for Exoplanets} \\
DE-MCMC & Differential Evolution Markov Chain Monte Carlo \\ %(Sect.~\ref{sec:method_mcmc}) \\
DPAC & Data Processing and Analysis Consortium \\
DR1 & \href{https://www.cosmos.esa.int/web/gaia/data-release-1}{Gaia Data Release 1} \\
DR2 & \href{https://www.cosmos.esa.int/web/gaia/data-release-2}{Gaia Data Release 2}\\
EDR3 & \href{https://www.cosmos.esa.int/web/gaia/early-data-release-3}{Gaia Early Data Release 3} \\
DR3 & \href{https://www.cosmos.esa.int/web/gaia/data-release-3}{Gaia Data Release 3} \\
%EB&Eclipsing Binary \\
%FLAME & Final Luminosity Age Mass Estimator \\
%FWHM&Full Width at Half-Maximum \\
GA & Genetic Algorithm \\ %(Sect.~\ref{sec:method_ga}) \\
%GALEX&GALaxy Evolution eXplorer \\
%GoF & Goodness of Fit \\
%GSPPhot&Generalised Stellar Parametriser PHOTometry \\
%GSPSpec&Generalised Stellar Parametriser SPECtroscopy \\
%GUCD & Gaia Ultra-cool Dwarf\\
%HealPix & Hierarchical Equal Area isoLatitude Pixelisation \\
%HPM & High Proper Motion \\
%HRD & Hertzsprung-Russell diagram\\
%IMF&Initial Mass Function \\
%LMC&Large Magellanic Cloud \\
%LPV & Long Period Variables \\
%LSF & Line Spread Function (one dimensional) \\
%MAD & Median Absolute Deviation \\
%MS&Main Sequence (star) \\
%NLS & Non linear spectro \\
%NS & Neutron star \\
NSS&Non-Single Star \\
%PMa&proper motion anomaly \\
%PSF & Point Spread Function (two dimensional) \\
%RGB&Red Giant Branch (star) \\
%RMS&Root-Mean-Square \\
%RP & Gaia Red Photometer \\
%RUWE&Re-normalised unit-weight error \\
%RV & Radial Velocity \\
%RVS&Radial Velocity Spectrometer \\
%SB & Spectroscopic Binary  \\
%SB2&Double-lined Spectroscopic Binary \\
%SNR & Signal to Noise ratio (also denoted SN and S/N)\\
%TBO & Two body orbits \\
%UCD&Ultra cool dwarf (star) \\
%UV&UltraViolet \\
%VIMF & variable-induced movers fixed \\
%WD & White dwarf\\
\hline
\end{tabular} 
}
%+++++++++++++++++++++++++++++++++++++++++++++++++++++++++++++++++++++++++++

%\clearpage
%\WORKSUGGESTION{
%\
%\
%\
%General items left to address before A\&A submission:
%\begin{itemize}
    %\item We describe the reason for our jitter term in several places, but not giving consistent description of what it for, we should make sure this is homogenised: Sect.~\ref{sec:orbitalModel} towards end of text, Sect.~\ref{ssec:mathModelDescr} towards end of text two times, Sect.~\ref{sssec:gof}, [add more places when identified]...
   % \item conclusions (and literature reference to predicted \gaia numbers of planets and BDs).
    %\item BH: (find out how to) add links to DR3 documentation and archive fields.
%    \item replace `geapre.esac.esa.int` (dr3.int6) with `gea.esac.esa.int` (the upcoming DR3 archive) in all links of this paper and `local.bib' (for documentation link) before publishing.
%    \item Sect.~\ref{sec:refSolParams} finalise little details of references list (can be done during DPAC review)]
%    \item check that all references in Table~\ref{tab:refSolParamsOrbTarNG} with "*" DACE link are actually publicly working/available.
%\end{itemize}

%}

\section{\gaia acknowledgements\label{ssec:appendixA}}

This work presents results from the European Space Agency (ESA) space mission \gaia. \gaia\ data are being processed by the \gaia\ Data Processing and Analysis Consortium (DPAC). Funding for the DPAC is provided by national institutions, in particular the institutions participating in the \gaia\ MultiLateral Agreement (MLA). The \gaia\ mission website is \url{https://www.cosmos.esa.int/gaia}. The \gaia\ archive website is \url{https://archives.esac.esa.int/gaia}.

The \gaia\ mission and data processing have financially been supported by, in alphabetical order by country:
\begin{itemize}
\item the Algerian Centre de Recherche en Astronomie, Astrophysique et G\'{e}ophysique of Bouzareah Observatory;
\item the Austrian Fonds zur F\"{o}rderung der wissenschaftlichen Forschung (FWF) Hertha Firnberg Programme through grants T359, P20046, and P23737;
\item the BELgian federal Science Policy Office (BELSPO) through various PROgramme de D\'{e}veloppement d'Exp\'{e}riences scientifiques (PRODEX) grants and the Polish Academy of Sciences - Fonds Wetenschappelijk Onderzoek through grant VS.091.16N, and the Fonds de la Recherche Scientifique (FNRS), and the Research Council of Katholieke Universiteit (KU) Leuven through grant C16/18/005 (Pushing AsteRoseismology to the next level with TESS, GaiA, and the Sloan DIgital Sky SurvEy -- PARADISE);  
\item the Brazil-France exchange programmes Funda\c{c}\~{a}o de Amparo \`{a} Pesquisa do Estado de S\~{a}o Paulo (FAPESP) and Coordena\c{c}\~{a}o de Aperfeicoamento de Pessoal de N\'{\i}vel Superior (CAPES) - Comit\'{e} Fran\c{c}ais d'Evaluation de la Coop\'{e}ration Universitaire et Scientifique avec le Br\'{e}sil (COFECUB);
\item the Chilean Agencia Nacional de Investigaci\'{o}n y Desarrollo (ANID) through Fondo Nacional de Desarrollo Cient\'{\i}fico y Tecnol\'{o}gico (FONDECYT) Regular Project 1210992 (L.~Chemin);
\item the National Natural Science Foundation of China (NSFC) through grants 11573054, 11703065, and 12173069, the China Scholarship Council through grant 201806040200, and the Natural Science Foundation of Shanghai through grant 21ZR1474100;  
\item the Tenure Track Pilot Programme of the Croatian Science Foundation and the \'{E}cole Polytechnique F\'{e}d\'{e}rale de Lausanne and the project TTP-2018-07-1171 `Mining the Variable Sky', with the funds of the Croatian-Swiss Research Programme;
\item the Czech-Republic Ministry of Education, Youth, and Sports through grant LG 15010 and INTER-EXCELLENCE grant LTAUSA18093, and the Czech Space Office through ESA PECS contract 98058;
\item the Danish Ministry of Science;
\item the Estonian Ministry of Education and Research through grant IUT40-1;
\item the European Commission’s Sixth Framework Programme through the European Leadership in Space Astrometry (\href{https://www.cosmos.esa.int/web/gaia/elsa-rtn-programme}{ELSA}) Marie Curie Research Training Network (MRTN-CT-2006-033481), through Marie Curie project PIOF-GA-2009-255267 (Space AsteroSeismology \& RR Lyrae stars, SAS-RRL), and through a Marie Curie Transfer-of-Knowledge (ToK) fellowship (MTKD-CT-2004-014188); the European Commission's Seventh Framework Programme through grant FP7-606740 (FP7-SPACE-2013-1) for the \gaia\ European Network for Improved data User Services (\href{https://gaia.ub.edu/twiki/do/view/GENIUS/}{GENIUS}) and through grant 264895 for the \gaia\ Research for European Astronomy Training (\href{https://www.cosmos.esa.int/web/gaia/great-programme}{GREAT-ITN}) network;
\item the European Cooperation in Science and Technology (COST) through COST Action CA18104 `Revealing the Milky Way with \gaia (MW-Gaia)';
\item the European Research Council (ERC) through grants 320360, 647208, and 834148 and through the European Union’s Horizon 2020 research and innovation and excellent science programmes through Marie Sk{\l}odowska-Curie grant 745617 (Our Galaxy at full HD -- Gal-HD) and 895174 (The build-up and fate of self-gravitating systems in the Universe) as well as grants 687378 (Small Bodies: Near and Far), 682115 (Using the Magellanic Clouds to Understand the Interaction of Galaxies), 695099 (A sub-percent distance scale from binaries and Cepheids -- CepBin), 716155 (Structured ACCREtion Disks -- SACCRED), 951549 (Sub-percent calibration of the extragalactic distance scale in the era of big surveys -- UniverScale), and 101004214 (Innovative Scientific Data Exploration and Exploitation Applications for Space Sciences -- EXPLORE);
\item the European Science Foundation (ESF), in the framework of the \gaia\ Research for European Astronomy Training Research Network Programme (\href{https://www.cosmos.esa.int/web/gaia/great-programme}{GREAT-ESF});
\item the European Space Agency (ESA) in the framework of the \gaia\ project, through the Plan for European Cooperating States (PECS) programme through contracts C98090 and 4000106398/12/NL/KML for Hungary, through contract 4000115263/15/NL/IB for Germany, and through PROgramme de D\'{e}veloppement d'Exp\'{e}riences scientifiques (PRODEX) grant 4000127986 for Slovenia;  
\item the Academy of Finland through grants 299543, 307157, 325805, 328654, 336546, and 345115 and the Magnus Ehrnrooth Foundation;
\item the French Centre National d’\'{E}tudes Spatiales (CNES), the Agence Nationale de la Recherche (ANR) through grant ANR-10-IDEX-0001-02 for the `Investissements d'avenir' programme, through grant ANR-15-CE31-0007 for project `Modelling the Milky Way in the \gaia era’ (MOD4Gaia), through grant ANR-14-CE33-0014-01 for project `The Milky Way disc formation in the \gaia era’ (ARCHEOGAL), through grant ANR-15-CE31-0012-01 for project `Unlocking the potential of Cepheids as primary distance calibrators’ (UnlockCepheids), through grant ANR-19-CE31-0017 for project `Secular evolution of galxies' (SEGAL), and through grant ANR-18-CE31-0006 for project `Galactic Dark Matter' (GaDaMa), the Centre National de la Recherche Scientifique (CNRS) and its SNO \gaia of the Institut des Sciences de l’Univers (INSU), its Programmes Nationaux: Cosmologie et Galaxies (PNCG), Gravitation R\'{e}f\'{e}rences Astronomie M\'{e}trologie (PNGRAM), Plan\'{e}tologie (PNP), Physique et Chimie du Milieu Interstellaire (PCMI), and Physique Stellaire (PNPS), the `Action F\'{e}d\'{e}ratrice \gaia' of the Observatoire de Paris, the R\'{e}gion de Franche-Comt\'{e}, the Institut National Polytechnique (INP) and the Institut National de Physique nucl\'{e}aire et de Physique des Particules (IN2P3) co-funded by CNES;
\item the German Aerospace Agency (Deutsches Zentrum f\"{u}r Luft- und Raumfahrt e.V., DLR) through grants 50QG0501, 50QG0601, 50QG0602, 50QG0701, 50QG0901, 50QG1001, 50QG1101, 50\-QG1401, 50QG1402, 50QG1403, 50QG1404, 50QG1904, 50QG2101, 50QG2102, and 50QG2202, and the Centre for Information Services and High Performance Computing (ZIH) at the Technische Universit\"{a}t Dresden for generous allocations of computer time;
\item the Hungarian Academy of Sciences through the Lend\"{u}let Programme grants LP2014-17 and LP2018-7 and the Hungarian National Research, Development, and Innovation Office (NKFIH) through grant KKP-137523 (`SeismoLab');
\item the Science Foundation Ireland (SFI) through a Royal Society - SFI University Research Fellowship (M.~Fraser);
\item the Israel Ministry of Science and Technology through grant 3-18143 and the Tel Aviv University Center for Artificial Intelligence and Data Science (TAD) through a grant;
\item the Agenzia Spaziale Italiana (ASI) through contracts I/037/08/0, I/058/10/0, 2014-025-R.0, 2014-025-R.1.2015, and 2018-24-HH.0 to the Italian Istituto Nazionale di Astrofisica (INAF), contract 2014-049-R.0/1/2 to INAF for the Space Science Data Centre (SSDC, formerly known as the ASI Science Data Center, ASDC), contracts I/008/10/0, 2013/030/I.0, 2013-030-I.0.1-2015, and 2016-17-I.0 to the Aerospace Logistics Technology Engineering Company (ALTEC S.p.A.), INAF, and the Italian Ministry of Education, University, and Research (Ministero dell'Istruzione, dell'Universit\`{a} e della Ricerca) through the Premiale project `MIning The Cosmos Big Data and Innovative Italian Technology for Frontier Astrophysics and Cosmology' (MITiC);
\item the Netherlands Organisation for Scientific Research (NWO) through grant NWO-M-614.061.414, through a VICI grant (A.~Helmi), and through a Spinoza prize (A.~Helmi), and the Netherlands Research School for Astronomy (NOVA);
\item the Polish National Science Centre through HARMONIA grant 2018/30/M/ST9/00311 and DAINA grant 2017/27/L/ST9/03221 and the Ministry of Science and Higher Education (MNiSW) through grant DIR/WK/2018/12;
\item the Portuguese Funda\c{c}\~{a}o para a Ci\^{e}ncia e a Tecnologia (FCT) through national funds, grants SFRH/\-BD/128840/2017 and PTDC/FIS-AST/30389/2017, and work contract DL 57/2016/CP1364/CT0006, the Fundo Europeu de Desenvolvimento Regional (FEDER) through grant POCI-01-0145-FEDER-030389 and its Programa Operacional Competitividade e Internacionaliza\c{c}\~{a}o (COMPETE2020) through grants UIDB/04434/2020 and UIDP/04434/2020, and the Strategic Programme UIDB/\-00099/2020 for the Centro de Astrof\'{\i}sica e Gravita\c{c}\~{a}o (CENTRA);  
\item the Slovenian Research Agency through grant P1-0188;
\item the Spanish Ministry of Economy (MINECO/FEDER, UE), the Spanish Ministry of Science and Innovation (MICIN), the Spanish Ministry of Education, Culture, and Sports, and the Spanish Government through grants BES-2016-078499, BES-2017-083126, BES-C-2017-0085, ESP2016-80079-C2-1-R, ESP2016-80079-C2-2-R, FPU16/03827, PDC2021-121059-C22, RTI2018-095076-B-C22, and TIN2015-65316-P (`Computaci\'{o}n de Altas Prestaciones VII'), the Juan de la Cierva Incorporaci\'{o}n Programme (FJCI-2015-2671 and IJC2019-04862-I for F.~Anders), the Severo Ochoa Centre of Excellence Programme (SEV2015-0493), and MICIN/AEI/10.13039/501100011033 (and the European Union through European Regional Development Fund `A way of making Europe') through grant RTI2018-095076-B-C21, the Institute of Cosmos Sciences University of Barcelona (ICCUB, Unidad de Excelencia `Mar\'{\i}a de Maeztu’) through grant CEX2019-000918-M, the University of Barcelona's official doctoral programme for the development of an R+D+i project through an Ajuts de Personal Investigador en Formaci\'{o} (APIF) grant, the Spanish Virtual Observatory through project AyA2017-84089, the Galician Regional Government, Xunta de Galicia, through grants ED431B-2021/36, ED481A-2019/155, and ED481A-2021/296, the Centro de Investigaci\'{o}n en Tecnolog\'{\i}as de la Informaci\'{o}n y las Comunicaciones (CITIC), funded by the Xunta de Galicia and the European Union (European Regional Development Fund -- Galicia 2014-2020 Programme), through grant ED431G-2019/01, the Red Espa\~{n}ola de Supercomputaci\'{o}n (RES) computer resources at MareNostrum, the Barcelona Supercomputing Centre - Centro Nacional de Supercomputaci\'{o}n (BSC-CNS) through activities AECT-2017-2-0002, AECT-2017-3-0006, AECT-2018-1-0017, AECT-2018-2-0013, AECT-2018-3-0011, AECT-2019-1-0010, AECT-2019-2-0014, AECT-2019-3-0003, AECT-2020-1-0004, and DATA-2020-1-0010, the Departament d'Innovaci\'{o}, Universitats i Empresa de la Generalitat de Catalunya through grant 2014-SGR-1051 for project `Models de Programaci\'{o} i Entorns d'Execuci\'{o} Parallels' (MPEXPAR), and Ramon y Cajal Fellowship RYC2018-025968-I funded by MICIN/AEI/10.13039/501100011033 and the European Science Foundation (`Investing in your future');
\item the Swedish National Space Agency (SNSA/Rymdstyrelsen);
\item the Swiss State Secretariat for Education, Research, and Innovation through the Swiss Activit\'{e}s Nationales Compl\'{e}mentaires and the Swiss National Science Foundation through an Eccellenza Professorial Fellowship (award PCEFP2\_194638 for R.~Anderson);
\item the United Kingdom Particle Physics and Astronomy Research Council (PPARC), the United Kingdom Science and Technology Facilities Council (STFC), and the United Kingdom Space Agency (UKSA) through the following grants to the University of Bristol, the University of Cambridge, the University of Edinburgh, the University of Leicester, the Mullard Space Sciences Laboratory of University College London, and the United Kingdom Rutherford Appleton Laboratory (RAL): PP/D006511/1, PP/D006546/1, PP/D006570/1, ST/I000852/1, ST/J005045/1, ST/K00056X/1, ST/\-K000209/1, ST/K000756/1, ST/L006561/1, ST/N000595/1, ST/N000641/1, ST/N000978/1, ST/\-N001117/1, ST/S000089/1, ST/S000976/1, ST/S000984/1, ST/S001123/1, ST/S001948/1, ST/\-S001980/1, ST/S002103/1, ST/V000969/1, ST/W002469/1, ST/W002493/1, ST/W002671/1, ST/W002809/1, and EP/V520342/1.
\end{itemize}

The GBOT programme  uses observations collected at (i) the European Organisation for Astronomical Research in the Southern Hemisphere (ESO) with the VLT Survey Telescope (VST), under ESO programmes
092.B-0165,
093.B-0236,
094.B-0181,
095.B-0046,
096.B-0162,
097.B-0304,
098.B-0030,
099.B-0034,
0100.B-0131,
0101.B-0156,
0102.B-0174, and
0103.B-0165;
%
% From Martin Altmann, 13 March 2019:
%  092.B-0165   01.10.13 - 31.03.14
%  093.B-0236   01.04.14 - 30.09.14
%  094.B-0181   01.10.14 - 31.03.15
%  095.B-0046   01.04.15 - 30.09.15
%  096.B-0162   01.10.15 - 31.03.16
%  097.B-0304   01.04.16 - 30.09.16
%  098.B-0030   01.10.16 - 31.03.17
%  099.B-0034   01.04.17 - 30.09.17
% 0100.B-0131   01.10.17 - 31.03.18
% 0101.B-0156   01.04.18 - 30.09.18
% 0102.B-0174   01.10.18 - 31.03.19
% 0103.B-0165   01.04.19 - 30.09.19
%
and (ii) the Liverpool Telescope, which is operated on the island of La Palma by Liverpool John Moores University in the Spanish Observatorio del Roque de los Muchachos of the Instituto de Astrof\'{\i}sica de Canarias with financial support from the United Kingdom Science and Technology Facilities Council, and (iii) telescopes of the Las Cumbres Observatory Global Telescope Network.

\end{appendix}

\end{document}